\documentclass[a4paper,10pt]{article}
\usepackage[margin=1in,footskip=0.5in]{geometry}
 \usepackage[bottom]{footmisc}%
 \usepackage{setspace}
 \usepackage{amsmath}
\usepackage{graphicx}  
\usepackage{float}  
\usepackage{subfigure}  
\usepackage[utf8]{inputenc} % allow utf-8 input
\usepackage[T1]{fontenc}    % use 8-bit T1 fonts
\usepackage{hyperref}       % hyperlinks
\usepackage{url}            % simple URL typesetting
\usepackage{booktabs}       % professional-quality tables
\usepackage{amsfonts}       % blackboard math symbols
\usepackage{nicefrac}       % compact symbols for 1/2, etc.
\usepackage{microtype}      % microtypography
\usepackage{lipsum}		% Can be removed after putting your text content
\usepackage{amsthm}
 \usepackage{multirow}
 \usepackage{appendix}
 \usepackage{mathrsfs}
 \usepackage{amssymb}
 \usepackage{xcolor}
  \usepackage{natbib}
  \usepackage{enumitem}

 \theoremstyle{plain}
\newtheorem{theorem}{Theorem}

\newtheorem{remark}{Remark}

\newtheorem{lemma}{Lemma}

\newtheorem{definition}{Definition}
\newtheorem*{definition*}{Definition}

\def\m{\mathcal}
\def\dd{{\rm d}}
\def\mb{\mathbb}

\def\ms{\mathscr}

\def\wt{\widetilde}
\def\wh{\widehat}
\def\sk{{k}}

\def\ov{\overline}

\def\sm{{k}}
\def\wh{\widehat}
\newcommand{\be}{\begin{equs}}
\newcommand{\ee}{\end{equs}}

\usepackage[plain,noend]{algorithm2e}

 \title{Estimating Distributions with Low-dimensional Structures Using Mixtures of Generative Models}
\author{Rong Tang and Yun Yang}
\date{ University of Illinois Urbana Champaign}

\begin{document}
 \maketitle
\begin{abstract}
There has been a growing interest in statistical inference from data satisfying the so-called manifold hypothesis, assuming data points in the high-dimensional ambient space to lie in close vicinity of a submanifold of much lower dimension. In machine learning,  encoder-decoder pair based generative modelling approaches have been successful in learning complicated high-dimensional distributions such as those over images and texts by explicitly imposing the low-dimensional manifold structure.  In this work,  we introduce a new approach for estimating distributions on unknown submanifolds via mixtures of generative models. We show that conventional generative modeling approaches using a single encoder-decoder pair are generally unable to capture data distributions under the manifold hypothesis, unless the underlying manifold admits a global parametrization; however, this issue can be solved by using a collection of encoder-decoder pairs for learning different local patches of the data supporting manifold.  A rigorous theoretical analysis is developed to demonstrate that the proposed estimator attains the minimax-optimal rate of convergence for the implicit estimation of data distributions with manifold structures.  Our experiments show that, by utilizing parameter sharing, the proposed method can significantly improve the performance of conventional auto-encoder based generative modelling approaches with minimal additional computational efforts.
\end{abstract}

\section{Introduction} 
Modelling and estimating complicated high-dimensional distributions with low-dimensional structures remains one of the major challenges in modern statistical learning. Suppose we observe $n$ i.i.d. samples $\{X_1,X_2,\ldots,X_n\}$ living in an ambient Euclidean space $\mb R^D$ according to some unknown distribution $\mu^*$. We wish to estimate $\mu^*$ based on the samples for conducting statistical inference and generating new samples.  One of the most popular nonparametric methods for distribution estimation is kernel density estimation (KDE). It has been shown that when $\mu^*$ admits a density function relative to the Lebesgue measure of $\mb R^D$, and the density function is $\beta$-smooth, then KDE can achieve the optimal rate $n^{-\frac{\beta}{2\beta+D}}$ for recovering the density value at any point in $\mb R^D$ ~\citep{silverman2018density,Tsybakov2009}. However, the non-parametric rate $n^{-\frac{\beta}{2\beta+D}}$ suffers from the curse of dimensionality as the ambient dimension $D$ appears in the rate exponent and can be enormous in machine learning applications involving images and texts~\citep{https://doi.org/10.48550/arxiv.1809.11096,https://doi.org/10.48550/arxiv.1609.03499}. In order to avoid this exponential blow-up of the dimension,  a common practice is to assume some additional structure in the data so that the effective dimension of the data space is relatively low.  
 
 One such structure that has attracted much attention recently is the so-called manifold hypothesis, which assumes the date to live on a $d$-dimensional submanifold $\m M$ embedded in the possibly high-dimensional ambient space $\mb R^D$.  Although submanifolds have more complicated geometry than the conventional Euclidean spaces, the manifold hypothesis is a natural assumption to make in a number of areas of science and technology.  For example, in computer vision and medical imaging, data are usually images represented as vectorized pixel intensities. 
 Although images may contain millions of pixels, it is usually determined by a comparatively smaller set of global characteristics such as  camera projection, lighting condition, texture, object position and orientation.  Other examples of high-dimensional complex data with low-dimensional manifold structures appear in natural language processing~\citep{luo2020neural,ling2017integrating}, protein-protein interaction detection~\citep{you2010using,terradot2004biochemical}, and astronomy and shape analysis~\citep{mardia1999directional,jupp2009directional}.
 
 Statistical theory and methodology for modeling manifold valued data have been developed in various contexts~\citep{lin2020robust,lin2017bayesian,https://doi.org/10.1111/rssc.12521,https://doi.org/10.1002/cjs.11601,Divol2022,tang2022minimax,berenfeld2022estimating}.
 Specifically, the problem of estimating a probability measure lying on an unknown low-dimensional Riemannian submanifold has been studied in a number of recent works. For example,~\citet{Divol2022} consider a kernel density type estimator based on a preliminary step of estimating the volume measure of the submanifold using local polynomial estimation techniques. They prove that the developed estimator can achieve the minimax-optimal error bound under the Wasserstein loss.~\citet{tang2022minimax} construct a two-step estimator: the first step estimates the data supporting submanifold; and the second step recovers the distribution on the estimated submanifold based on wavelet type estimators. They also show that such an estimation is minimax-optimal with respect to certain adversarial loss functions.
 ~\citet{berenfeld2022estimating} develop a Bayesian procedure based on location-scale mixtures of Gaussians for estimating the density of data living close to an unknown submanifold with theoretical guarantees.  However, although these existing methods are theoretically appealing, they usually have poor computational scalability with the ambient dimensionality and the sample size, making them costly to implement for modeling massive and high-dimensional real data, such as images and texts.

Auto-encoder based deep generative
modeling approaches in the machine learning literature, such as variational auto-encoder (VAE)~\citep{kingma2013autoencoding,rezende2014stochastic,NIPS2016_ddeebdee}, Wasserstein auto-encoder (WAE)~\citep{tolstikhin2019wasserstein}, InfoVAE~\citep{Zhao_Song_Ermon_2019} and inferential Wasserstein generative adversarial networks (iWGAN)~\citep{https://doi.org/10.1111/rssb.12476}, have achieved great successes in generating synthetic realistic-looking images and texts, and are usually very efficient to implement. However, despite their empirical successes, a general theoretical framework explaining whether and how these generative modelling approaches
benefit from the low-dimensional manifold structure is lacking, and it is also not clear whether these existing methods are theoretically optimal in the minimax sense. For example, the key step in the auto-encoder is the extraction of $d$ ($d\ll D$) latent features (via an encoder $Q:\,\mb R^D\to\mb R^d$) that can be used for accurately reconstructing the original data (via a decoder $G:\,\mb R^d\to \mb R^D$). In other words, these auto-encoder based methods implicitly assume data $X$ to have a low-dimensional structure so that they can be accurately reconstructed in the sense that $X\approx G(Q(X))$.  Moreover, it is often the case that real-world data falls on a manifold that does not admit a global parametrization. For example, when the data space is a boundaryless manifold such as a sphere, or disconnected~\citep{NEURIPS2018_2b346a0a}. This lack of global parametrization makes conventional auto-encoder methods equipped with a single encoder/decoder pair incapable of recovering the entire data space without incurring distortions. 
Our empirical results (c.f.~Fig.~\ref{fig:sphere_smoothing}) also suggest that conventional auto-encoder based generative modelling approaches tend to generate off real-manifold samples with unrealistic appearances.

In this article, we propose a new generative modelling approach for learning manifold-supported distributions that is theoretically minimax-optimal, computationally efficient, and empirically promising in generating complicated yet realistic-looking data.  Unlike most existing generative modelling procedures that rely on the strong assumption of the existence of a global parametrization of the data space,  we employ multiple encoder/decoder pairs, where each pair corresponds to the parametrization of a local patch of the data supporting manifold. Moreover, we utilize the partition of unity technique for gluing local probability measures estimated in the patches to form a global estimation of the probability measure on the manifold. 
In addition, most existing methods simply plug in the data empirical distribution in constructing the objective function for defining a GAN \citep{goodfellow2014generative} type estimator, which may lead to theoretical deficiency due to the failure of taking the smoothness of the target distribution into account. 
We instead propose to plug in a smoothness-regularized version that provably improves the estimation accuracy.  Concretely, we show that when the target distribution is $\alpha$-smooth and lies in a $\beta$-smooth $d$-dimensional submanifold in $\mb R^D$, then the corresponding estimator $\wh \mu$ based on $n$ data points achieves a non-asymptotic error bound of order $O\big(\frac{\log n}{\sqrt{n}}\vee n^{-\frac{\alpha\wedge (\beta-1)+1}{2(\alpha\wedge(\beta-1))+d}}\big)$ (here $a\wedge b$ denotes $\min\{a,b\})$ under the $1$-Wasserstein distance, which corresponds to the minimax rate modulo a logarithmic factor when $\alpha\leq \beta-1$. The implied rate of convergence does not suffer from the ``curse of dimensionality'' and only depends on the intrinsic dimensionality $d$ of the data.  Our numerical results also show that the proposed method tends to be more accurate than conventional auto-encoder based generative modelling approaches and classic kernel density estimators for learning target distributions with low-intrinsic dimensional structures.

 \subsection{Notation}\label{Notation}
We summarize some necessary notations and definitions here.  For any positive integer $k$,  we use the shorthand $[k]:=\{1,2,\cdots,k\}$. We use $\|\cdot\|_p$ to denote the usual vector $\ell_p$ norm, and reserve $\|\cdot\|$ for the $\ell_2$ norm. We use $\mb S_1^{d}=\{x\in \mb R^{d+1}\,:\, \|x\|=1\}$ to denote the $d$-dimensional unit sphere in $\mb R^{d+1}$. For a probability measure $\mu$, we use ${\rm supp}(\mu)$  to denote its support. For any measure $\nu$  and map $G$, the push-forward measure $\mu = G_{\#}\nu$ is defined as the unique measure  such that $\mu(A)=\nu(G^{-1}(A))$ holds for any measurable set $A$.
For two probability measures $\mu,\nu$, the $1$-Wasserstein distance  between $\mu$ and $\nu$ is defined as $W_{1}(\mu, \nu)=\inf\big\{\int \|x-T(x)\|_1\,\dd \mu(x)\,:\, T_{\#}\mu=\nu\big\}=\sup \big\{\int f(x) \dd(\mu-\nu)\,: \operatorname{Lip}(f) \leq 1\big\}$, where $\operatorname{Lip}(f)$ denotes the minimal Lipschitz constant for $f$. When no ambiguity arises, for an absolutely continuous probability measure $\nu$,  we may also use $\nu$ to refer to its density function. We use  $\m P(\Omega)$ to denote the set of probability  measures on $\Omega$.  We use $\mb B_r(x)$ to denote the closed ball centered at $x$ with radius $r$ under the $\ell_2$ distance. We use $C^{\alpha}_r(\Omega)$ to denote the set of all $\alpha$-H\"{o}lder smooth functions with H\"{o}lder norm $\|\cdot\|_{C^{\alpha}(\Omega)}$ being bounded by $r$ (see for example,~\cite{evans2010partial}). Similarly, we use $C^{\alpha}_r(\Omega; \mb R^D)=\big\{f=(f_1,\ldots,f_D):\, \Omega\to \mb R^D\,\big|\, \forall \,j \in [D],\, f_j\in C^{\alpha}_{r}(\Omega)\big\}$ to denote the vector valued function space counterpart. 
%   Note that the manifold structure is an intrinsic property that does not rely on the choice of the atlas.
\subsection{Organization}

The rest of the paper is organized as follows. In Section 2, we give a brief introduction to the  auto-encoder based generative modelling approaches and define smooth distributions on manifolds.  Our proposed model is introduced in Section 3, and its implementation and theoretical properties are described in Section 4 and Section 5, respectively.  Simulations and a real data application are provided in Section 6 and Section 7.

\section{Background}\label{background}
 
 \subsection{Auto-encoder based generative modelling approaches}

Assume i.i.d.~data $X^{(n)}=\{X_1,X_2,\cdots,X_n\}$ sampled from an unknown target distribution $\mu^\ast$ over data space $\m X\subset\mb R^D$ are available.  In the literature of generative modelling, the target distribution $\mu^*$ is implicitly specified by its sampling scheme, represented by a generative model. Mathematically, a generative model is defined as a pair $(\nu_0, G)$, where $\nu_0$  is a distribution on a low-dimensional latent space $\m Z\subset \mb R^d$, called \emph{generative distribution}, that is easy to sample from; and $G: \m Z\to \m X$ is a map from $\m Z$ to $\m X$, called \emph{generative map}, so that if $Z \sim \nu_0$, then $G(Z) \sim \mu^\ast$.  The goal of generative modelling is to fit a generative model that specifies a stochastic process whose simulated data look indistinguishable from real data. In particular, auto-encoder based generative modelling approaches introduce a family of encoders $Q$ that send the data $X$ to the (low-dimensional) latent variables $Z$, and a family of decoders $G$ that reconstruct the data from the latent variables, so that they jointly minimizes the following objective: 
\begin{align*}
   \frac{1}{n}\sum_{i=1}^n c\big(X_i,G(Q(X_i))\big)+{\rm Penalty \, term},\\[-2em]
\end{align*}
where  $c(\cdot,\cdot)$  is a cost function. The first component $ n^{-1}\sum_{i=1}^n c(X_i,G(Q(X_i)))$ of the objective function corresponds to the reconstruction cost: a common choice is the squared loss $c(x,y)=\|x-y\|^2_2$.  This reconstruction cost enforces the push-forward measure of  $(G\circ Q)_{\#}\mu^*$ to be as close to $\mu^*$ as possible. On the other hand, the second component involves a penalty term for regularizing the encoder/decoder pair. For example, in VAE~\citep{kingma2013autoencoding,rezende2014stochastic,NIPS2016_ddeebdee}, the penalty term is chosen to be an averaged Kullback-Leibler (KL) divergence between the latent variable distribution induced by the (probabilistic) encoder and a prior distribution $\nu_0$; in iWGAN~\citep{https://doi.org/10.1111/rssb.12476}, the penalty term is chosen to be an approximation to the $1$-Wasserstein distance between the reconstructed data distribution $(G\circ Q)_{\#}\mu^*$ and induced distribution from the generative model $G_{\#}\nu_0$ using prior $\nu_0$. Other approaches choose penalty terms for directly matching the encoder-induced latent variable distribution and a given prior in the latent space (for example, WAE,~\cite{tolstikhin2019wasserstein}; infoVAE,~\cite{Zhao_Song_Ermon_2019}; Sliced WAE,~\cite{kolouri2018sliced}), which leads to the following training objective:
\begin{equation}\label{train_obj}
  \frac{1}{n}\sum_{i=1}^n c\big(X_i,G(Q(X_i))\big)+\lambda\cdot \m D(Q_{\#}\widehat{\mu}_{\rm em},\nu_0),
\end{equation}
where $\widehat\mu_{\rm em}:\,=n^{-1}\sum_{i=1}^n\delta_{X_i}$ denotes the discrete empirical distribution of the data $X^{(n)}$, and $\m D$ is a generic discrepancy metrics characterizing closeness between distributions over the latent space.   We will call any method whose training objective takes the form as~\eqref{train_obj} a latent distribution matched auto-encoder  (LDMAE) for future reference. Ideally, the learned decoder $\wh G$ from minimizing~\eqref{train_obj} has the property that  $\wh\mu=\wh G_{\#}\nu_0 \approx \mu^\ast$. Comparing with approaches directly dealing with distributions on the ambient space~\citep{goodfellow2014generative,arjovsky2017wasserstein,NEURIPS2018_2b346a0a}, the latent distribution matching schemes bring several computational benefits. First, the choice of $\m D$ for quantifying the discrepancy between distributions in the latent space $\m Z$ is more flexible since these distributions usually admit a density function relative to the Lebesgue measure over $\m Z$; in contrast, many commonly used discrepancies metrics such as the total variation distance, the Hellinger distance and the KL divergence are known to be unsuitable for characterizing closeness between nearly singular measures over the ambient space~\citep{li2017mmd,xu2018empirical}. 
Second, computing a discrepancy between distributions in the relatively low-dimensional latent space is much more efficient and does not suffer from the curse of dimensionality, make the training process more stable and less time-consuming. In addition, the extra flexibility of $\m D$ allows one to employ those metrics that have simple and explicit computational formulas, such as the squared maximum mean discrepancy (MMD)~\citep{tolstikhin2019wasserstein,Zhao_Song_Ermon_2019} and the sliced Wasserstein distance~\citep{kolouri2018sliced}.

 At the end of this subsection, we describe two important limitations of LDMAE, which motivate our proposed method to be introduced in Section~\ref{MLDMAE}.
 Concretely, based on the aforementioned decomposition perspective of the training objective~\eqref{train_obj}, the estimation error of $\wh\mu$ from the target distribution $\mu^\ast$ depends on two terms: (1) the ``distance'' between $\mu^\ast$ and $(\wh G \circ\wh Q)_{\#}\mu^*$; (2) the ``distance'' between $\wh Q_{\#}\mu^*$ and $\nu_0$, where $\wh Q$ denotes the learned encoder from minimizing~\eqref{train_obj}.  
 Let $\m M$ denote the support of the true data generating distribution $\m \mu^\ast$ as a $d$-dimensional submanifold embedded in $\mb R^D$.
 To  control the first distance, one needs $\m M$ to have a global parametrization, that is, we can find some continuous maps $G^\ast:\mb R^d\to \mb R^D$ and $Q^\ast:\mb R^D\to \mb R^d$ so that for any $x\in \m M$, $G^\ast(Q^\ast(x))=x$. This global-parametrization condition does not hold for many common manifolds, such as disconnected manifolds and boundaryless manifolds like spheres and torus. The second distance depends on how well the empirical data distribution $\widehat\mu_{\rm em}$ induced empirical latent distribution $\wh Q_{\#}\wh{\mu}_{\rm em}$ can approximate the population level distribution $\wh Q_{\#}\mu^\ast$. However, even though we assume that manifold $\m M$ admits a global parameterization such that $\mu^*=G^*_{\#}\nu_0$ and $Q^*=(G^*)^{-1}$ for some $G^\ast$, $Q^\ast$ in the decoder and encoder families, the discrete empirical distribution $\widehat\mu_{\rm em}$ may suffer from statistical deficiency for approximating a smooth measure~\citep{liang2020generative,tang2022minimax}. As a result, simply plugging-in $Q_{\#}\wh{\mu}_{\rm em}$ in the penalty term may lead to overfitting.

 \subsection{Partition of unity and distributions on manifolds}
 Intuitively speaking, a manifold is a topological space that locally resembles the Euclidean
space. Formally, we have the following mathematical definition of a manifold.
  \begin{definition}
  A $d$-dimensional manifold $\m M$ is defined as a topological space satisfying:(1) There exists an atlas on $\m M$ consisting of a collection of $d$-dimensional charts $\ms A = \{(U_\lambda, \varphi_\lambda)\}_{\lambda\in \Lambda}$ covering $\m M$, that is, $\m M = \bigcup_{\lambda\in\Lambda} U_\lambda$. (2) Each chart \footnote{Subscript $\lambda$ is suppressed for the simplicity of notation.}$(U, \varphi)$ in atlas $\ms A$ consists of a homeomorphism $\varphi:\, U \to \widetilde U$, called coordinate map, from an open set $U \subset \m M$ to an open set $\widetilde U \subset \mb R^d$. 
 \end{definition}
We call a manifold $\m M$ a ($\beta$-smooth) submanifold embedded on $\mb R^D$ if $\m M\subset \mb R^D$, and the coordinate map $\varphi$ and its inverse $\varphi^{-1}$ in each chart are $\beta$-smooth maps when identified as  functions  defined on subsets of Euclidean spaces.  Another useful notion related to the manifold is \emph{partition of unity}.
\begin{definition}
 A partition of unity of a manifold $\mathcal{M}$ is a collection of functions $\{\rho_\lambda\}_{\lambda\in \Lambda}$ satisfying
 
\quad{1. $0\leq \rho_\lambda\leq 1$ for all $\lambda\in \Lambda$, and $\sum_{\lambda\in \Lambda}\rho_\lambda(x)=1$ for all $x\in \mathcal{M}$.}

\vspace{0.5em}
\quad{2. Each point $x\in \mathcal{M}$ has a neighborhood which intersects ${\rm supp}(\rho_\lambda)$ for only finitely many $\lambda\in \Lambda$.}
  
 \end{definition}

\noindent Using the partition of unity, one can glue constructions in the local charts to form a global construction on the manifold. A partition of unity can be constructed from any open cover $\{U_{\lambda}\}_{\lambda\in\Lambda}$ of the manifold in a way where the partition $\{\rho_\lambda\}_{\lambda\in\Lambda}$ is indexed over the same set and ${\rm supp}(\rho_\lambda)\subset U_{\lambda}$ for any $\lambda\in\Lambda$. Such a partition of unity is said to be \emph{subordinate to} the open cover $\{U_{\lambda}\}_{\lambda\in\Lambda}$.  
  
  For a manifold $\m M$ with atlas $\ms A = \{(U_\lambda, \varphi_\lambda)\}_{\lambda\in \Lambda}$, suppose $\Lambda$ is finite and we write it as $\Lambda=[K]$.  Given a partition of unity $\{\rho_k\}_{k\in [K]}$ subordinate to the open cover $\{U_k\}_{k\in [K]}$, one can decompose any distribution $\mu^\ast$ on $\m M$ as 
\begin{equation}\label{eqnpou}
    \begin{aligned}
 \mu^\ast  = \sum_{k\in [K]} \rho_{k} \mu^\ast =\sum_{k\in [K]} (\varphi_k^{-1})_{\#}\big[(\varphi_k)_\# (\rho_k \mu^\ast)\big],
    \end{aligned}
\end{equation}
% where $\rho_k\mu$ stands for the non-negative measure whose Radon-Nikodym derivative relative to $\mu$ is $\rho_k$
where the first inequality uses $\sum_{k\in [K]}\rho_k(x)=1$ for all $x\in \m M$, and the second inequality uses the fact that ${\rm supp}(\rho_k)\subset U_k$ and $\varphi_k$ is a homeomorphism on $U_k\to \varphi_k(U_k)$.\footnote{If $(\varphi_k)_\# (\rho_k \mu^\ast)$ admits an $\alpha$-smooth density function for $\alpha\in [0,\beta-1]$  relative to the
Lebesgue measure on $\mb R^d$ for each $k\in [K]$, then $\mu^\ast$ is said to be an $\alpha$-smooth distribution on $\m M$.} Then if we write $\nu^\ast_{k}=(\varphi_k)_\# \big(\frac{\rho_k \mu^\ast}{\mb{E}_{\mu^\ast}[\rho_{k}]}\big)$, $\mu^\ast$ can be expressed as the following (mixture of) generative models:
\begin{equation}\label{eqngen}
    \mu^\ast=\sum_{k\in [K]} \mb{E}_{\mu^\ast}[\rho_{k}]\cdot (\varphi_k^{-1})_{\#}\nu_{k}^\ast,
\end{equation}
% which is characterized by generative maps $\{\varphi_k^{-1}\}_{k\in [K]}$, generative distributions $\{\nu^\ast_k\}_{k\in [K]}$ and weights $\{\mb{E}_{\mu^*}[\rho_k]\}_{k\in [K]}$. The decomposition~\eqref{eqngen} is not unique. For example, consider any set of diffeomorphisms $\{\phi_k\}_{k\in [K]}$  of $\mb R^d$, then  decomposition~\eqref{eqngen} can be rewritten as 
%     \begin{equation*} 
%     \mu^\ast=\sum_{k\in [K]} \mb{E}_{\mu^\ast}[\rho_{k}]\cdot (\varphi_k^{-1}\circ \phi _k^{-1})_{\#}((\phi _k)_{\#}\nu^\ast_{k})=\sum_{k\in [K]} \mb{E}_{\mu^\ast}[\rho_{k}]\cdot (\wt\varphi_k^{-1})_{\#}(\wt\nu_{k}),
% \end{equation*}
%     with $\wt \nu_k=(\phi _k)_{\#}\nu^\ast_{k}$ and $\wt\varphi_k=\phi _k\circ \varphi_k$. 
Decomposition~\eqref{eqngen} suggests that any distribution $\mu^\ast$ lying on a $d$-dimensional submanifold embedded on $\mb R^D$ whose atlas composed of at most $K$-number of charts belongs to the following mixture of generative models class:
 \begin{equation*}
     \m S^*=\Big\{\mu=\sum_{k\in[K]} p_k \cdot (G_k)_{\#}\nu_k\,\Big|\, G_k:\mb R^d\to \mb R^D, \nu_k\in \m P(\mb R^d), 0\leq p_k\leq 1, \sum_{k\in [K]} p_k=1\Big\}.
 \end{equation*}
This space forms our model space representing distributions on manifolds.

\section{Mixture of latent distribution matched auto-encoder}\label{MLDMAE}
From discussions in Section~\ref{background}, we see that conventional auto-encoder based generative modelling approaches may suffer from low representation power when the target distribution to be estimated lies on a general submanifold without global parametrization. However, the property that any manifold-supported distribution can be expressed in the form of a mixture of generative models (c.f. decomposition~\eqref{eqngen}) motivates us to employ multiple encoder/decoder pairs, and to use the partition of unity to glue them together with proper weights.
 
Recall that we have a set of $n$ i.i.d observations $X^{(n)}=\{X_1,\ldots,X_n\}$ sampled from the target distribution $\mu^*$ lying on a $d$-dimensional submanifold $\m M$ embedded in $\mb R^D$ with $d\leq D$.  Let $\{S_k\}_{k\in [K]}$  be a suitably chosen open cover to $\m M={\rm supp}(\mu^\ast)\subset \mb R^D$, fix a partition of unity $\{\rho_k\}_{k\in [K]}$ subordinate to $\{S_k\}_{k\in [K]}$,\footnote{Here we may consider any function $\rho_k: \mb R^D\to [0,1]$ so that $\text{supp}(\rho_k)\subset S_k\subset \mb R^D$ and $\sum_{k\in [K]}\rho_k(x)=1$ for any $x\in \cup_{k\in [K]} S_k$. Note that $\{\rho_k|_{S_k\cap \m M}\}_{k\in [K]}$ would form a partition of unity to $\m M$.} which can be chosen without the knowledge of $\m M$ (c.f.~Section~\ref{computation}). For any generic approximation family $\m G$ consists of  $(\bold{G},\bold{Q},\bold{v})$, where $\bold{G}=(G_{1}, G_{2},\cdots,G_{K})$  with $G_{\sk}:\mb R^d\to \mb R^D$, $\bold{Q}=(Q_{1}, Q_{2},\cdots,Q_{K})$ with $Q_{\sk}:\mb R^D\to \mb R^d$, and $\bold{v}=(\nu_{1}, \nu_{2},\cdots,\nu_{K})$ with $\nu_{\sk}\in \m P(\mb R^d)$, we define the following estimator, which we call mixture of latent distribution matched auto-encoder (MLDMAE) estimator:
\begin{equation}\label{MLDMAEest}
    \begin{aligned}
    &\qquad \wh\mu=\sum_{k\in[K]} \wh{p}_k \cdot (\wh{G}_{k})_{\#}\wh{\nu}_k,\qquad 
    \text{with} \quad\wh{p}_k=\frac{1}{n}\sum_{i=1}^n \rho_k(X_i)\quad\mbox{and}\\
    & (\wh{\bold{G}},\wh{\bold{Q}},\wh{\bold{v}})=\underset{(\bold{G},\bold{Q},\bold{v})\in \m G}{\arg\min}\sum_{k=1}^K \bigg\{\frac{1}{n}\sum_{i=1}^n c\big(X_i,G_k(Q_k(X_i))\big)\cdot\rho_k(X_i)+\lambda_k \cdot \m D\big(\wt{\nu}_{k,Q_k},\nu_k\big)\bigg\},
    % \cdot\Big[\underset{f\in\m F}{\sup}\,\Big(\int_{\mb R^d} f(z)\nu_k(z)\,\dd z -\int_{\mb R^d} f(z)\wt{\nu}_{k,Q_k}(z)\,\dd z\Big)\Big]^2\bigg\},\\
    % &\wh{\bold{G}}=(\wh G_{1}, \wh G_{2},\cdots,\wh G_{[K]}), \wh{\bold{Q}}=(\wh Q_{1}, \wh Q_{2},\cdots,\wh Q_{[K]}), \wh{\bold{v}}=(\wh \nu_{1}, \wh \nu_{2},\cdots,\wh \nu_{[K]}),\\
    \end{aligned}
\end{equation}
 where recall that $c(\cdot,\cdot)$ is the cost function, $\wt{\nu}_{k, Q_k}$ is a (smoothness-regularized) estimator to the density of $(Q_k)_{\#}(\frac{\mu^*\cdot\rho_k}{\mb{E}_{\mu^\ast}[\rho_k]})$ (the precise definition is available in Appendix C), and $\m D(\cdot,\cdot)$ is a generic discrepancy measure between distributions on the latent space. 
 Different from conventional LDMAE estimators, MLDMAE can also use empirical Bayes method to select data-dependent prior distributions for local latent variables, which adds extra flexibility in the modeling and may potentially reduce the approximation error.
 We show in Theorem~\ref{upperbound} that for some carefully chosen approximation family $\m G$, cost function $c$, discrepancy metric $\m D$, and smoothness-regularized estimator $\wt{\nu}_{k, Q_k}$, the resulting estimator $\wh \mu$ 
 attains the minimax rate of convergence under the $1$-Wasserstein distance as $W_1(\wh \mu, \mu^\ast)\leq C\,n^{-\frac{\alpha\wedge  (\beta-1)+1}{2(\alpha\wedge  (\beta-1))+d}} \vee\frac{\log n}{\sqrt{n}}$ when $\mu^\ast$ is an $\alpha$-smooth distribution on an unknown $\beta$-smooth $d$-dimensional submanifold.
 
\begin{figure}[h]
\vspace{-1.5em}
     \subfigure[Real data]
        { \includegraphics[width=0.3\textwidth]{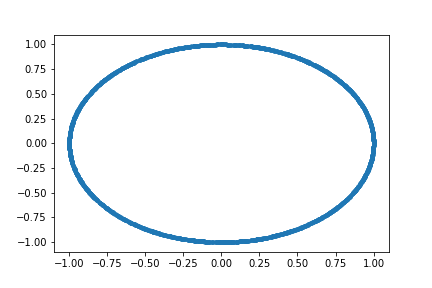}}
         \subfigure[LDMAE]
        { \includegraphics[width=0.3\textwidth]{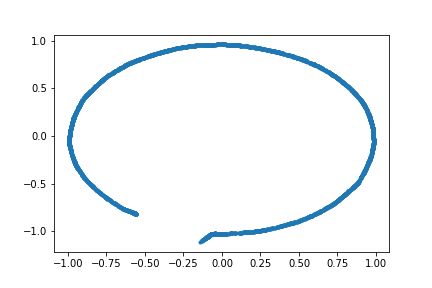}}
        \subfigure[MLDMAE]
        { \includegraphics[width=0.3\textwidth]{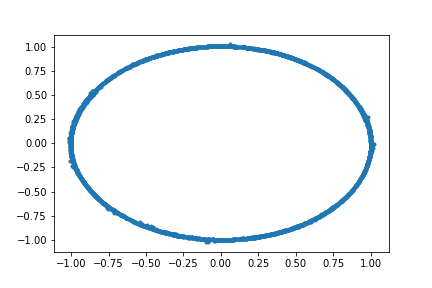}}
  
        \caption{Comparison between LDMAE and MLDMAE when the target distribution is the uniform distribution on a sphere. Figure (a) plots the real data, and Figures (b), (c) plot the randomly generated samples from the MLDMAE and LDMAE estimators respectively, based on $10000$ training samples. The partition of unity chosen in MLDMAE is the smooth partition of unity described in Section~\ref{computation} with $K=10$ and $\gamma=10$. The discrepancy metric $\m D(\cdot,\cdot)$ is chosen to be the MMD with Gaussian kernel.  We can see that LDMAE fails to capture the correct shape of a sphere. The reason is that the sphere (or any boundaryless  manifold) requires at least two covering charts in its describing atlas. The LDMAE model uses a single pair of encoder/decoder, and thus it returns a curve that has start/end points. On the contrary, our estimator is able to learn  general manifolds that can not be globally  parametrized. }
        \label{fig:sphere_smoothing}
\end{figure}

The objective function of MLDMAE can be decomposed into two parts: the reconstruction  cost $n^{-1}\sum_{k=1}^K \sum_{i=1}^n c\big(X_i,G_k(Q_k(X_i))\big)\cdot\rho_k(X_i)$ and the penalty  $\sum_{k=1}^K\lambda_k \cdot \m D\big(\wt{\nu}_{k,Q_k},\nu_k\big)$. We may also allow the latent dimension $d$ to be different across encoder/decoder pairs over $k\in[K]$. The reconstruction cost aims to learn local parametrizations of the supporting manifold of $\mu^\ast$ by enforcing the encoder/decoder pair $(\wh Q_k,\wh G_k)$ to represent some coordinate system $(\varphi_k,\varphi_k^{-1})$
of local patch $U_k=\m M\cap S_k$ of $\m M$ in decomposition~\eqref{eqngen}. Therefore, it corresponds to the support recovery of $\mu^\ast$.  By employing multiple encoder/decoder pairs,  the MLDMAE estimator avoids the restrictive global-parametrization assumption that is implicitly assumed in conventional LDMAE estimators. As a result, MLDMAE is suitable for a wider range of problems (see Fig.~\ref{fig:sphere_smoothing} for an illustration). On the other  hand,  the penalty term aims to enforce the reweighted local latent distribution $(\wh Q_k)_{\#}\big(\frac{\mu^\ast\cdot\rho_k}{\mb{E}_{\mu^*}[\rho_k]}\big)$ to match some member $\wh \nu_k$ in the pre-specified prior family, so that $\wh \nu_k$ is close to the $\nu_k^\ast$ in decomposition~\eqref{eqngen}. 
     
In practice,  instead of selecting the best data-dependent priors, we can also fix the prior as a simple distribution $\nu_0$ such as an isotropic Gaussian. 
% \textcolor{red}{what is the purpose of the next sentence? it seems break the flow and not providing much information} Due to the non-uniqueness of decomposition~\eqref{eqngen}, by fixing the priors,  we then require the MLDMAE estimator $(\wh Q_k, \wh G_k)$ to match  coordinate maps $(\varphi_\sk,\varphi_\sk^{-1})$ satisfying $\nu^\ast_{k}=(\varphi_k)_\# \big(\frac{\rho_k \mu^\ast}{\mb{E}_{\mu^\ast}[\rho_{k}]}\big)\approx \nu_0$. 
Moreover, %even though selecting the discrepancy metric $\m D(\cdot,\cdot)$ as some adversarial losses with complicated discriminator classes, such as Wasserstein distance as suggested in Theorem~\ref{upperbound}, can lead to a theoretically optimal estimator, it  usually requires solving a min-max problem that may make  the training process time-consuming and unstable.  Therefore 
 $\m D(\cdot,\cdot)$ can be chosen as certain squared maximum mean discrepancy (MMD) loss\footnote{ For a positive-definite reproducing kernel $k$, the MMD loss is defined as ${\rm MMD}^2(\mu_1,\mu_2)=\mb{E}_{X,X'\in \mu_1}[k(X,X')]+\mb{E}_{Y,Y'\in \mu_2}[k(Y,Y')]-2\mb{E}_{X\in \mu_1,Y\in \mu_2}[k(X,Y)]$} that can be efficiently computed in a closed-form formula. The $k$-th smoothness-regularized distribution $\wt{\nu}_{k, Q_k}$ in~\eqref{MLDMAEest} can be constructed by applying kernel-smoothing to its (weighted) empirical counterpart $(Q_{\sk})_{\#}\wh{\mu}_{n}^{\sk}$ with $\wh{\mu}_{n}^{\sk}=(n\wh{p}_k)^{-1} \sum_{i=1}^n\rho_k(X_i)\, \delta_{X_i}$, leading to $\wt{\nu}_{k, Q_k}(z)=(n\wh{p}_k)^{-1} \sum_{i=1}^n \wt k(z,Q_k(X_i))\rho_k(X_i)$ for a suitable kernel $\wt k$.
%an adjusted empirical latent distribution $(Q_{\sk})_{\#}\widehat \mu_{\rm em}$, leading to $(Q_{\sk})_{\#}\wh{\mu}_{n}^{\sk}$ with $\wh{\mu}_{n}^{\sk}$ denoting the empirical measure of $\mu^\ast$ reweighted by $\rho_k$ (i.e., $\wt{\nu}_{k, Q_k}(z)=\frac{1}{n\wh{p}_k} \sum_{i=1}^n \wt k(z,Q_k(X_i))\rho_k(X_i)$ for some kernels $\wt k$).  
Note that when kernel $\wt k$ is the Gaussian kernel $\wt k(x,y)= (2\pi h)^{-\frac{d}{2}}\exp(-\frac{\|x-y\|^2}{2h})$ with bandwidth parameter $h$, then $\wt{\nu}_{k, Q_k}$ corresponds to the Gaussian-smoothed distribution $(\wt Q_{\sk})_{\#}\wh{\mu}_{n}^{\sk}$, where $\wt Q_{\sk}$ is the randomly perturbed encoder defined by $\wt Q_k(X)=Q_k(X)+\sqrt{h}\cdot \m N(0,I_d)$.\footnote{Here for randomized map $Q$, the push forward measure $Q_{\#}\mu$ is defined as the measure so that for any measureable function $f$, $\int f(x)\, \dd[Q_{\#}\mu]=\mb {E} \big[\int f(Q(x))\, \dd\mu\big]$ where the expectation is with respect to the randomness of $Q$.}  Employing such a smoothness-regularized distribution can be viewed as applying a randomized data augmentation to increase the variability of the encoded training samples, which mitigates potential overfitting to data and improves the generalization ability of the resulting estimator.

% This kind of randomized encoder is commonly-used in VAEs. However, the difference is that in our framework, the randomness is only for smoothing the empirical distribution and thus the bandwidth (noise level) $h$ will go to zero as the sample size increase. Moreover, the encoders and decoders are deterministic in the reconstruction part, so it does not suffer from model misspecification as VAEs when the target measure lies exactly on a submanifold,\footnote{Note that VAE models require the target measure to be absolutely continuous  with respect to the Lebesgue measure of  $\mb R^D$.} and thus our model will not generate samples as noisy (blurred) as VAEs. 
  
Introducing the encoder-decoder structure as in our estimator brings several benefits. Computationally, the encoders turn the high-dimensional data into low-dimensional latent variables so that we only need to compute a penalty term over low-dimensional distributions.
Therefore, the MLDMAE framework brings less computational burden compared with generative modelling approaches (e.g.~iWGAN) that directly deal with distributions in the ambient space. Theoretically, when $d\ll D$, the data distribution $\mu^*$ becomes a singular measure in $\mb R^D$. As a consequence, with the information about the supporting manifold of $\mu^\ast$, which is explicitly induced by the encoder-decoder pairs, it is possible to utilize classical techniques of nonparametric density estimation, such as wavelet truncation, to construct a minimax-optimal estimator.
Specifically, the underlying true latent variable distribution $(Q_k)_{\#}(\frac{\mu^*\cdot\rho_k}{\mb{E}_{\mu^\ast}[\rho_k]})$ defined in the mixture of generative models~\eqref{eqngen} is, with high probability, absolutely continuous with respect to the Lebesgue measure on $\mb R^d$ (c.f. Lemma 4 in Appendix C),  which enables us to develop smoothness-regularized estimators by borrowing techniques from~\cite{liang2020generative} and~\cite{NEURIPS2018_4996dcc4}. Indeed, as suggested by~\cite{liang2020generative} and~\cite{tang2022minimax}, the rate $O\big(n^{-\frac{\alpha+1}{2\alpha+d}} \vee\frac{\log n}{\sqrt{n}}\big)$ achieved by the MLDMAE estimator when $\alpha\leq \beta-1$ is minimax-optimal up to logarithmic factor relative to the $1$-Wasserstein distance.

\section{Computation}\label{computation}
     
\noindent In this section, we discuss some important computational aspects of the proposed method.
    
\smallskip
\noindent \textbf{Choice of partition of unity:} One issue we need to address is how to choose a reasonable partition of unity $\{\rho_k\}_{k\in [K]}$. To do this, we can first run a clustering algorithm such as (mini-batch) $K$-means to the data using a sufficiently large cluster number $K$. Based on the clustering result, one straightforward choice of $\rho_k$ is the indicator function $\bold{1}(x\in \text{$k$-th cluster})$. We can also choose a smooth partition of unity by the following:  firstly we record the centroid of the $k$-th cluster as $a_k$ and the smallest radius $r_k$ so that data points in the $k$-th cluster are included in $\mb B_{r_k}(a_k)$.  Then we can construct an open cover $\{S_k=\mb B_{r_k+\varepsilon}(a_k)^\circ\}_{k \in[K]}$  where $\varepsilon$ is a small positive number  so that $\{S_k\}_{k \in[K]}$ can cover the unknown support of $\mu^\ast$ with high probability. Given the open cover, for each $k\in [K]$, we define a local partition function as $ \widetilde{\rho}_{k}(x)=((r_k+\varepsilon)^2-\|x-a_k\|^2)^\gamma\cdot \bold{1}(x\in S_k)$, where $\gamma>1$ is a tuning parameter. The resulting $\{{\rho}_k\}_{k\in[K]}$ forms a partition of unity for $\mathcal{M}$ with $\rho_k={\widetilde\rho}_k/\big(\sum_{k'=1}^K\widetilde{\rho}_{k'}\big)$ for $k\in[K]$. Note that  $\wt\rho_k$ tends to give less weight to points away from the centroid $a_k$ for large $\gamma$.
 
 \smallskip
\noindent \textbf{Choice of penalty terms:}
 We choose the smoothness-regularized distribution $\wt{\nu}_{k, Q_k}$ as the Gaussian kernel-smoothed version of  $(Q_{\sk})_{\#}\wh{\mu}_{n}^{\sk}$ as described in Section~\ref{MLDMAE}. This $\wt{\nu}_{k, Q_k}$ relates to the commonly-used Gaussian encoder in VAE, and thus we can utilize the reparametrization trick in VAE to optimize the desired objective function.  For the priors $\bold{v}$, we can consider a simple distribution $\nu_0$ such as standard Gaussian $\m N(0, I_d)$ as is usually done in conventional generative modelling approaches. Moreover, to ensure the smoothness of the learned manifold, we can consider  data-driven priors described in Remark~\ref{remark:prior} in Section~\ref{sec:theory} below,  so that $\nu_k$ and $\wt{\nu}_{\sk, \wh Q_k}$ can be ensured to have matching tails. For the discrepancy metric $\m D(\cdot,\cdot)$, to prevent instability in the adversarial training,  we can: (1) choose $\m D(\cdot,\cdot)$ to be  the (squared) MMD characterized by a positive-definite kernel $k$, such as the inverse multiquadratics (IMQ) kernel  $k(x,y)=C_{im}/(C_{im}+\|x-y\|^2_2)$ and the RBF kernel $k(x,y)=\exp(-\|x-y\|^2/C)$; (2)  when the intrinsic (latent) dimension is of order $O(1)$, we can choose $\m D(\cdot,\cdot)$ as the $1$-Wasserstein distance computed by the ``POT'' package~\citep{flamary2021pot}, which returns the optimal transport map between two discrete measures using network simplex algorithm~\citep{bonneel2011displacement}.

 \smallskip
\noindent \textbf{Construction of decoders and encoders:}       The decoders $\bold{G}$ and encoders $\bold{Q}$ can be  realized through neural networks. However,
for a large $K$,  we will have a large number of parameters to train if we see each decoder/encoder as an independent neural network. To address this issue, we enable parameter sharing inside the set of decoders $\bold{G}$ and  the set of encoders  $\bold{Q}$. Specifically,  for the set of encoders, we set the last (output) layer to be free among the encoders $\{Q_{\sk}\}_{k\in[K]}$ while other layers to be tied. Moreover, since the decoder-encoder structure aims to reconstruct  the data, we want the decoder $G_k$ to be close to the inverse of the encoder $Q_k$. To achieve this, for the set of decoders, we oppositely set the first (input) layer to be free among the decoders $\{G_{\sk}\}_{k\in[K]}$ and tie other layers.
% This parameter sharing idea we employed is motivated by  transfer learning~\citep{https://doi.org/10.48550/arxiv.1811.08883,https://doi.org/10.48550/arxiv.1409.1556}, which uses a pre-trained model or a model with known weights as the base, and only the last few layers is to be re-trained. 
The rationale behind our parameter sharing scheme is that in the encoder, the first (convolutional) layer focuses on ``low-level'' features extraction and other layers extract ``high-level'' features. Therefore, the described parameter sharing scheme enables the networks to leverage the common ``low-level'' information among different clusters of the data, hence improves the model training efficiency.

\smallskip
\noindent \textbf{Optimization of objective function}: Recall that $\wt{\nu}_{k, Q_k}=\wt Q_{\sk}(\wh{\mu}_{n}^{\sk})$, where $\wt Q_{\sk}$ is a randomized perturbed encoder $\wt Q_k(X)=Q_k(X)+\sqrt{h}\cdot \m N(0,I_d)$ and $\wh{\mu}_{n}^{\sk}$ is the re-weighted empirical measure $\wh{\mu}_{n}^{\sk}=\frac{\wh{\mu}_n\cdot\rho_k}{\wh {p}_k}$.  Given a discrepancy metric $\m D(\cdot,\cdot)$ and cost function $c(\cdot,\cdot)$ defining the penalty and reconstruction cost, we can rewrite the objective function as
\begin{align*}
 \sum_{k=1}^K \Big\{\wh{p}_k\int c\big(x,G_k(Q_k(x))\big) \,\dd \wh{\mu}_n^{\sk} +\lambda_k \cdot\m D\big(\nu_k,(\wt{Q}_\sk)_{\#}\wh{\mu}_n^{\sk} \big)\Big\}.\\[-1em]
 \end{align*}
To approximate the gradient of this objective function, we sample from the measure $\wh{\mu}_n^{\sk}$ by introducing auxiliary random variables $u\in {\rm Unif}(0,1)$. Specifically, we include data $X_i$ into the empirical measure $\wh{\mu}_{n}^{\sk}$ if $u<\rho_k(X_i)$ and otherwise we exclude the data from $\wh{\mu}_{n}^{\sk}$.  Moreover, the penalty term can be estimated using finite samples from $\nu_k$ and $(\wt{Q}_k)_{\#}\wh \mu_n^m$.

 \RestyleAlgo{boxruled}
\begin{algorithm}[ht!]
\caption{Algorithm for implementing MLDMAE}\label{algorithm}
\SetAlgoLined
\SetKwRepeat{Do}{do}{while}%
 \textbf{Input}: Regularization coefficient $\{\lambda_k\}_{k\in [K]}$,  partition of unity $\{\rho_k\}_{k\in [K]}$, discrepancy metric $\m D(\cdot,\cdot)$ and cost function $c(\cdot,\cdot)$, priors $\{\nu_k\}_{k\in [K]}$, latent (intrinsic) dimension $d$\;
 \textbf{Data}:$X^{(n)}=\{X_1,X_2\cdots,X_n\}$\;

    \Repeat{$(\phi,\theta)$ converges }{
      Sample a mini-batch dataset $\ms D$ from $X^{(n)}$\;
       \For{$k \leftarrow 1 \,\,to\,\, K$}
         { Initialize an empty dataset $\wt{\ms D}_k$\;
         \For{$X\in \ms D$}{ Generate random variable $u$ form ${\rm Unif}(0,1)$\;
  \If{$u\leq \rho_k(X)$}{Add $X$ to dataset $\wt{\ms D}_k$\;
  }}
          Generate dataset $\ms L_k$ from prior $\nu_k$\;
         Generate dataset $\ms E_k$ from $\m N(Q_{\phi,\sk}(X),h\,I_d)$ for $X$ uniformly picked in $\wt{\ms D}_k$.
         }
         Update $\phi$ and $\theta$ by one step first-order method (e.g., Adam,~\cite{kingma2014adam}) with objective function:
        %  \begin{equation*}
        %  \begin{aligned}
        %   \sum_{k=1}^K \Big\{  &\frac{\wh{p}_k}{|\wt{\m D}_k|}\sum_{X\in \wt{\m D}_k} \|X-G_{\theta,\sk}(Q_{\phi,\sk}(X))\|^2 +\lambda_k \cdot \Big(\frac{1}{|\m L_k|\cdot(|\m L_k|-1)}\sum_{z\in \m L_k,\atop z'\in \m L_k, z\neq z'} k(z,z')\\
        %   &+\frac{1}{|\m E_k|\cdot(|\m E_k|-1)}\sum_{z\in \m E_k,\atop z'\in \m E_k, z\neq z'} k(z,z')+\frac{1}{|\m E_k|\cdot|\m L_k|}\sum_{z\in \m E_k,\atop z'\in \m L_k} k(z,z')
        %   \Big)\Big\}
        %               \end{aligned}
        %  \end{equation*}
         \begin{equation*}
         \begin{aligned}
          &\sum_{k=1}^K \Big\{   \frac{\wh{p}_k}{|\wt{\ms D}_k|}\sum_{X\in \wt{\ms D}_k} c\big(X,G_{\theta,\sk}(Q_{\phi,\sk}(X))\big) +\lambda_k \cdot \m D\Big(\frac{1}{|\ms L_k|}\sum_{z\in \ms L_k} \delta_{z},\quad\frac{1}{|\ms E_k|}\sum_{z\in \ms E_k} \delta_{z}\Big)\Big\},\\
          &\text{where } \delta_z \text{ is the Dirac measure on point } z.
                      \end{aligned}
         \end{equation*}
       }
\end{algorithm}

Based on the above discussion, we can now develop the algorithm as described in Algorithm~\ref{algorithm} for implementing the MLDMAE estimation, where we use $\bold{G}_{\theta}=\{G_{\theta,1}, G_{\theta,2},\cdots,G_{\theta, M}\}$ and $\bold{Q}_{\phi}=\{Q_{\phi,1}, Q_{\phi,2},\cdots,Q_{\phi, M}\}$ to denote the decoders and encoders parametrized by $\theta$ and $\phi$ respectively.

     \section{Theoretical Analysis}\label{sec:theory}
     In this section, we derive the finite sample error of the MLDMAE estimator. We first state the following assumptions on the target distribution $\mu^\ast$ and the approximation family used in defining the estimator~\eqref{MLDMAEest}.
     
 \vspace{0.5em}
     \noindent\textbf{Assumption A (Target distribution):} The target distribution $\mu^\ast$ on manifold $\m M$ satisfies that: (1) $\m M\subset {\cup}_{k\in [K]} S_k$;  (2) for any $k\in [K]$, there exists $G_{\sk}^\ast\in C^{\beta}_L(\mb R^d;\mb R^D)$ and $Q_{\sk}^\ast\in C^{\beta}_L(\mb R^D;\mb R^d)$ so that  $x=G^\ast_{\sk}(Q^\ast_{\sk}(x))$ holds for any  $x\in \m M\cap S_k$; (3)  for any $m \in [K]$,  let $p_k=\mb {E}_{\mu^*}[\rho_k]$, then $p_k>0$ and $\nu^\ast_{\sk}=(Q^\ast_\sk)_{\#}(\frac{\mu^*\cdot\rho_k}{p_k})\in C^{\alpha}_L(\mb R^d)$; let $\Omega_k=Q_{k}^\ast(\m M\cap S_k)$,  there exists a function $g_k:\mb R^+\to \mb R^+$, so that for any $r>0$ and $z\in \Omega_k$, there exists $z'\in \Omega_k$ such that $z\in B_r(z')$ and $\nu^\ast_{\sk}(z')\geq g_k(r)$.
     
 \vspace{0.5em}
     \noindent\textbf{Assumption B (Approximation family)}: The approximation family $\m G$ satisfies that (1) $(\bold{G}^\ast,\bold{Q}^\ast,\bold{v}^\ast)\in \m G$ with $\bold{G}^\ast=(G_{1}^\ast, G_{2}^\ast,\cdots,G_{K}^\ast)$,  $\bold{Q}^\ast=(Q_{1}^\ast, Q_{2}^\ast,\cdots,Q_{K}^\ast)$ and $\bold{v}^\ast=(\nu_{1}^\ast, \nu_{2}^\ast,\cdots,\nu_{K}^\ast)$; (2) for any $(\bold{G},\bold{Q},\bold{v})\in \m G$, it holds that $G_{\sk}\in C^{\beta}_L(\mb R^d;\mb R^D)$ and $Q_{\sk}\in C^{\beta}_L(\mb R^D;\mb R^d)$ for any $k\in [K]$.

\begin{figure}[H]
     \subfigure[MLDMAE: truncated normal prior]
        { \includegraphics[width=0.3\textwidth]{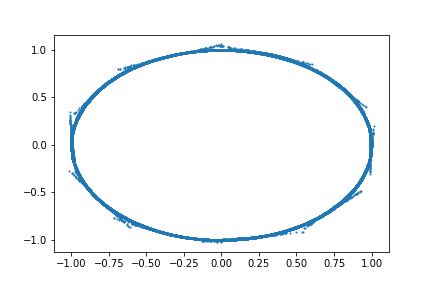}}
          \subfigure[MLDMAE: data-driven prior (once update) ]
        { \includegraphics[width=0.3\textwidth]{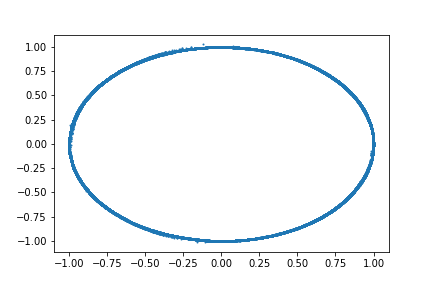}}
        \subfigure[MLDMAE: data-driven prior (twice updates)]
        { \includegraphics[width=0.3\textwidth]{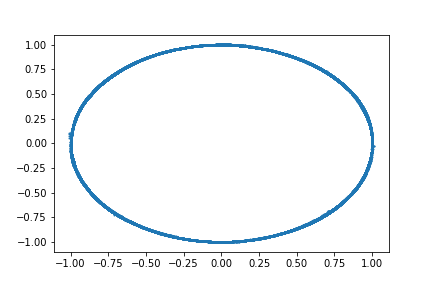}}
        \caption{Performance of MLDMAE with different choices of priors when the target measure is the uniform distribution on a sphere. Figure (a) plots the generated samples from MLDMAE estimator when the priors are truncated normal. Figures (b) and (c) plot the generated samples from MLDMAE estimator with data-driven priors described in Remark~\ref{remark:prior} under once and twice updates respectively, where $\nu_0$ is the truncated normal as in Figure (a).  We can see with a simple truncated normal prior, the generated plot tends to be non-smooth at the intersection of different partition functions. While  once update of priors using the strategy described in Remark~\ref{remark:prior} can lead to much better performance.}
        \label{fig:sphere_prior}
\end{figure}

  \vspace{0.5em}
  \noindent\textbf{Example (Manifold-supported distributions):}  For any  $\alpha$-smooth distribution $\mu^*$ on a $\beta$-smooth $d$-dimensional boundaryless compact submanifold embedded in $\mb R^D$ and with a positive density,   we can find a suitable open cover $\{S_k\}_{k\in [K]}$ and partition of unity $\{\rho_k\}_{k\in [K]}$ so that Assumption A holds.  Moreover, the (mixture of) generative model class induced by the approximation family $\m G=\big\{(\bold{G},\bold{Q},\bold{v})\,:\,\forall k\in [K], G_k\in C^{\beta}_L(\mb R^d;\mb R^D), Q_k\in C^{\beta}_L(\mb R^D; \mb R^d), \nu_k=\frac{(V_{k\#}\nu_0)\cdot \rho_k(G_k(z))}{\mb{E}_{\nu_0}[\rho_k(G_k(V_k(z)))] }, V_k\in C^{\alpha+1}_{L}(\mb B_1^d; \mb B_1^d)\big\}$ with $\nu_0$ being any fixed $\alpha$-smooth distribution on $\mb B_1^d$ whose density value bounded away from zero (e.g., uniform distribution), suffices to model the manifold-supported distributions $\mu^*$ (i.e., Assumption B holds for the approximation family $\m G$). In particular, when $\alpha+ 1=\beta$, we can consider the compositions $G_k\circ V_k$ as  $\beta$-smooth encoders for preventing the estimation of priors, that is, we can use the approximation family $\m G=\big\{(\bold{G},\bold{Q},\bold{v})\,:\,\forall k\in [K], G_k\in C^{\beta}_L(\mb R^d;\mb R^D), Q_k\in C^{\beta}_L(\mb R^D; \mb R^d), \nu_k=\frac{\nu_0\cdot \rho_k(G_k(z))}{\mb{E}_{\nu_0}[\rho_k(G_k(z))] }\big\}$. Further details are available in Appendix B.
  \begin{remark}\label{remark:prior}
  The choice of the approximation family suggests that, for learning a manifold-supported distribution, instead of taking the priors $\nu_k$ to be some fixed simple distribution $\nu_0$, we may rescale $\nu_0$ by the weight $\rho_k(G_k(\cdot))$ so that the resulting distribution has a matching tail  as $Q_{k\#}(\rho_k\mu^*/\mb{E}_{\mu^*}[\rho_k])$ after convergence. However, incorporating such a prior family in MLDMAE may lead to an unstable training due to the high irregularity of functions $\rho_k$. To address this issue, we can consider  ``data-driven'' priors as follows: we first fix $\nu_k$ to be some simple fixed distribution $\nu_0$, such as uniform distribution or truncated normal, then we run the MLDMAE algorithm to obtain estimators of encoder/decoder pairs $\{(\wh G_{k}^{[1]}, \wh Q_{k}^{[1]})\}_{k\in [K]}$.  Now we fix the priors $\nu_k$ to be $\nu_0$ rescaled by $\rho_k(G_{k}^{[1]}(\cdot))$,  and run the MLDMAE algorithm with initialization $\{(\wh G_{k}^{[1]}, \wh Q_{k}^{[1]})\}_{k\in [K]}$  to obtain estimators of encoder/decoder pairs $\{(\wh G_{k}^{[2]}, \wh Q_{k}^{[2]})\}_{k\in [K]}$. The above steps can continue by fixing the priors $\nu_k$ to be $\nu_0\cdot\rho_k(G_{k}^{[l]}(\cdot))$ and obtaining estimators $\{(\wh G_{k}^{[l+1]}, \wh Q_{k}^{[l+1]})\}_{k\in [K]}$, and stop until no improvement in validation error is seen.  Using such a data-driven prior can largely improve the performance of MLDMAE at the intersections of the support of different partition functions, see Fig.~\ref{fig:sphere_prior} for an illustration.  
  \end{remark}
  
  \noindent\textbf{Example (Distributions with clustering structures):}  Another example is a distribution induced by the mixture of generative models $\mu^\ast=\sum_{k=1}^K p_k\cdot (G_k^\ast)_{\#}\nu^\ast_\sk$, where supports of  generative models are disjoint. In this case, the supporting manifold of $\mu^\ast$ is a disconnected manifold, and we can simply choose $\{S_k\}_{k\in [K]}$ to be any disjoint sets that can cover each support of the generative model (i.e., ${\rm supp}((G_k^\ast)_{\#}\nu^\ast_\sk)\subset S_k$ for $k\in [K]$), and take $\rho_k$ to be the indicator function $\bold{1}(x\in S_k)$. With such choices, Assumption A holds if  for each $k\in [K]$, $\nu^\ast_\sk$ is $\alpha$-smooth with a compact support, and $G^\ast_\sk$ is $\beta$-smooth with a $\beta$-smooth inverse.

  \begin{theorem}\label{upperbound}
Fix $\alpha\geq 0$, $\beta\geq 1$ and a partition of unity $\{\rho_k\}_{k\in [K]}$ subordinate to the open cover $\{S_k\}_{k\in [K]}$. Suppose the target distribution $\mu^\ast$ satisfies Assumption A and the approximation family $\m G$ satisfies Assumption B.  If we choose the discrepancy metric $\m D(\cdot,\cdot)$ to be the $1$-Wasserstein distance $W_1$ and cost function $c(\cdot,\cdot)$ to be the squared $\ell_2$ loss, then there exists a choice of regularization coefficients  $\{\lambda_k\}_{k\in [K]}$ so that for large enough $n$,
\begin{enumerate}
    \item if $\alpha=0$, then by choosing the re-weighted empirical measure $\wt \nu_{\sk, Q_k}=\frac{1}{\wh{p}_kn}\sum_{i=1}^n \delta_{Q(X_i)}\rho_k(X_i)$ with $\wh{p}_k=\frac{1}{n}\sum_{i=1}^n \rho_k(X_i)$ as the plug-in, the resulting estimator $\wh \mu$ satisfies with probability at least $1-n^{-1}$ that 
\begin{equation} \label{WAE}
  W_1(\wh{\mu},\mu^\ast)\leq C\,n^{-\frac{1}{d}}  \vee\frac{\log n}{\sqrt{n}}.
\end{equation}
\item if $\alpha>0$, $\beta>1$, then there exists a smoothness-regularized empirical measure $\wt \nu_{\sk, Q_k}$ as the plug-in, so that the resulting estimator $\wh \mu$ satisfies with probability at least $1-n^{-1}$ that 
\begin{equation}\label{upperbound1}
  W_1(\wh{\mu},\mu^\ast)\leq C\,n^{-\frac{(\alpha\wedge (\beta-1))+1}{2(\alpha\wedge (\beta-1))+d}} \vee\frac{\log n}{\sqrt{n}}.
\end{equation}
\end{enumerate}
\end{theorem}
 
\noindent The smoothness-regularized empirical measure $\wt \nu_{\sk, Q_k}$ adopted in the proof of Theorem~\ref{upperbound} can either be based on wavelet truncation or kernel density estimator of the measure $Q_{k\#}(\frac{\mu^*\cdot \rho_k}{p_k})$. 
%  In practice, we may use the kernel density estimator with proper bandwidth to regularize the empirical distribution of local latent variables;  
 Based on the minimax lower bound developed in~\cite{liang2020generative} and~\cite{tang2022minimax}, the convergence rate in Theorem~\ref{upperbound} is minimax-optimal relative to the $1$-Wasserstein distance when $\alpha\leq \beta-1$. If Assumption A holds for $K=1$, then statement~\eqref{WAE} can provide a theoretical guarantee to the LDMAE estimator. The ambient dimension $D$ does not appear in the exponent of the developed rate; thus Theorem~\ref{upperbound} shows the
adaptiveness of MLDMAE to low-dimensional submanifold structures since the bound does not suffer from the ``curse of dimensionality''. Moreover, the MLDMAE estimator can take advantage of the smoothness of the target measure to further enhance the estimation accuracy by regularizing the empirical measure in the penalty.

\section{Simulation}
\begin{figure}[H]
    \centering
    \begin{minipage}{0.18\linewidth}
    \vspace{4pt}
    \centerline{  \includegraphics[trim={1.15cm, 0.3cm 1.15cm 1.15cm}, clip,width=1\textwidth]{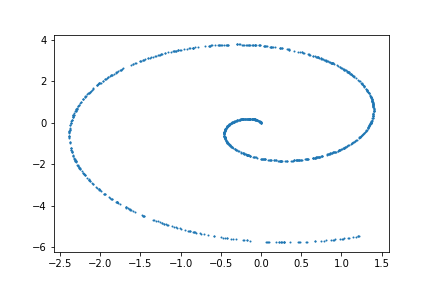}}
    \vspace{4pt}
    \centerline{ \includegraphics[trim={3.15cm, 0.5cm 1.1cm 1.4cm}, clip,width=1\textwidth]{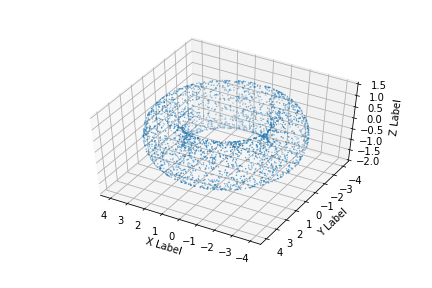}}
    \vspace{4pt}
    \centerline{Training sample}
    \end{minipage}
    \begin{minipage}{0.18\linewidth}
    \vspace{4pt}
    \centerline{ \includegraphics[trim={1.15cm, 0.3cm 1.15cm 1.15cm}, clip,width=1\textwidth]{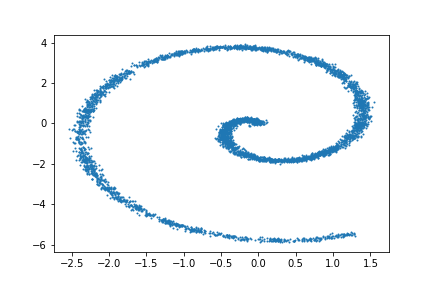}}
    \vspace{4pt}
    \centerline{ \includegraphics[trim={3.15cm, 0.5cm 1.1cm 1.4cm}, clip,width=1\textwidth]{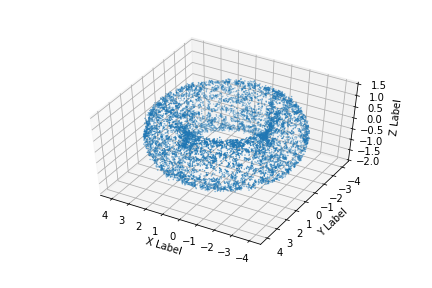}}
    \vspace{4pt}
    \centerline{KDE}
    \end{minipage}
    \begin{minipage}{0.18\linewidth}
    \vspace{4pt}
    \centerline{ \includegraphics[trim={1.15cm, 0.3cm 1.15cm 1.15cm}, clip,width=1\textwidth]{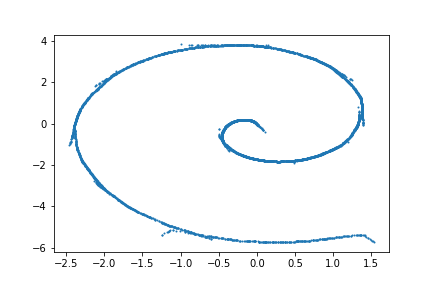}}
    \vspace{4pt}
    \centerline{ \includegraphics[trim={3.15cm, 0.5cm 1.1cm 1.4cm}, clip,width=1\textwidth]{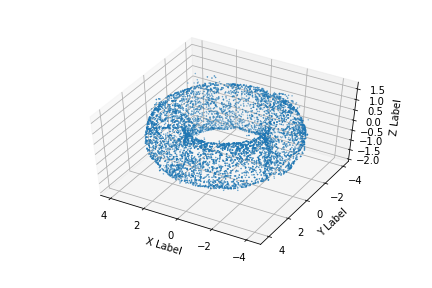}}
    \vspace{4pt}
    \centerline{MLDMAE (1-NN)}
    \end{minipage}
    \begin{minipage}{0.18\linewidth}
    \vspace{4pt}
    \centerline{\includegraphics[trim={1.15cm, 0.3cm 1.15cm 1.15cm}, clip,width=1\textwidth]{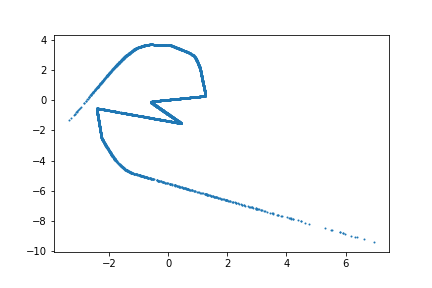}}
    \vspace{4pt}
    \centerline{ \includegraphics[trim={3.15cm, 0.5cm 1.1cm 1.4cm}, clip,width=1\textwidth]{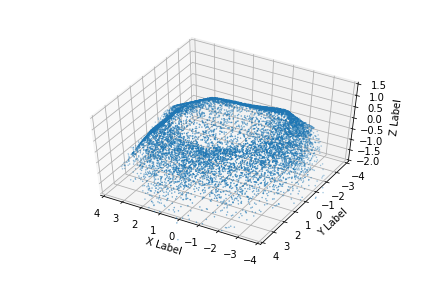}}
    \vspace{4pt}
    \centerline{LDMAE (1-NN)}
    \end{minipage}
    \begin{minipage}{0.18\linewidth}
    \vspace{4pt}
    \centerline{\includegraphics[trim={1.15cm, 0.3cm 1.15cm 1.15cm}, clip,width=1\textwidth]{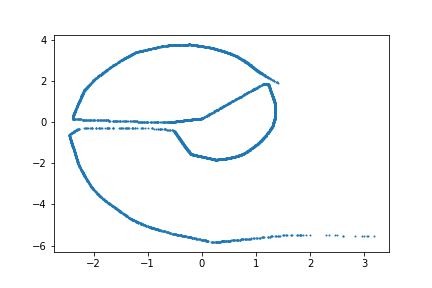}}
    \vspace{4pt}
    \centerline{ \includegraphics[trim={3.15cm, 0.5cm 1.1cm 1.4cm}, clip,width=1\textwidth]{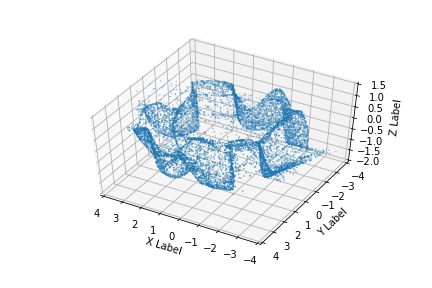}}
    \vspace{4pt}
    \centerline{LDMAE (2-NN)}
    \end{minipage}
    \caption{The figure illustrates the performance of MLDMAE, LDMAE and KDE when the target measures lying on a spiral and torus. We observe $n=1000$ and $n=3000$ training points for the example of  spiral (Top row) and torus (Bottom row) respectively.  The first column plots the training samples. The second column plots the generated samples from the classic kernel density estimator, which corresponds to adding Gaussian noises to the original training samples. The third column plots the  generated samples from our proposed MLDMAE estimator ($K=10$ for spiral and $K=15$ for torus),   the encoders and decoders are parameterized by  one-hidden layer neural networks with hidden layer size being  $128$, the partition of unity is chosen to be the smooth partition of unity described in Section~\ref{computation} with $\gamma=10$, the priors are selected as described in Remark~\ref{remark:prior} with $\nu_0$ being a truncated normal.  The fourth and fifth columns plot the  generated samples from the LDMAE estimator where the encoder-decoder pair are parameterized by one-hidden layer and  two-hidden layer neural networks respectively.}
    \label{fig:sp_torus}
\end{figure}

In this section, we present some visual results of the MLDMAE approach when apply to common manifolds: $2$D-spiral and torus. The precise data generating distributions are given in Appendix A. The penalty term $\m D(\cdot,\cdot)$ is chosen to be the $1$-Wasserstein distance computed by the ``POT'' package~\citep{flamary2021pot} and the cost function $c(\cdot,\cdot)$ is the squared $\ell_2$ loss. As benchmarks, we also consider (1) the classic kernel density estimator (KDE) with Gaussian kernel  that is commonly employed in statistics literature for density estimation; (2) the LDMAE estimator with the same kind of cost function and discrepancy metric as MLDMAE. The generated samples from the generators learned by different approaches are given in Fig.~\ref{fig:sp_torus}.  The $1$-Wasserstein distance between the estimated distribution and the true distribution are given in Table~\ref{tab:simu_wass}. We can see that employing multiple encoders and decoders can lead to much better performance than employing a single pair of encoder and decoder. In particular, we can see even though we increase the complexity of the encoder/decoder family, LDMAE still can not capture the correct shape of these standard manifolds in  statistics. On the other hand, by allowing multiple encoders/decoders in MLDMAE, the manifold structure can be correctly learned  with  a relatively simple encoder/decoder structure and a smaller number of  training parameters. Moreover, MLDMAE can beat the classic KDE in both examples. 

\begin{table}[H]
    \centering
    \begin{tabular}{ccccccc}
    \hline
   & &KDE & MLDMAE &LDMAE (1-NN)&LDMAE (2-NN)  \\
      \hline
   \multirow{2}{*}{Spiral}& $W_1$ distance &0.2119&0.1936&0.6315&0.4666\\
      &Number of training parameters&/ &4492&1027&34051\\
      \hline
      \hline
      \multirow{2}{*}{Torus}&$W_1$ distance&0.2837&0.2335&0.9101&0.6193\\
 & Number of training parameters & / &10529&1412&34565\\
     \hline
    \end{tabular}
    \caption{The table gives the $1$-Wasserstein distance between the target measure and the distribution estimators of different approaches for the spiral and torus examples. We also provide numbers of training parameters for different methods.}
    \label{tab:simu_wass}
\end{table}

\section{Real Data Application}
In this section, we empirically evaluate the proposed MLDMAE approach using three real-world datasets: MNIST handwritten digit~\citep{lecun1995learning},   Fashion-MNIST~\citep{xiao2017fashion} and CelebA $64\times 64$~\citep{krizhevsky2009learning}.  For comparison, we consider LDMAE approach and
 variational auto-enocoder (VAE)~\citep{kingma2013autoencoding,rezende2014stochastic,NIPS2016_ddeebdee}, which are commonly used auto-encoder based generative modelling approaches. In all reported experiments, we set the reconstruction cost to be the squared $\ell_2$ loss, and fix the priors $\nu_{\sk}$ to be a standard Gaussian $\m N(0,I_d)$. The partition of unity is chosen to be the indicator functions  described in Section~\ref{computation}. The encoders and decoders are modelled by convolutional deep neural networks with  parameter sharing as described in Section~\ref{computation}, and further details  are available in Appendix A.   We consider two kinds of discrepancy metric $\m D(\cdot,\cdot)$ in the penalty terms of LDMAE and MLDMAE, one is the $1$-Wasserstein distance computed through ``POT'' package, and the other one is MMD with the inverse multiquadratics (IMQ) kernel  $k(x,y)=2d/(2d+\|x-y\|^2_2)$. As described in~\cite{tolstikhin2019wasserstein}, the inverse multiquadratics kernel  has a much heavier tail than the conventional RBF kernel $k(x,y)=\exp(-\|x-y\|^2/C)$, so it can provide more meaningful gradients for outliers. The numbers $K$ of clusters for MLDMAE are selected so that the number of free parameters is around $\frac{1}{5}\sim\frac{1}{2}$  of the number of sharing parameters, and it turns out $K=5$ works well for all the examples. Another important factor  is the latent dimension, which is not explicit for the real dataset. Selecting a too small latent dimension would lead to large reconstruction errors and thus result in noisy generated samples. On the contrary, selecting a too large latent dimension would lead to the singularity of the encoded distribution and thus result in numerical instabilities.  We  use $d=4$ for Fashion-MNIST, $d=8$ for MNIST handwritten digit and $d=64$ for CelebA,  which seems to work reasonably well, and trainings of MLDMAE are stable and robust to  initialization.

\begin{figure}[H]
\vspace{-1em}
    \centering
    \begin{minipage}{0.18\linewidth}
    \vspace{3pt}
    \centerline{   \includegraphics[trim={3.45cm, 3.15cm 2.8cm 2.8cm}, clip,width=1\textwidth]{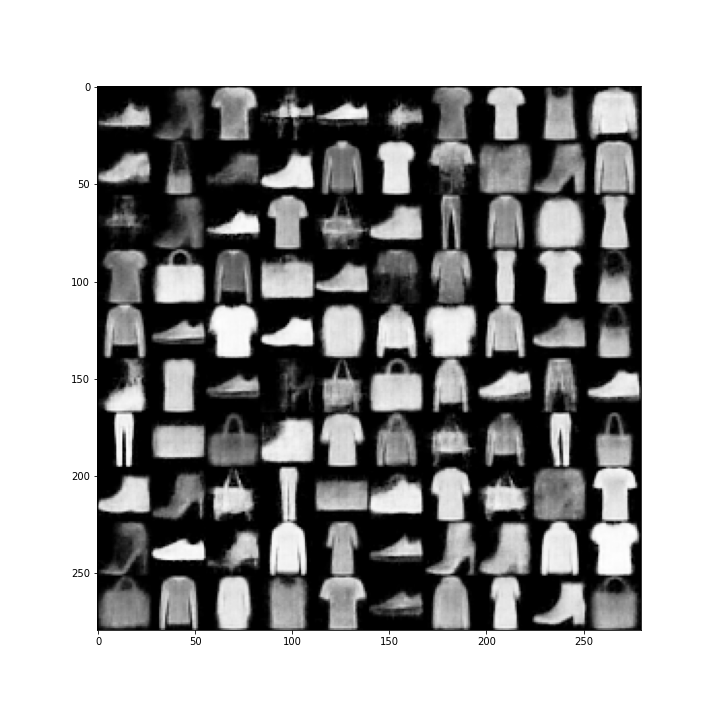}}
    \vspace{4pt}
    \centerline{  \includegraphics[trim={3.45cm, 3.15cm 2.8cm 2.8cm}, clip,width=1\textwidth]{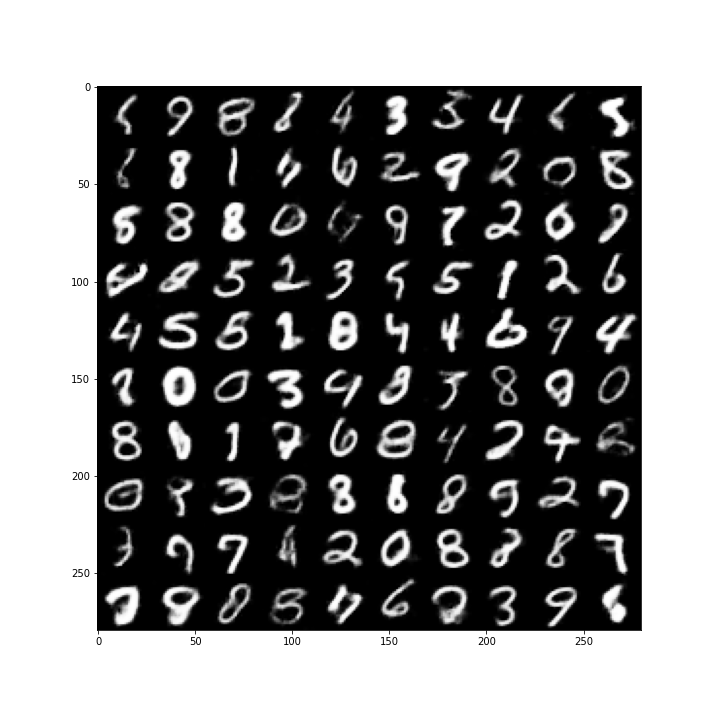}}
    \vspace{4pt}
    \centerline{VAE}
    \end{minipage}
    \begin{minipage}{0.18\linewidth}
    \vspace{4pt}
    \centerline{  \includegraphics[trim={3.45cm, 3.15cm 2.8cm 2.8cm}, clip,width=1\textwidth]{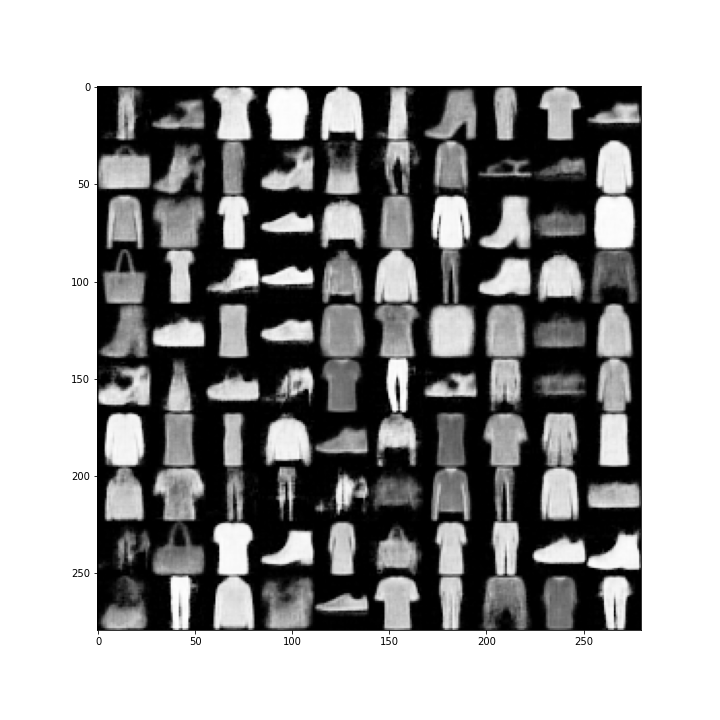}}
    \vspace{4pt}
    \centerline{  \includegraphics[trim={3.45cm, 3.15cm 2.8cm 2.8cm}, clip,width=1\textwidth]{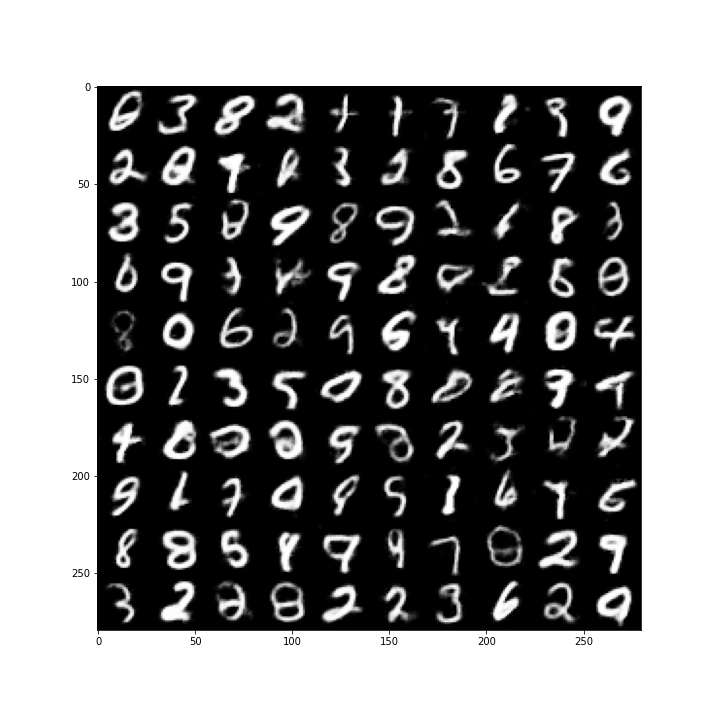}}
    \vspace{4pt}
    \centerline{LDMAE+MMD}
    \end{minipage}
    \begin{minipage}{0.18\linewidth}
    \vspace{4pt}
    \centerline{ \includegraphics[trim={3.45cm, 3.15cm 2.8cm 2.8cm}, clip,width=1\textwidth]{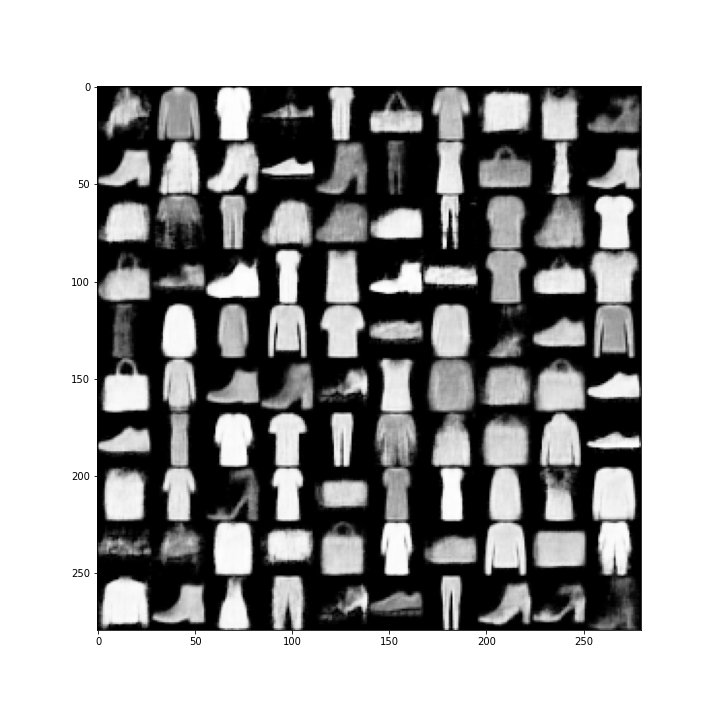}}
    \vspace{4pt}
    \centerline{ \includegraphics[trim={3.45cm, 3.15cm 2.8cm 2.8cm}, clip,width=1\textwidth]{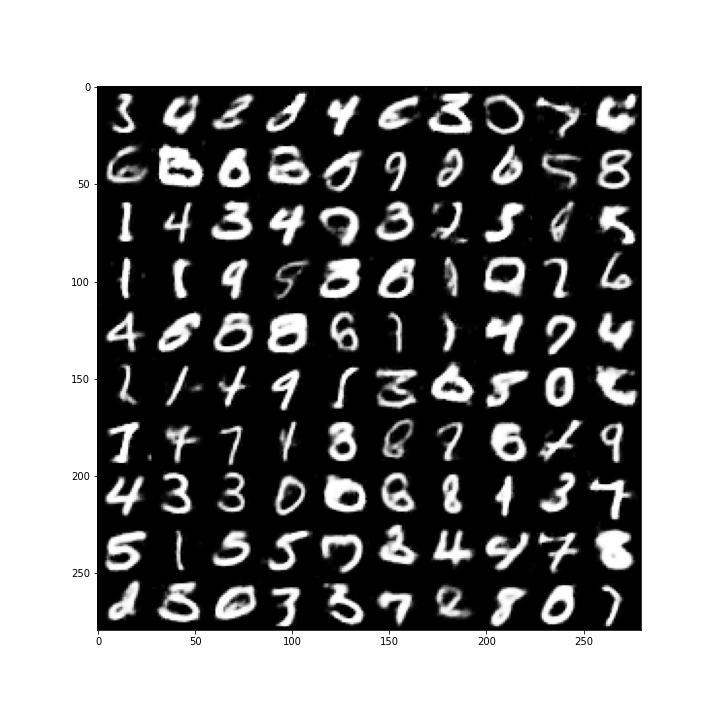}}
    \vspace{4pt}
    \centerline{LDMAE+$W_1$}
    \end{minipage}
    \begin{minipage}{0.18\linewidth}
    \vspace{4pt}
    \centerline{ \includegraphics[trim={3.45cm, 3.15cm 2.8cm 2.8cm}, clip,width=1\textwidth]{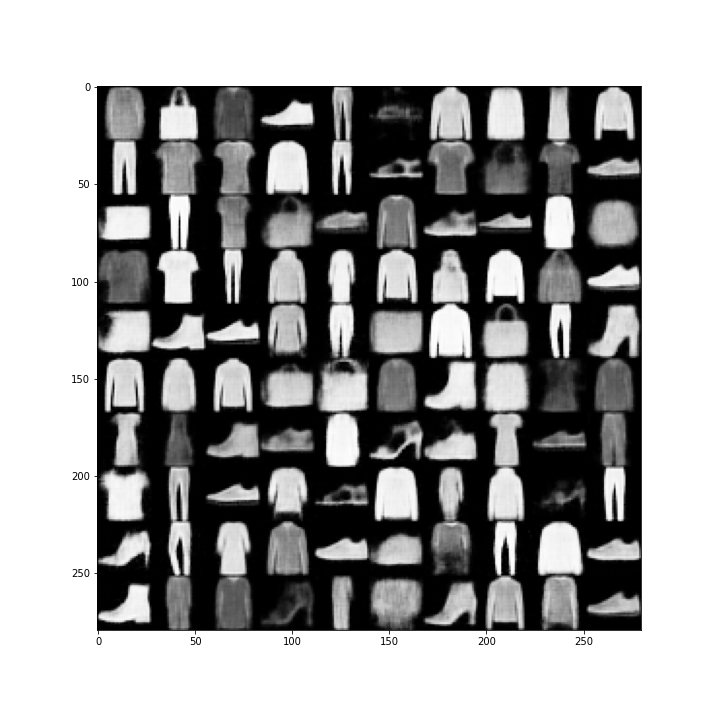}}
    \vspace{4pt}
    \centerline{  \includegraphics[trim={3.45cm, 3.15cm 2.8cm 2.8cm}, clip,width=1\textwidth]{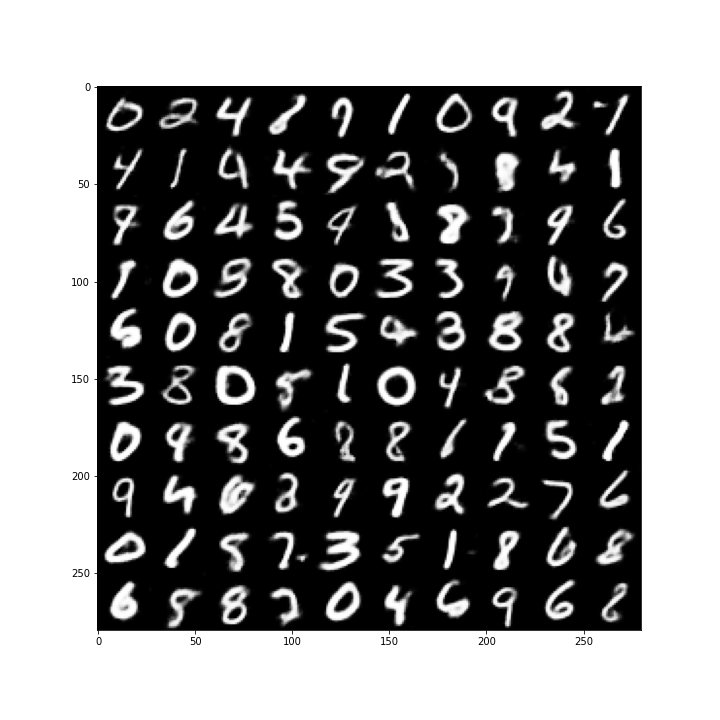}}
    \vspace{4pt}
    \centerline{MLDMAE+MMD}
    \end{minipage}
    \begin{minipage}{0.18\linewidth}
    \vspace{4pt}
    \centerline{ \includegraphics[trim={3.45cm, 3.15cm 2.8cm 2.8cm}, clip,width=1\textwidth]{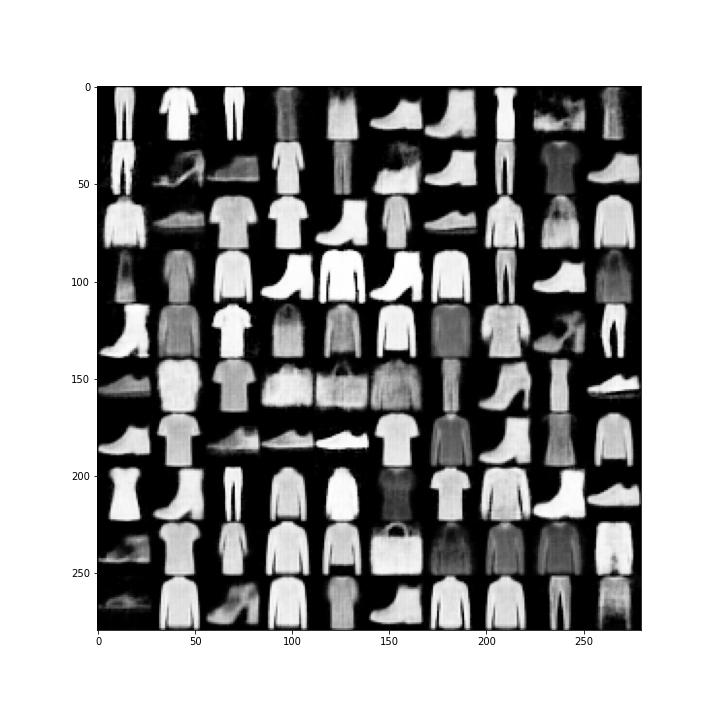}}
    \vspace{4pt}
    \centerline{   \includegraphics[trim={3.45cm, 3.15cm 2.8cm 2.8cm}, clip,width=1\textwidth]{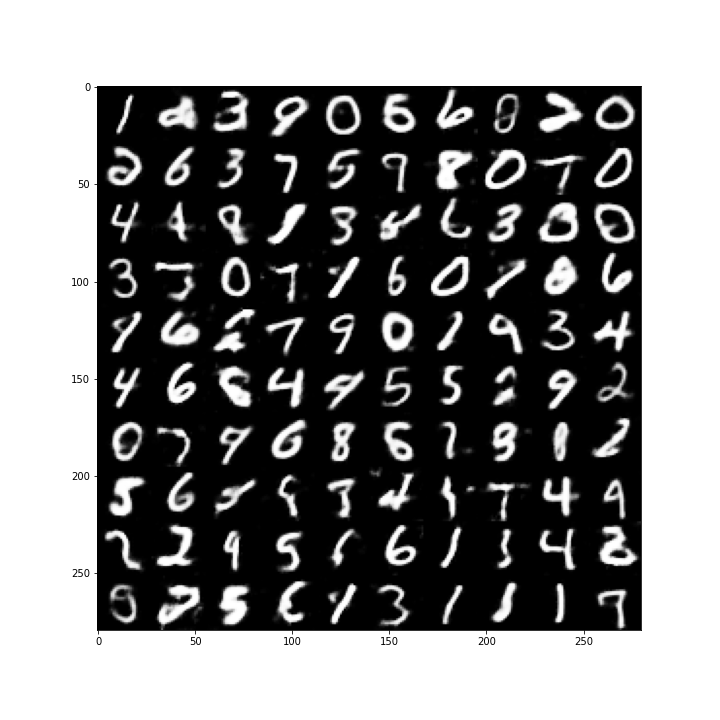}}
    \vspace{4pt}
    \centerline{MLDMAE+$W_1$}
    \end{minipage}
    \caption{The generated samples from different approaches for Fashion-MNIST (Top row) and MNIST handwritten digit dataset (Bottom row). The first column corresponds to the VAE estimator; the second and third columns  correspond to the LMDAE estimator with MMD and $W_1$ penalty respectively; the fourth and firth columns correspond to the MLMDAE estimator with MMD and $W_1$ penalty respectively.}
    \label{fig:mnist}
\end{figure}

\begin{figure}[H]
    \centering
		\includegraphics[width=0.5\textwidth]{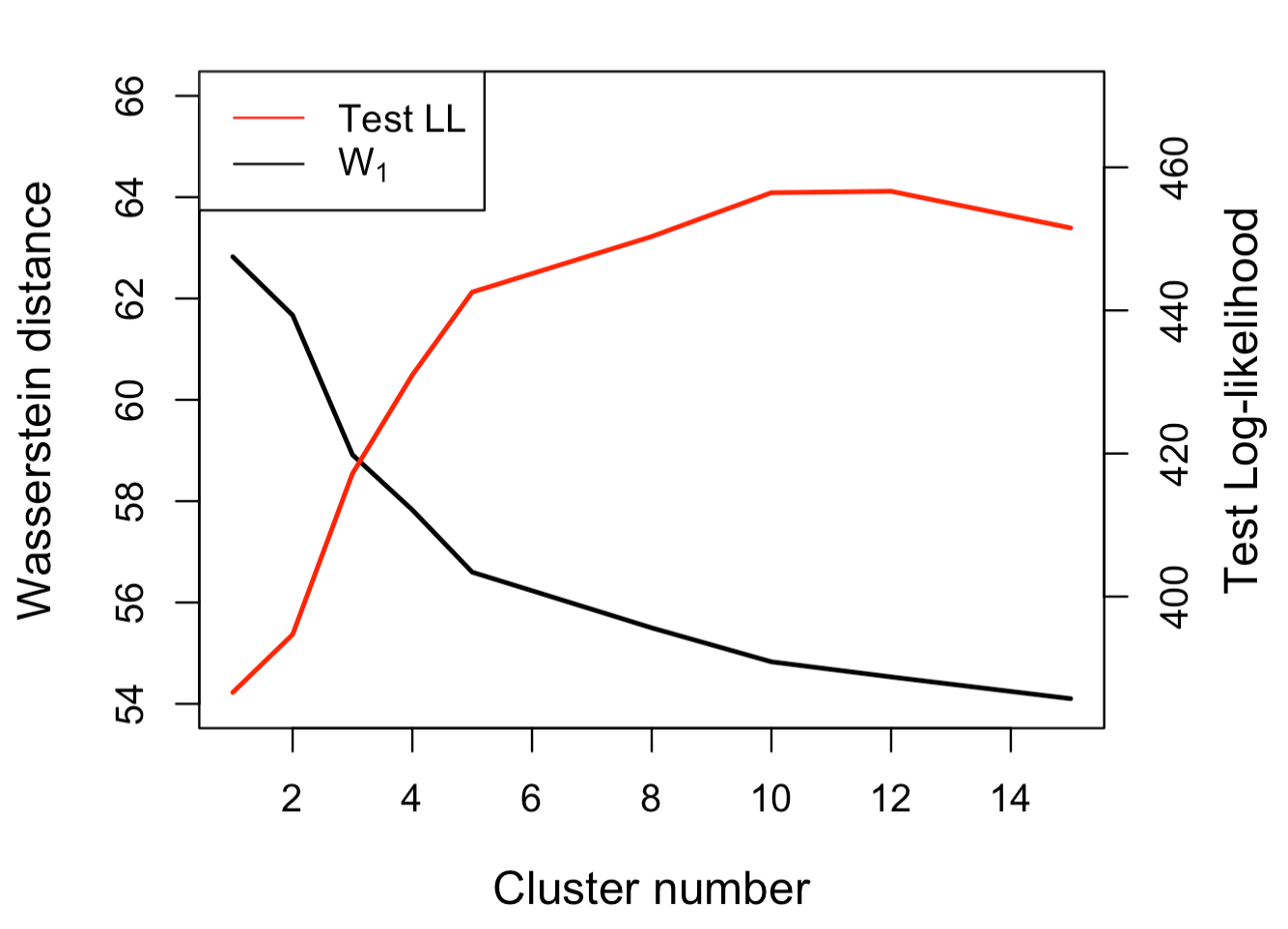}
 \caption{Negative test log-likelihood (red) and $W_1$ distance (black) of MLDMAE with MMD penalty and different cluster numbers $K$ for the MNIST handwritten digit dataset.}
    \label{fig:mnist_M_T}
\end{figure}

To quantitatively assess the MLDMAE estimator, for the dataset of MNIST handwritten digit and Fashion-MNIST,  we consider two kinds of evaluation metrics: one is the test log-likelihood (Test LL) used in~\citet{goodfellow2014generative}  by fitting a Gaussian Parzen window to the generated samples and reporting the log-likelihood evaluated at the test samples, and the other one is the $1$-Wasserstein distance ($W_1$) between the test samples and generated samples. The generated samples are shown in Fig.~\ref{fig:mnist} and the Test LL and $W_1$ distance are provided in Table~\ref{tab:MNIST}. We can see for the Fashion-MNIST dataset, MLDMAE with $W_1$ or MMD penalty obviously outperforms VAE and LDMAE. For the MNIST handwritten digit dataset, MLDMAE with MMD penalty outperforms the Wasserstein penalty in the $W_1$ metric, and it may attribute to the fact that the MNIST digit dataset has a relatively larger latent dimension $d=8$, so we need a very large batch size for accurately estimating the Wasserstein penalty,  which will reduce the number of gradient updating. In addition, Fig.~\ref{fig:mnist_M_T} gives the trends of the Test LL and $W_1$ distance
as the cluster number $K$ increases for the MNIST digit dataset. We can see the trends of both metrics become smooth when $M\geq 10$, this is consistent with the  underlying fact that the MNIST digit dataset contains $10$ digits (clusters).  Furthermore, we can see an obvious improvement in both metrics when increasing $K$ in the range of $[1,10]$, while the total number of training parameters only increases by $7\%$ when cluster number $K$ increases by $1$. 

\begin{table}[H]

	\begin{minipage}{0.6\linewidth}
	\caption{MNIST handwritten digit and Fashion-MNIST dataset: Test LL and $W_1$ distance for different approaches, ``+MMD'' and ``+$W_1$'' represent the choices of the discrepancy metric $\m D(\cdot,\cdot)$ in the penalty terms.}
	\label{tab:MNIST}
	\begin{tabular}{c|cc|cc}
	\hline
	&\multicolumn{2}{c|}{Fashion-MNIST  }&\multicolumn{2}{c}{MNIST}\\
	\hline
	    &Test LL $\uparrow$&$W_1$ $\downarrow$&Test LL $\uparrow$ &$W_1$ $\downarrow$\\
	   \hline
	  VAE   & 533&77.7 &393&63.2\\
LDMAE+MMD  & 549&75.5&381&63.0\\
	 LDMAE+$W_1$ &562  &74.6&400&68.1\\
	   MLDMAE+MMD &557&66.0&443&\textbf{56.6}\\
	   	MLDMAE+$W_1$ &\textbf{562}&\textbf{65.8}&\textbf{456}&63.1\\
	   \hline
	\end{tabular}
	\end{minipage}\hfill
		\begin{minipage}{0.36\linewidth}
		 	\centering
		{%
 \caption{CelebA dataset: FID and KID  for different approaches. The penalty term in LDMAE and MLDMAE are chosen to be MMD penalties.}
	\label{tab:CelebA}
	\begin{tabular}{ccc}
	\hline
	    &FID $\downarrow$&KID $\downarrow$\\
	   \hline
	  VAE   & 63.0&0.063 \\
LDMAE+MMD  & 55.5&0.057\\
  LDMAE+$SW_1$  & 62.8&0.064\\
	   MLDMAE+MMD &\textbf{51.1}&\textbf{0.051}\\
	   	   MLDMAE+$SW_1$  & 52.0&0.052\\
	   \hline
	  
	\end{tabular}
	
		}
\end{minipage} 
\end{table}


\begin{figure}[H]
 \centering
 \begin{subfigure}[VAE] {
       \includegraphics[trim={3.45cm, 3.15cm 2.8cm 2.8cm}, clip,width=0.18\textwidth]{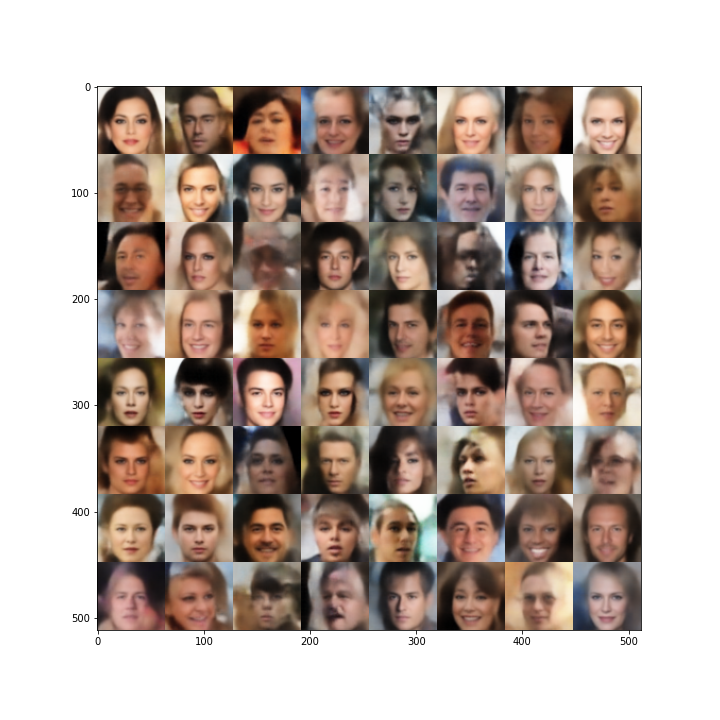}}
 \end{subfigure}
\begin{subfigure}[LDMAE+MMD]{
       \includegraphics[trim={3.45cm, 3.15cm 2.8cm 2.8cm}, clip,width=0.18\textwidth]{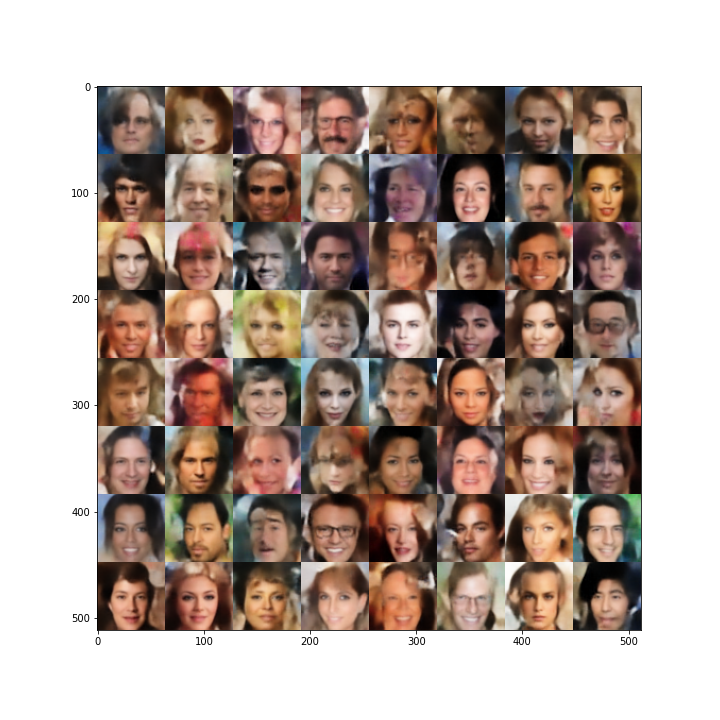}}
\end{subfigure}
\begin{subfigure} [LDMAE+$SW_1$]{
     \includegraphics[trim={3.45cm, 3.15cm 2.8cm 2.8cm}, clip,width=0.18\textwidth]{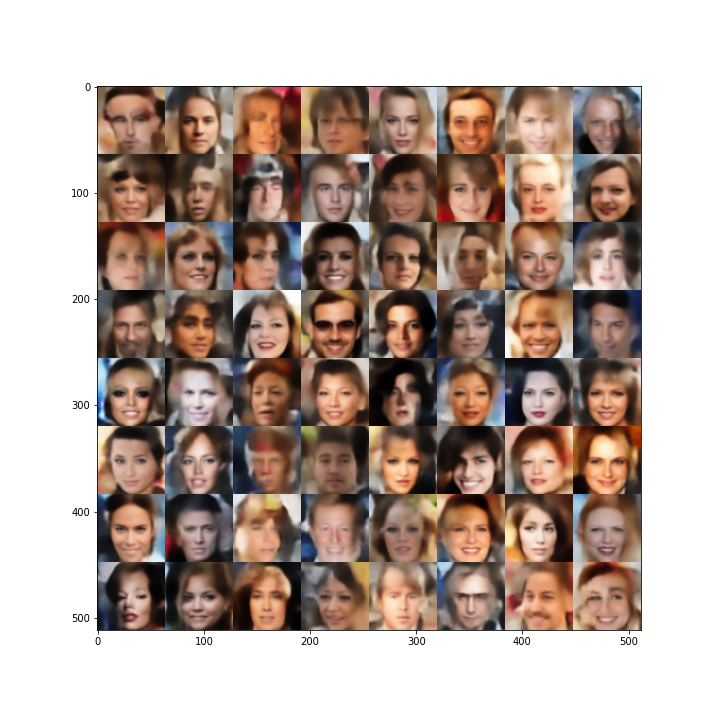}}
      \end{subfigure}
       \begin{subfigure} [MLDMAE+MMD]{
     \includegraphics[trim={3.45cm, 3.15cm 2.8cm 2.8cm}, clip,width=0.18\textwidth]{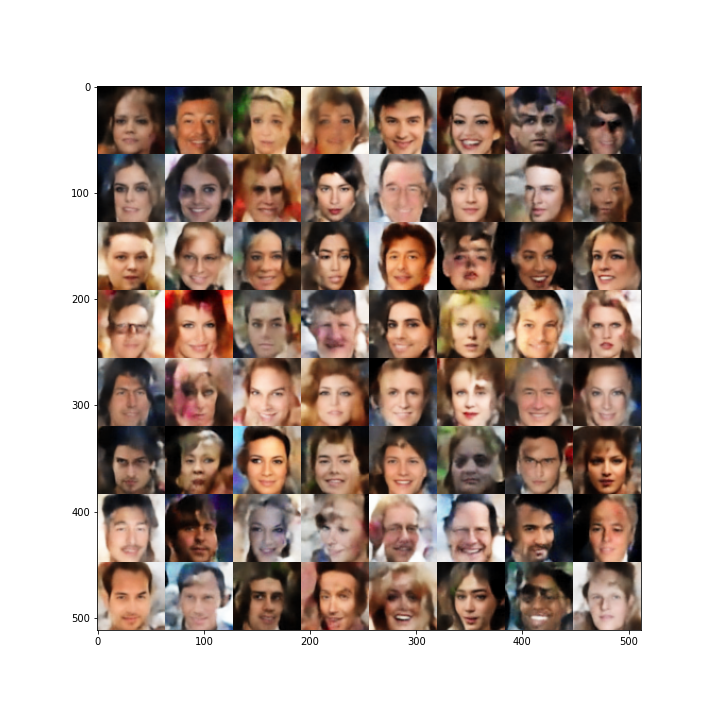}}
      \end{subfigure}
        \begin{subfigure} [MLDMAE+$SW_1$]{
     \includegraphics[trim={3.45cm, 3.15cm 2.8cm 2.8cm}, clip,width=0.18\textwidth]{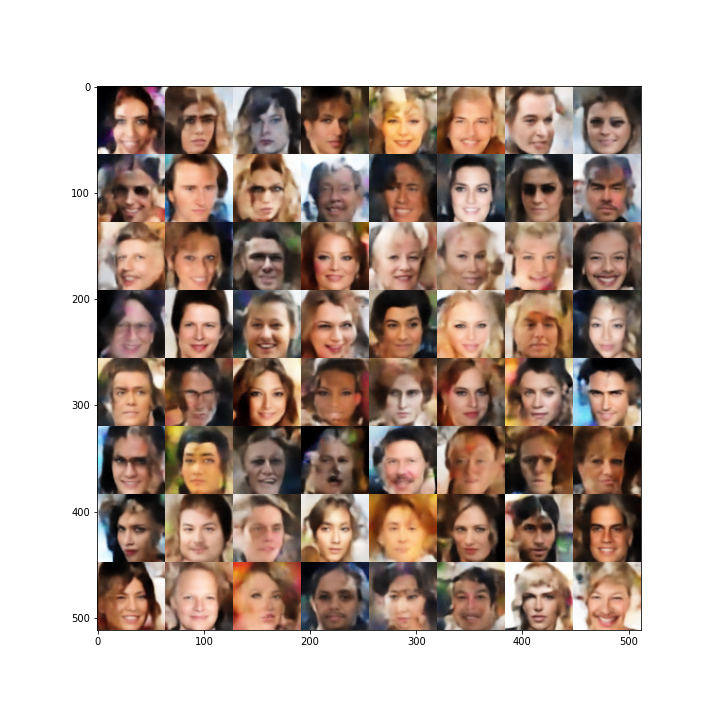}}
      \end{subfigure}
        
          \caption{The generated samples from different approaches (first column: VAE; second and third columns: LDMAE with MMD and $SW_1$ penalty respectively; fourth and fifth columns: MLDMAE with MMD  and $SW_1$ penalty respectively) for the CelebA dataset.}
        \label{fig:celebA}
\end{figure}

For the celebA dataset, since the ambient dimension $D=64\times 64\times 3$ is extremely large. Instead of using test log-likelihood or $W_1$ distance, which are evaluated in the ambient space; we consider two commonly-used metrics for color image data: FID~\citep{https://doi.org/10.48550/arxiv.1706.08500} and KID~\citep{binkowski2018demystifying}, in which the original high-dimensional data is fed into an ImageNet-pretrained inception network to obtain $2048$-dimensional inception (feature) representations,  and the FID and KID are the fr\'{e}chet distance and the squared MMD between inception representations of generated samples and test samples, respectively. Moreover, as described previously, the $W_1$ penalty is unsuitable for large intrinsic dimensions, for  avoiding the curse of dimensionality, we consider the so-called sliced Wasserstein distance: $SW_1(\mu,\nu):=\mb E_{\theta\sim{\rm Unif}(\mb S_1^{d-1})}\big[W_1({\rm Proj}_{\theta\#}\mu,{\rm Proj}_{\theta\#}\nu)\big]$, where ${\rm Proj}_{\theta}$ denotes the projection function to the direction $\theta$ and ${\rm Unif}(\mb S_1^{d-1})$ denotes the uniform distribution on $\mb S^{d-1}_1$.  The expectation over ${\rm Unif}(\mb S^{d-1}_1)$ can be estimated by Monte Carlo method. The sliced-Wasserstein distance  slices  high-dimensional
probability densities into sets of one-dimensional marginal distributions and compare these marginal
distributions via the Wasserstein distance, it
 has similar qualitative properties to the Wasserstein distance, but is much easier to compute.
 The generated samples and FID, KID are given in Fig.~\ref{fig:celebA} and Table~\ref{tab:CelebA}. We can see  that the MLDMAE estimator can achieve the best performance  under all evaluation metrics. The  MLDMAE  with MMD penalty performs slightly better than the $SW_1$ penalty, while it also requires more computation time (the computation time for MLDMAE+MMD is 150s per epoch using NVIDIA A100-SXM4-40GB GPU, while that is 120s per epoch for MLDMAE+$SW_1$).

%   \begin{table}[H]\label{table2}
%  \caption{Computation time (/s) per epoch}
%     \centering
%   \begin{tabular}{cccccc}
%   \hline
% \multicolumn{2}{c}{Fashion-MNIST}&\multicolumn{2}{c}{MNIST digit}&\multicolumn{2}{c}{CIFAR-10}\\
% \hline
%  WAE&MLDMAE& WAE&MLDMAE&WAE&MLDMAE\\
%  \hline
%  3.42&4.08&5.01&6.42&7.20&8.00\\
%  \hline
%   \end{tabular}
%           \end{table}
\section{Conclusion}
In this work, we proposed a new approach, mixture of latent distribution matched auto-encoder (MLDMAE), to improve the conventional auto-encoder based generative modelling approaches for learning manifold-supported distributions.  We showed theoretically that the proposed estimator can learn manifold-supported distributions with a minimax-optimal convergence rate. Moreover, we conducted experiments to show that by employing multiple encoder/decoder pairs, the estimators derived from MLDMAE can substantially boost the target distribution estimation accuracy.  In our theoretical analysis, we consider the case where the penalty term is chosen to be the Wasserstein distance. We leave the theoretical analysis for some other adversarial losses, such as the MMD distance considered in our experiments, to future work. 
 \newpage
\bibliographystyle{biometrika}
\bibliography{main_arxiv}
  
%%%%%%%%%%%%%%%%%%%%%%%%%%%%%%%%%%%%%%%%%%%%%%%%%%%%%%%%%%%%%%%%%%%%%%%%%%%%%%%%%%%%%%%%%%%%%%%%%%%%%%%%%%%%%%%%%%%%%%
%%%%%%%%%%%%%%%%%%%%%%%%%%%%%%%%%%%%%%%%%%%%%%%%%%%%%%%%%%%%%%%%%%%%%%%%%%%%%%%%%%%%%%%%%%%%%%%%%%%%%%%%%%%%%%%%%%%%%%
%%%%%%%%%%%%%%%%%%%%%%%%%%%%%%%%%%%%%%%%%%%%%%%%%%%%%%%%%%%%%%%%%%%%%%%%%%%%%%%%%%%%%%%%%%%%%%%%%%%%%%%%%%%%%%%%%%%%%%
%%%%%%%%%%%%%%%%%%%%%%%%%%%%%%%%%%%%%%%%%%%%%%%%%%%%%%%%%%%%%%%%%%%%%%%%%%%%%%%%%%%%%%%%%%%%%%%%%%%%%%%%%%%%%%%%%%%%%%
%%%%%%%%%%%%%%%%%%%%%%%%%%%%%%%%%%%%%%%%%%%%%%%%%%%%%%%%%%%%%%%%%%%%%%%%%%%%%%%%%%%%%%%%%%%%%%%%%%%%%%%%%%%%%%%%%%%%%%
%%%%%%%%%%%%%%%%%%%%%%%%%%%%%%%%%%%%%%%%%%%%%%%%%%%%%%%%%%%%%%%%%%%%%%%%%%%%%%%%%%%%%%%%%%%%%%%%%%%%%%%%%%%%%%%%%%%%%%
%%%%%%%%%%%%%%%%%%%%%%%%%%%%%%%%%%%%%%%%%%%%%%%%%%%%%%%%%%%%%%%%%%%%%%%%%%%%%%%%%%%%%%%%%%%%%%%%%%%%%%%%%%%%%%%%%%%%%%
%%%%%%%%%%%%%%%%%%%%%%%%%%%%%%%%%%%%%%%%%%%%%%%%%%%%%%%%%%%%%%%%%%%%%%%%%%%%%%%%%%%%%%%%%%%%%%%%%%%%%%%%%%%%%%%%%%%%%%
%APPENDIX
\newpage
\appendix
\begin{center}
{\bf\Large Appendix}
\end{center}
 
\textbf{Notations}: We adopt the notations in the manuscript, and further introduce the following additional notations for technical proofs. We use $\bold{N}(\mathcal{F},\,\widetilde{d}, \,\epsilon)$ to denote the $\epsilon$-covering number of function space $\mathcal{F}$ with respect to pseudo-metric $\widetilde{d}$.  We use $\mb B_r(x)$ to denote the closed ball centered at $x$ with radius $r$ under the $\ell_2$ distance; in particular, we use $\mb B_r^d$
denote $\mb B_r(\bold{0}_d)$ when no ambiguity may arise. We denote $\mb S_1^{d-1}=\{x\in \mb R^d\,:\, \|x\|=1\}$. For a function $f: \Omega\to \mb R^d$, we use $\bold{J}_f(x)$ to denote the $d\times m$ Jacobian matrix of $f$ at $x\in \Omega$. For a function $f:\,\mb R^d\to\mb R$, we use $f^{(a)}$ to denote its mixed partial derivative $\partial^{|a|} f/ \partial x_{1}^{a_{1}} \cdots \partial x_{d}^{a_{d}}$. We define the $\alpha$-smooth H\"{o}lder (function) class (see e.g.,~\cite{evans10}) with radius $r>0$ over $\Omega$ as $C^{\alpha}_r(\Omega):=\big\{f:\, \Omega \rightarrow \mathbb{R}\,\big|\,\|f\|_{C^{\alpha}(\Omega)}=\sum_{|a| \leq\lfloor\alpha\rfloor}\max_{x\in \Omega}| f^{(a)}(x)|+\sum_{|a| =\lfloor\alpha\rfloor}\max _{x, y\in\Omega,\,x\neq y} \left|f^{(a)}(x)-f^{(a)}(y)\right| /\|x-y\|^{\tilde\alpha-\lfloor\alpha\rfloor} \leq r \big\}$.
Similarly, we use $C^{\alpha}_r(\Omega; \mb R^D)=\big\{f=(f_1,\ldots,f_D):\, \Omega\to \mb R^D\,\big|\, \forall \,j \in [D],\, f_j\in C^{\alpha}_{r}(\Omega)\big\}$ to denote the vector valued function space counterpart. For an $f\in C^{\alpha}_r(\Omega; \mb R^D)$ and a multi-index $a\in \mb N_0^d$, we denote $f^{(a)}$ as the $D$ dimensional vector whose $j$-th component is the mixed partial derivative $[f_j]^{(a)}$ of $f_j$ for $j\in[D]$. Throughout, $C$, $c$, $C_0$, $c_0$, $C_1$, $c_1$, $C_2$, $c_2$,\ldots are generically used to denote positive constants whose values might change from one line to another, but are independent from everything else.

\section{Remaining implementation details}
\subsection{Simulation}
The training points $\ms D$ in spiral are generated via the following steps: (1) generate $\phi_0\sim \m N(0,1)$; (2) set $\phi=3\pi\phi_0$; (3) generate data point $X$ though $X=[\frac{\cos(\phi+2)\cdot \phi}{\pi},\frac{2\sin(\phi+2)\cdot \phi}{\pi} ]$. The training points $\ms D$ in torus are generated via the following steps: (1) generate $\phi_0,\phi_1\sim \m N(0,1)$; (2) set $\phi=2\pi \phi_0$ and $\theta=2\pi\phi_1$; (3) generate data point $X$ through $X=[(3+\cos(\theta))\cos(\phi), (3+\cos(\theta))\sin(\phi), \sin(\theta)]$.   The cluster number $M$ in MLDMAE is $M=10$ for the dataset of spiral and $M=15$ for the dataset of torus.  The partition of unity is given by $\rho_k={\widetilde\rho}_k/\big(\sum_{k'=1}^K \widetilde{\rho}_{k'}\big)$ with 
${\widetilde\rho}_k=(r_k^2-\|x-a_k\|^2)^{10}\cdot \bold{1}(x\in S_k)$, where $\{a_k\}_{k\in [K]}$ are the centers returned by the K-means algorithm, $r_k=\sup\{\|x-a_k\|\,:\, x\in \m D; \, \forall k_1\in [K],\, \|x-a_k\|\leq \|x-a_{k_1}\|\}$, and $S_k=\mb B_{r_k}(a_k)$.

\subsection{Real data application}
The specification of our models trained on MNIST handwritten digit, Fashion-MNIST and CelebA are described in Table~\ref{MNIST},~\ref{Fashion-MNIST}, and~\ref{CelebA}.  ``Shared'' is short for parameter sharing among encoders or among decoders.  All  models are optimized using Adam optimization  with learning rate $0.001$, $\beta_1=0.9$, and $\beta_2=0.999$.  The partition of unity for all datasets is chosen as the indicator function $\rho_k(x)=\bold{1}(x\in \text{cluster } m)$ for $k\in [K]$.  The codes for reproducing the experiments are available in \url{https://github.com/rtang1997/MLDMAE}.
 
\begin{table}[H]
    \centering
    \begin{tabular}{cccccc}
    \hline
      Operation   & Kernel &Strides& Feature maps& Activation & Shared? \\
      \hline
      Decoder $G_k(z): k\in [K], z\in \m N(0,I_d)$ & &&8&&\\
    Fully connected&&&$3\times3\times 128$&ReLU&No\\
    Transposed convolution & $3\times3$&$2\times2$& $7\times7\times 64$&ReLU&Yes\\
     Transposed convolution & $3\times3$&$2\times2$& $14\times14\times 32$&ReLU&Yes\\

      Transposed convolution & $3\times3$&$2\times2$& $28\times28\times 1$&ReLU&Yes\\
       \hline
        Encoder $Q_k(x): k\in [K]$ & &&$28\times28\times1$&&\\
    Convolution & $3\times3$&$2\times2$& $26\times26\times 3$&LeakyReLU&Yes\\
   Convolution & $3\times3$&$2\times2$& $12\times12\times 32$&LeakyReLU&Yes\\
Convolution & $3\times3$&$2\times2$& $5\times5\times 64$&LeakyReLU&Yes\\
Convolution & $3\times3$&$2\times2$& $2\times2\times 128$&LeakyReLU&Yes\\
Fully connected&&&$8$&&No\\
\hline
Cluster number $K$ for MLDMAE &\multicolumn{5}{l}{5}\\
Batch size &\multicolumn{5}{l}{$256$ for MLDMAE, and $128$ for WAE and VAE}\\
Number of epochs&\multicolumn{5}{l}{50}\\
Leaky ReLU slope&\multicolumn{5}{l}{0.1}\\
Regularization coefficients ($\lambda_k$)&\multicolumn{5}{l}{100 for MMD penalty and 10 for $W_1$ penalty}\\
Bandwidth ($h$) for MLDMAE &\multicolumn{5}{l}{0.01}\\
Number of training samples &\multicolumn{5}{l}{60k}\\
  \hline
    \end{tabular}
    \caption{Encoder/decoder Network architecture and hyperparameters for the MNIST handwritten digit  dataset.}
    \label{MNIST}
\end{table}
\begin{table}[H]
    \centering
    \begin{tabular}{cccccc}
    \hline
      Operation   & Kernel &Strides& Feature maps& Activation & Shared? \\
      \hline
      Decoder $G_k(z): k\in [K], z\in \m N(0,I_d)$ & &&4&&\\
    Fully connected&&&$3\times3\times 128$&ReLU&No\\
    Transposed convolution & $3\times3$&$2\times2$& $7\times7\times 64$&ReLU&Yes\\
     Transposed convolution & $3\times3$&$2\times2$& $14\times14\times 32$&ReLU&Yes\\
      Transposed convolution & $3\times3$&$2\times2$& $28\times28\times 1$&ReLU&Yes\\
       \hline
        Encoder $Q_k(x): k\in [K]$ & &&$28\times28\times1$&&\\
    Convolution & $3\times3$&$2\times2$& $26\times26\times 3$&LeakyReLU&Yes\\
   Convolution & $3\times3$&$2\times2$& $12\times12\times 32$&LeakyReLU&Yes\\
Convolution & $3\times3$&$2\times2$& $5\times5\times 64$&LeakyReLU&Yes\\
Convolution & $3\times3$&$2\times2$& $2\times2\times 128$&LeakyReLU&Yes\\
Fully connected&&&$4$&&No\\
\hline
Cluster number $K$ for MLDMAE &\multicolumn{5}{l}{5}\\
Batch size &\multicolumn{5}{l}{$256$ for MLDMAE, and $128$ for WAE and VAE}\\
Number of epochs&\multicolumn{5}{l}{50}\\
Leaky ReLU slope&\multicolumn{5}{l}{0.1}\\
Regularization coefficients ($\lambda_k$)&\multicolumn{5}{l}{100 for MMD penalty and 10 for $W_1$ penalty}\\
Bandwidth ($h$) for MLDMAE &\multicolumn{5}{l}{0.01}\\
Number of training samples  &\multicolumn{5}{l}{60k}\\
  \hline
    \end{tabular}
    \caption{Encoder/decoder architecture and hyperparameters for the Fashion-MNIST dataset.}
    \label{Fashion-MNIST}
\end{table}

\begin{table}[H]
    \centering
    \begin{tabular}{cccccc}
    \hline
      Operation   & Kernel &Strides& Feature maps& Activation & Shared? \\
      \hline
      Decoder $G_k(z): k\in [K], z\in \m N(0,I_d)$ & &&64&&\\
    Fully connected&&&$8\times8\times 1024$&ReLU&No\\
    Transposed convolution & $5\times5$&$2\times2$& $16\times16\times 512$&ReLU&Yes\\
          Batch normalization &&&&&\\
     Transposed convolution & $5\times5$&$2\times2$& $32\times32\times 256$&ReLU&Yes\\
           Batch normalization &&&&&\\
      Transposed convolution & $5\times5$&$2\times2$& $64\times64\times 128$&ReLU&Yes\\
            Batch normalization &&&&&\\
       Transposed convolution & $3\times3$&$1\times1$& $64\times64\times 3$&Tanh&Yes\\
       \hline
        Encoder $Q_k(x): k\in [K]$ & &&$64\times64\times3$&&\\
    Convolution & $5\times5$&$2\times2$& $32\times32\times 128$&ReLU&Yes\\
          Batch normalization &&&&&\\
   Convolution & $5\times5$&$2\times2$& $16\times16\times 256$&ReLU&Yes\\
         Batch normalization &&&&&\\
Convolution & $5\times5$&$2\times2$& $8\times8\times 512$&ReLU&Yes\\
      Batch normalization &&&&&\\
Convolution & $5\times5$&$2\times2$& $4\times4\times 1023$&ReLU&Yes\\
      Batch normalization &&&&&\\
Fully connected&&&$64$&&No\\
\hline
Cluster number $K$ for MLDMAE &\multicolumn{5}{l}{5}\\
Batch size &\multicolumn{5}{l}{$256$ for MLDMAE, and $128$ for WAE and VAE}\\
Number of epochs&\multicolumn{5}{l}{50}\\
Regularization coefficients ($\lambda_k$)&\multicolumn{5}{l}{100}\\
Bandwidth ($h$) for SWAE and MLDMAE &\multicolumn{5}{l}{0.01}\\
Number of training samples &\multicolumn{5}{l}{180k}\\
  \hline
    \end{tabular}
    \caption{Encoder/decoder architecture and hyperparameters for the CelebA dataset.}
    \label{CelebA}
\end{table}

\section{Generative modelling of Distributions on submanifolds}
 A submanifold in the ambient space $\mb R^D$ can be viewed as a nonlinear ``subspace''. Borrow the definition in~\citet{tang2022minimax}, we   define the family of smooth distributions on $d$-dimensional smooth compact submanifolds without boundaries on $\mb R^D$ as the set $\m P^\ast = \m P^\ast(d,D,\alpha,\beta,L^\ast)$ with $d\leq D$, $\beta>1$ and $\alpha\in(0,\beta-1]$ composed of all probability measures $\mu\in \m P(\mb R^D)$ satisfying:
 
 \vspace{0.5em}
  1. $\mu$ is an $\alpha$-smooth distribution on  a $\beta$-smooth $d$-dimensional compact submanifold $\m M$ embedded in $\mb R^D$.

  2. The density $\mu$ relative to the volume measure of $\m M$ is uniformly bounded from below by $1/L^\ast$ on $\m M$.

  3. $\m M$ is covered by an atlas  $\ms{A}=\{(U_{\lambda},\phi_{\lambda})\}_{\lambda\in \Lambda}$ on $\m M$ such that: a)~each chart $(U,\phi)$ in atlas $\ms A$ satisfies $\|\phi^{-1}\|_{C^\beta(\phi(U))}\leq L^\ast$ and $\|\mu\circ \phi^{-1}\|_{C^\alpha(\phi(U))} \leq L^\ast$; b)~for any $z\in \phi(U)$, the Jacobian of $\phi^{-1}(z)$ is full rank and all its singular values are lower bounded by $1/L^\ast$ in absolute values. Moreover, for any $x\in \m M$, there exists a $\lambda\in \Lambda$ such that $U_{\lambda}$ and $\phi_{\lambda}(U_{\lambda})$ covers $\mb B_{1/L^\ast}(x)\cap \m M$ and $\mb B_{1/L^\ast}(\phi_{\lambda}(x))$ respectively.
 
  \vspace{0.5em}
We have the following lemma describing the mixture of generative model classes that can model the submanifold-supported distributions.

\begin{lemma}
\label{lemmaB.1}
  Consider $\ms{O}_K=\{S_k=B^{\circ}_{r_k}(a_k)\}_{m=1}^K$, for $k\in [K]$, choosing $\rho_k(x)=\frac{\wt \rho_k(x)}{\sum_{k\in [K]}\wt \rho_k(x)}$ with $\wt\rho_k(x)=(r_k^2-\|x-a_k\|^2)^{\gamma}\cdot \bold{1}(x\in S_k)$ for $\gamma\geq \alpha+1$. There exists a constant $r^*$ that only depends on $(L^*,\beta,\alpha,d,D)$ so that for any $\mu^*\in \m P^*(d,D,\alpha,\beta,L^*)$, if (1) $\m {\rm supp}(\mu^*)\subset\cup_{k\in [K]} S_k$; (2) for any $k\in [K]$, $r_k\leq r^*$ and $a_k\in \m M$; (3) there exists some positive constants $L^*_1$ so that $\min_{k\in [K]} r_k\geq L_1^*$ and $\inf_{x\in \m M}\sum_{k\in [K]}\wt \rho_k(x)\geq L_1^*$. Then: 
  
    \vspace{0.5em}
  1. there exist some universal constants $(L, c)$ that only depend on $(L^*,L^*_1,\beta,\alpha,d,D,
  \gamma)$ so that Assumption A holds for $\mu^*$ with  upper bound $L$ and function $g_k(r)=c\, (r^{\gamma}\wedge 1)$ for any $k\in [K]$;
 
  2. consider  $\nu_0\in \m P(\mb B_1^d)$ whose density being $\alpha$-smooth  and bounded below from zero, and approximation families
  
  \vspace{0.5em}
       \quad(a)$\quad\m G_1=\big\{(G,Q,v)\,:\,\forall k\in [K], G_k\in C^{\beta}_L(\mb R^d;\mb R^D), Q_k\in C^{\beta}_L(\mb R^D; \mb R^d), \nu_k\in\m P(\mb B_1^d) {\,\rm with\,} \nu_k\in C^{\alpha}_L(\mb B_1^d) \big\}$;
    
    \quad(b).$\quad\m G_2=\big\{(G,Q,v)\,:\,\forall k\in [K], G_k\in C^{\beta}_L(\mb R^d;\mb R^D), Q_k\in C^{\beta}_L(\mb R^D; \mb R^d), v_k=\frac{(V_{m\#}\nu_0)\cdot \rho_k(G_k(z))}{\mb{E}_{\nu_0}[\rho_k(G_k(V_k(z)))] }, V_k\in C^{\alpha+1}_{L}(\mb B_1^d; \mb B_1^d)\big\}$;
      
       \quad(c).$\quad\m G_3=\big\{(G,Q,v)\,:\,\forall k\in [K], G_k\in C^{\beta}_L(\mb R^d;\mb R^D), Q_k\in C^{\beta}_L(\mb R^D; \mb R^d), v_k=\frac{\nu_0\cdot \rho_k(G_k(z))}{\mb{E}_{\nu_0}[\rho_k(G_k(z))] }\big\}$;
       
 \vspace{0.5em}
 \noindent then for sufficiently large $L$, we have Assumption B holds for $\m G_1$ or $\m G_2$, that is, the approximation families $\m G_1$ and  $\m G_2$ are both sufficient to model distributions inside $\m P^*(d,D,\alpha,\beta,L^*)$. Moreover, if $\alpha=\beta-1$, then Assumption B holds for $\m G_3$.
  \end{lemma}

  \noindent The next lemma shows that $\ms O_K$ satisfying conditions of Lemma~\ref{lemmaB.1} can be found based on a small portion of data.
\begin{lemma}\label{lemmapou}
Consider any $\mu^*\in \m P^*(d,D,\alpha,\beta,L^*)$ with support $\m M$, fix $r^*$ being an arbitrary positive constant and  let $n_1\leq n$ be a positive integer.  Then  for any positive constant $c$, there exist constants $C,c_1$ that only depend on $(d,D,\beta,L^*,r^*,c)$ so that when $C\leq n_1\leq n$, let $I_1$ be any subset of $[n]$ with $|I_1|=n_1$, it holds with probability larger than $1-n_1^{-c}$ that

  \vspace{0.5em}
   (1). $\m M\subset \bigcup_{i\in I_1}\mb B_{c_1(\frac{\log n_1}{n_1})^{\frac{1}{d}}}(x_i)$;
   
   (2). there exists a constant $K$ that only depends on $(d,D,\beta,L^*,r^*)$ and a subset $\{a_k\}_{k=1}^K\subset \{X_i\}_{i\in I_1}$ so that
   
   \vspace{0.5em}
   
        \quad (a). $\bigcup_{i\in I_1}\mb B_{c_1(\frac{\log n_1}{n_1})^{\frac{1}{d}}}(x_i)\subset \bigcup_{k=1}^K \mb B_{r^*}(a_k)$. 
         
     \quad (b).  $\inf_{x\in \m M}\sum_{k\in [K]}\wt \rho_k(x)> (\frac{r^*}{\sqrt{2}})^{2\gamma}$, where $\wt \rho_k(x)=((r^*)^2-\|x-a_k\|^2)^{\gamma}\cdot \bold{1}(x\in \mb B_{r^*}(a_k))$.
   
  \end{lemma}

\section{Proof of Theorem 5.1}\label{Proof5.1}
 We first consider the case $\alpha>0$ and $\beta>1$. To simplify the notation,  we write $\tilde\alpha=\alpha\wedge (\beta-1)$.  We consider two kinds of smoothness-regularized empirical measure $\wt \nu_{\sm, Q_\sm}$, one is based on kernel density estimator and one is based on wavelet estimator. \\

\noindent\textbf{Kernel density estimator:}
 Define 
 \begin{equation}\label{KDE}
      \wt{\nu}_{\sm, Q_\sm}(y)=\frac{1}{n\wh p_kh^d}\sum_{i=1}^n \wt k\big(\frac{y-Q_k(X_i)}{h}\big)\rho_k(X_i),\quad \wh p_k=\frac{1}{n}\sum_{i=1}^n \rho_k(X_i),
 \end{equation}
 with $h=n^{-{1}/{(2{\widetilde {\alpha}}+d)}}$ and $ \wt k: \mb R^d\to \mb R $ satisfies that  
 
 \vspace{0.5em}
   \quad{1.$ \wt k(\cdot)$ is $\lceil \tilde\alpha\rceil \vee \lceil\frac{d}{2}\rceil$ smooth in $\mb R^d$ and has support contained in $[-1,1]^d$;}
   
     \vspace{0.5em}
       \quad{2. $\int_{\mb R^d}  \wt k(z)\, \dd z=1$ and for any $j\in \mb N_0^d$ with $1\leq |j|\leq \lfloor\alpha\rfloor +1$, $\int  \wt k(z)\cdot z^j\,\dd z=0$;}
    
    \vspace{0.5em}
    \quad{3. for any $z\in \mb R^d$, $\wt k(z)= \wt k(-z)$.}\\

 \noindent\textbf{Wavelet estimator:}
 Define $ \wt{\nu}_{\sm, Q_\sm}(y)$  as

\begin{equation}\label{WE}
 \wt{\nu}_{\sm, Q_\sm}(y)=\frac{1}{\wh{p}_k}\Big(\sum_{m \in  \mathbb{S}}\wt{a}^{Q_\sm}_m \phi_m(y) +\sum_{l=1}^{2^{d}-1}\sum_{j=0}^J\sum_{m\in \mathbb{S}_{lj}}\wt{\theta}_{ljm}^{Q_\sm} \psi_{ljm}(y)\Big),\\
 \end{equation}
 with
\begin{equation*}
\begin{aligned}
&\wh p_k=\frac{1}{n}\sum_{i=1}^n \rho_k(X_i);\\
& \mathbb{S}=\{m\in\mathbb{Z}^d\, |\, \text{supp}(\phi_m)\cap[-L,L]^d\neq \emptyset\};\\
& \mathbb{S}_{lj}=\{m \in \mathbb{Z}^d\,| \, \text{supp}(\psi_{ljm})\cap [-L,L]^d\neq \emptyset\};
 \end{aligned}
 \end{equation*}

\begin{equation*}
\begin{aligned}
\wt{a}^{Q_k}_{m}=\frac{1}{n}\sum_{i=1}^n \phi_m(Q_k(X_i))\rho_k(X_i);\\
\wt{\theta}_{ljm}^{Q_k}=\frac{1}{n}\sum_{i=1}^n \psi_{ljm}(Q_k(X_i))\rho_k(X_i),
 \end{aligned}
\end{equation*}
where $2^{dJ}\asymp n^{\frac{d}{2\alpha+d}}$ and $\{\phi_m, \psi_{ljm}:\, l=1,\cdots, 2^{d}-1,j \in \mathbb{N}, m\in \mathbb{Z}^d\}$ is the orthonormal wavelet basis  for Besov space on $\mathbb{R}^d$ defined as $\phi_m(y)=\phi(y-m)$ and $\psi_{ljm}(y)=2^{\frac{jd}{2}}\psi_l (2^jy-m)$, and it holds that $\phi(\cdot)$ and $\psi_l(\cdot)$  are compactly supported and have bounded $ \lceil\alpha\vee (\frac{d}{2}-\alpha)\rceil$ order derivatives for any $1\leq l\leq 2^d-1$~\citep{doi:10.1080/03610926.2015.1019144}.\\
\quad\\
We will show both choices of $\wt\nu_{k,Q_k}$ can lead to the desired result.  By Assumption  A of $\mu^*$, for any $k\in [K]$, there exist $G^*_\sm\in C^\beta_L(\mb R^d; \mb R^D)$ and $Q^*_\sm\in C^\beta_L(\mb R^D; \mb R^d)$ so that and for any $x\in \m M\cap S_k$, $G_\sm^*(Q_\sm^*(x))=x$.   By the optimality of $\wh{\bold{G}}$, $\wh{\bold{Q}}$ and $\wh{\bold{v}}$ for the training objective, we can get that 
\begin{equation}\label{eqn1}
\begin{aligned}
&\sum_{k=1}^K \bigg\{\frac{1}{n}\sum_{i=1}^n\|X_i-\wh G_\sm(\wh Q_\sm(X_i))\|^2\rho_k(X_i)+\lambda_k \cdot \underset{f\in{\rm Lip}_1(\mb R^d)}{\sup}\Big(\int f(z)\wh \nu_\sm(z)\,\dd z -\int f(z)\wt{\nu}_{\sm,\wh Q_\sm}(z)\,\dd z\Big) \bigg\}\\
&\leq\sum_{k=1}^K \lambda_k \cdot \underset{f\in{\rm Lip}_1(\mb R^d)}{\sup}\Big(\int f(z) \nu^\ast_\sm(z)\,\dd z -\int f(z)\wt{\nu}_{\sm,Q^\ast_\sm}(z)\,\dd z\Big),
\end{aligned}
\end{equation}
where recall $\nu^\ast_k=(Q^\ast_\sm)_{\#}(\frac{\mu^*\rho_k}{p_k})$. Then we have the following lemma.
\begin{lemma}\label{leup1}
For any fixed $c_1$ and $c_2$, define
\begin{equation*}
 \wt{\mathcal{Q}}_k=\{Q\in C^{\beta}_L(\mb R^D;\mb R^d): \text{the density }\nu^*_{k,Q} \text{ of } Q_{\#}\big(\frac{\mu^*\cdot \rho_k}{p_k}\big) \text{ exists and }\nu^*_{\sm, Q}\in C^{\tilde\alpha}_{c_2} (\mathbb{R}^d) \}.
 \end{equation*}
 Then there exists a constant $c_3$ such that it holds with probability larger than $1-\frac{1}{n^2}$ that  for any $k\in [K]$,
     \begin{equation*}
     \underset{Q\in \wt{\mathcal{Q}}_\sm}{\sup}   \underset{f\in{\rm Lip}_1(\mb R^d)}{\sup}\Big(\int f(z)\nu^\ast_{\sm,Q}(z)\,\dd z -\int f(z)\wt{\nu}_{\sm,Q}(z)\,\dd z\Big)\leq c_4\,\Big( n^{-\frac{\tilde\alpha+1}{2\tilde\alpha+d}} +\frac{\log n}{\sqrt{n}} \Big),
     \end{equation*}
     where $\wt\nu_{k,Q}$ can either be the kernel density estimator in~\eqref{KDE} or wavelet estimator in~\eqref{WE}.
%      \item For any $k\in [K]$ and $Q\in \wt{\mathcal{Q}}_\sm$, there exists a constant $c_4$ such that it holds with probability larger than $1-\frac{1}{n^2}$ that 
%      \begin{equation*}
%   \underset{f\in{\rm Lip}_1(\mb R^d)}{\sup}\Big(\int f(z)d\nu^\ast_{\sm,Q} -\int f(z)\wt{\nu}_{\sm,Q}(z)\,\dd z\Big)\leq c_4\,\Big( \big(\frac{\log n}{n}\big)^{\frac{\tilde\alpha+1}{2\tilde\alpha+d}} +\sqrt{\frac{\log n}{n}}\Big).
%      \end{equation*}

%  \end{enumerate}
\end{lemma}
So we choose $\lambda_k=\lambda= \big(n^{-\frac{\tilde\alpha+1}{2\tilde\alpha+d}} +\frac{\log n}{\sqrt{n}}\big)^{-1}\cdot n^{-\frac{2\beta}{d}-1}$ for any $k\in [K]$, then by the second statement of Lemma~\ref{leup1} and $Q^\ast_\sm\in \wt{\m Q}_k$, it holds with probability larger than $1-M\,n^{-2}$ that, 
\begin{equation*}
\begin{aligned}
&\sum_{k=1}^K \bigg\{\frac{1}{n}\sum_{i=1}^n\|X_i-\wh G_\sm(\wh Q_\sm(X_i))\|^2\rho_k(X_i)+\lambda \cdot\underset{f\in{\rm Lip}_1(\mb R^d)}{\sup}\Big(\int f(z)\wh \nu_\sm(z)\,\dd z -\int f(z)\wt{\nu}_{\sm,\wh Q_\sm}(z)\,\dd z\Big) \bigg\}\\
&\leq C\, \lambda\cdot  \big(n^{-\frac{\tilde\alpha+1}{2\tilde\alpha+d}} +\frac{\log n}{\sqrt{n}}\big) .
\end{aligned}
\end{equation*}
So it holds with probability larger than $1-M\,n^{-2}$ that for any $k\in [K]$,
\begin{equation}\label{error:manifold}
\frac{1}{n}\sum_{i=1}^n\|X_i-\wh G_\sm(\wh Q_\sm(X_i))\|^2\rho_k(X_i)\leq C \,n^{-\frac{2\beta}{d}-1},
\end{equation}
and
\begin{equation}\label{error:density1}
 \underset{f\in{\rm Lip}_1(\mb R^d)}{\sup}\Big(\int f(z)\wh \nu_\sm(z)\,\dd z -\int f(z)\wt{\nu}_{\sm,\wh Q_\sm}(z)\,\dd z\Big)\leq C\, \Big( n^{-\frac{\tilde\alpha+1}{2\tilde\alpha+d}} +\frac{\log n}{\sqrt{n}}\Big).
\end{equation}
Then we use the following lemma for bounding the population level reconstruction error.
\begin{lemma}\label{leup2}
For the estimator $\wh{\bold{G}}$ and $\wh{\bold{Q}}$, there exist positive constants $N$, $c_1$, $c_2$ and $c_3$ such that when $n\geq N$, for any $k \in [K]$,

 \vspace{0.5em}
\quad(1). it holds with probability larger than $1-c_1\,n^{-3}$ that  $\mathbb{E}[\|X-\wh{G}_\sm(\wh{Q}_\sm(X))\|_2\cdot \rho_k(X)]\leq c_2\,n^{-\frac{\beta}{d}}\vee \frac{\log n}{\sqrt{n}}$;

\quad(2). if $\tilde\alpha>0$, then it holds with probability larger than $1-c_1\,n^{-3}$ that the density $\nu^\ast_{\sm, \wh{Q}_\sm}$ of $(\wh{Q}_\sm)_{\#}\big(\frac{\mu^\ast\cdot \rho_k}{p_k}\big)$ exists and  belongs to $C^{\tilde\alpha}_{c_3}(\mathbb{R}^d)$.
\end{lemma}
Then by Lemma~\ref{leup1} and second statement of Lemma~\ref{leup2}, there exist constants $c,c_1$ such that it holds with probability larger than $1-c\,n^{-2}$ that 
\begin{equation}\label{error:density2}
   \underset{f\in{\rm Lip}_1(\mb R^d)}{\sup}\Big(\int f(z)\nu^\ast_{\sm,\wh Q_\sm}(z)\,\dd z -\int f(z)\wt{\nu}_{\sm,\wh Q_\sm}(z)\,\dd z\Big)\leq c_1\,\Big( n^{-\frac{\tilde\alpha+1}{2\tilde\alpha+d}} +\frac{\log n}{\sqrt{n}}\Big).
\end{equation}
 So combined with equation~\eqref{error:manifold},~\eqref{error:density1} and~\eqref{error:density2},  it holds with probability larger than $1-\frac{1}{n}$ that   
  \begin{equation}\label{finalbound}
\begin{aligned}
W_1(\wh\mu, \mu^\ast)&=\underset{f\in {\rm Lip}_1(\mb R^D)}{\sup} \Big( \sum_{k=1}^K \int  {p}_k\cdot f(X)\,\dd\Big(\frac{\mu^\ast\cdot \rho_k}{p_k}\Big) -  \sum_{k=1}^K \int \wh{p}_k \cdot f(X) \,\dd(\wh{G}_{\sm})_{\#}\wh{\nu}_{\sm}\Big)\\
&\overset{(i)}{\leq} C\,\sqrt{\frac{\log n}{n}}+\underset{f\in {\rm Lip}_1(\mb R^D)}{\sup} \Big( \sum_{k=1}^K \int  \wh{p}_k\cdot f(X)\,\dd\Big(\frac{\mu^\ast\cdot \rho_k}{p_k}\Big) -  \sum_{k=1}^K \int \wh{p}_k \cdot f(X) \,\dd(\wh{G}_{\sm})_{\#}\wh{\nu}_{\sm}\Big)\\
&\leq C\,\sqrt{\frac{\log n}{n}}+\sum_{k=1}^K \underset{f\in {\rm Lip}_1(\mb R^D)}{\sup} \Big(\int \wh{p}_k\cdot f(X)\,\dd\Big(\frac{\mu^\ast\cdot \rho_k}{p_k}\Big) -   \int  \wh{p}_k\cdot f(x) \,\dd(\wh{G}_{\sm})_{\#}\wh{\nu}_{\sm}\Big)\\
&\leq C\,\sqrt{\frac{\log n}{n}}+\sum_{k=1}^K \underset{f\in {\rm Lip}_1(\mb R^D)}{\sup} \Big(\int \frac{\wh p_k}{p_k}\cdot f(X) \rho_k(X)\dd \mu^\ast -\int \frac{\wh p_k}{p_k}\cdot f(\wh{G}_\sm(\wh Q_\sm(X)) \rho_k(X)\dd \mu^\ast \\
&\qquad\qquad+\int \frac{\wh p_k}{p_k}\cdot f(\wh{G}_\sm(\wh Q_\sm(X)) \rho_k(X)\dd \mu^\ast -  \int \wh{p}_k\cdot f(x) \,\dd(\wh{G}_{\sm})_{\#}\wh{\nu}_{\sm}\Big)\\
&\leq  C\,\sqrt{\frac{\log n}{n}}+2\,\sum_{k=1}^K \mb {E}_{\mu^\ast}\Big[\|X-\wh G_\sm (\wh Q_\sm(X))\|\cdot \rho_k(X)\Big]\\
&\qquad\qquad+\sum_{k=1}^K \underset{f\in {\rm Lip}_1(\mb R^D)}{\sup} \Big(\int f(\wh G_\sm (z))\nu^*_{\sm,\wh{Q}_\sm} (z) \,\dd z-\int f(\wh G_\sm (z))\,\wh \nu_\sm(z)\,\dd z\Big)\\
&\leq C_1\, n^{-\frac{\beta}{d}}\vee \frac{\log n}{\sqrt{n}} +\sum_{k=1}^K \bigg[\underset{f\in {\rm Lip}_1(\mb R^D)}{\sup} \Big(\int f(\wh G_\sm (z))\nu^*_{\sm,\wh{Q}_\sm} (z) \,\dd z-\int f(\wh G_\sm (z)) \wt\nu_{\sm,\wh{Q}_\sm} (z) \,\dd z\Big)\\
&\qquad\qquad +\underset{f\in {\rm Lip}_1(\mb R^D)}{\sup} \Big(\int f(\wh G_\sm (z)) \wt\nu_{\sm,\wh{Q}_\sm} (z) \,\dd z-\int f(\wh G_\sm (z))\,\wh \nu_\sm(z)\,\dd z\Big)\bigg]\\
&\overset{(ii)}{\leq}  C_1\, n^{-\frac{\beta}{d}}\vee \frac{\log n}{\sqrt{n}} +C_2\,\sum_{k=1}^K \bigg[\underset{f\in {\rm Lip}_1(\mb R^d)}{\sup} \Big(\int f(z)\nu^*_{\sm,\wh{Q}_\sm} (z) \,\dd z-\int f(z) \wt\nu_{\sm,\wh{Q}_\sm} (z) \,\dd z\Big)\\
&\qquad\qquad +\underset{f\in {\rm Lip}_1(\mb R^d)}{\sup} \Big(\int f(z) \wt\nu_{\sm,\wh{Q}_\sm} (z) \,\dd z-\int f(z)\,\wh \nu_\sm(z)\,\dd z\Big)\bigg]\\
&\leq C_2\,n^{-\frac{\tilde\alpha+1}{2\tilde\alpha+d}} \vee\frac{\log n}{\sqrt{n}},
\end{aligned}
\end{equation}
where $(i)$ uses Bernstein's inequality to obtain that  $\big|\wh p_k-p_k\big|\leq  C\, \sqrt{\frac{ \log n}{n}}$ holds with probability at least $1-n^{-3}$, and $(ii)$  uses 
the fact that $\beta \geq 1$. 
 
Then we consider the case $\alpha=0$. Define $\wt{\nu}_{k,Q}=\frac{1}{\wh{p}_kn}\sum_{i=1}^n \delta_{Q(X_i)}\rho_k(X_i)$. For any $f\in {\rm Lip}_1(\mb R^d)$, we have 
\begin{equation*}
    \int f(z)\,\dd \wt{\nu}_{k,Q}=\frac{1}{\wh{p}_kn}\sum_{i=1}^n f(Q(X_i))\rho_k(X_i).
\end{equation*}
Then we have the following lemma.
\begin{lemma}\label{empiricalmean}
 There exists a constant $c$ so that it holds with probability larger than $1-\frac{1}{n^2}$ that 
  \begin{equation*}
      \underset{Q\in C^{\beta}_L(\mb R^D;\mb R^d)}{\sup}\underset{f\in {\rm Lip}_1(\mb R^d)}{\sup}\Big(\frac{1}{p_k}\int f(Q(x))\rho_k(x)\, \dd \mu^*-\int f(z)  \wt{\nu}_{k,Q}(z)\,\dd z \Big)\leq c\, \Big({\frac{\log n}{\sqrt n}}+n^{-\frac{1}{d}} \Big).
  \end{equation*}
\end{lemma}

Then by Lemma~\ref{empiricalmean} and equation~\eqref{eqn1}, choose $\lambda_k=\lambda= \big(n^{-\frac{1}{d}}  +{\frac{\log n}{\sqrt n}}\big)^{-1}\cdot n^{-\frac{2\beta}{d}-1}$, we have that statement~\eqref{error:manifold} holds and 

\begin{equation*} 
 \underset{f\in{\rm Lip}_1(\mb R^d)}{\sup}\Big(\int f(z)\wh \nu_\sm(z)\,\dd z -\int f(z)\wt{\nu}_{\sm,\wh Q_\sm}(z)\,\dd z\Big)\leq C\, \Big( n^{-\frac{1}{d}} + \frac{\log n}{\sqrt n}\Big).
\end{equation*}
Then by Lemma~\ref{leup2}, we have 
\begin{equation*}
    \mb{E}[\|X-\wh G_\sm(\wh Q_\sm(X))\|_2\rho_k(X)]\leq c_2 n^{-\frac{\beta}{d}}\vee \frac{\log n}{\sqrt{n}}.
\end{equation*}
So combined with Lemma~\ref{empiricalmean}, following equation~\eqref{finalbound}, we have 
\begin{equation*}
    \begin{aligned}
W_1(\wh\mu, \mu^\ast)&\leq  C\,\sqrt{\frac{\log n}{n}}+2\,\sum_{k=1}^K \mb {E}_{\mu^\ast}\Big[\|X-\wh G_\sm (\wh Q_\sm(X))\|\cdot \rho_k(X)\Big]\\
&\qquad\qquad+\sum_{k=1}^K \underset{f\in {\rm Lip}_1(\mb R^D)}{\sup} \Big(\int f(\wh G_\sm (z))\nu^*_{\sm,\wh{Q}_\sm}(z)\,\dd z -\int f(\wh G_\sm (z))\,\wh \nu_\sm(z)\,\dd z\Big)\\
&\leq C_1\, n^{-\frac{\beta}{d}}\vee \frac{\log n}{\sqrt{n}} +C_2\sum_{k=1}^K \bigg[\underset{f\in {\rm Lip}_1(\mb R^d)}{\sup} \Big(\int f(z)\nu^*_{\sm,\wh{Q}_\sm}(z)\,\dd z  -\int f(z)\wt\nu_{\sm,\wh{Q}_\sm}(z)\,\dd z \Big)\\
&\qquad\qquad +\underset{f\in {\rm Lip}_1(\mb R^D)}{\sup} \Big(\int f(z) \wt\nu_{\sm,\wh{Q}_\sm}(z)\,\dd z -\int f(z)  \,\wh \nu_\sm(z)\,\dd z\Big)\bigg]\\
&\leq C_2\,n^{-\frac{1}{d}} \vee\frac{\log n}{\sqrt{n}}.
\end{aligned}
\end{equation*}
\subsection{Proof of Lemma~\ref{leup1}: kernel density estimator}
 
Fix an arbitrary $k\in [K]$. Since $\wt{\mathcal{Q}}_k\subseteq C^{\beta}_L(\mb R^D;\mb R^d)$, it holds that for any $Q\in \wt{\mathcal{Q}}_k$, $\nu^*_{\sm,Q}\in C^{\tilde\alpha}_{L} (\mathbb{R}^d)$ and ${\rm supp}(\nu^*_{\sm,Q})\in[-L, L]^d$, where $\nu^*_{\sm,Q}$ is the density of the  push-forward measure of $\frac{\mu^*\cdot\rho_k}{p_k}$ by map $Q$. Recall that 
\begin{equation*}
 \wh{p}_k\cdot\wt{\nu}_{\sm,Q}(y)=\frac{1}{nh^d}\sum_{i=1}^n \wt k(\frac{y-Q(X_i)}{h})\rho_k(X_i).
 \end{equation*}
 Since $\nu^*_{\sm ,Q}$ and $\wt\nu_{\sm ,Q}$ are both compactly supported,  there exists a constant $C$ so that for any $Q\in \wt{\m Q}_k$,
 \begin{align*}
\underset{f\in {\rm Lip}_1(\mb R^d)}{\sup}\Big(\int f(y) \nu^*_{\sm,Q}(y) \,\dd y-\int f(y) \wt{\nu}_{\sm ,Q}(y)\,\dd y\Big)\leq C\, \underset{f\in \bold{C}_1^1(\mb R^d)}{\sup}\Big(\int f(y) \nu^*_{\sm,Q}(y) \,\dd y-\int f(y) \wt{\nu}_{\sm ,Q}(y)\,\dd y\Big),\\
\end{align*}
where $ \bold{C}_1^1(\mb R^d)=\big\{f:\mb R^d\to \mb R\,|\, {\sup}_{z\in \mb R^d}\sum_{j\in \mb N_0^d, |j|\leq 1} |f^{(j)}(z)|\leq 1\big\}$. Then we consider $f\in \bold{C}_1^{1}(\mathbb{R}^d)$, we can get 
\begin{equation} 
\begin{aligned}
&\int f(y) \nu^*_{\sm,Q}(y) \,\dd y-\int f(y) \wt{\nu}_{\sm ,Q}(y)\,\dd y\\
&=\frac{1}{\wh p_k}\Big(\int p_k\cdot f(y) \nu^*_{\sm, Q}(y) \,\dd y-\int \wh p_k \cdot f(y) \wt{\nu}_{\sm,Q}(y)\,\dd y\Big)+\int f(y) \nu^*_{\sm,Q}(y) \,\dd y\cdot\big(1-\frac{p_k}{\wh p_k}\big)\\
&\leq\underbrace{\frac{1}{\wh p_k}\Big|\int f(y) \cdot p_k\cdot\nu^*_{\sm, Q}(y) \,\dd y-\int f(y)\cdot \mb E_{X^{(n)}}\big[\wh p_k\cdot \wt \nu_{\sm ,Q}(y)\big]\,\dd y\Big|}_{(A)}\\
&\quad+\underbrace{\frac{1}{\wh p_k}\Big|\int f(y)\cdot \mb E_{X^{(n)}}\big[\wh p_k\cdot\wt \nu_{\sm ,Q}(y)\big]\,\dd y-\int f(y)\cdot \wh p_k\cdot \wt \nu_{\sm ,Q}(y)\,\dd y\Big|}_{(B)} +\underbrace{\big|1-\frac{p_k}{\wh p_k}\big|}_{(C)}.
\end{aligned}
\end{equation}
First for term $(C)$,by Bernstein's inequality, it holds with probability at least $1-n^{-3}$ that $|p_k-\wh p_k|\leq C\, \sqrt{\frac{\log n}{n}}$, then by $p_k>0$, for large enough $n$, we have  $\big|1-\frac{p_k}{\wh p_k}\big|\leq  C\, \sqrt{\frac{ \log n}{n}}$. For bounding the term $(A)$,
 we use a similar strategy as in the proof of Lemma 4.3 of~\cite{Divol2022}. Recall $\nu^*_{k,Q}={Q}_{\#}[\frac{\mu^*\cdot \rho_k}{p_k}]$, we can write 
 \begin{equation*}
     \begin{aligned}
     \int f(y) \cdot \mb {E}_{X^{(n)}}\big[\wh p_k\cdot \wt \nu_{k,Q}(y)\big]\,\dd y&=  \int f(y) \cdot\frac{1}{h^d}\cdot \mb {E}_{\mu^*}\big[\wt k(\frac{y-Q(X)}{h})\cdot \rho_k(X)\big]\,\dd y\\
     &=\int \int f(y)\cdot \frac{1}{h^d}\cdot \wt k(\frac{y-z}{h})\cdot p_k\cdot \nu^*_{k,Q}(z)\,\dd z\,\dd y
     \end{aligned}
 \end{equation*}
Denote $ \upsilon(\cdot)=p_k\cdot \nu^*_{k,Q}(\cdot)$, we can obtain 
\begin{equation*}
    \begin{aligned}
   & \Big|\int f(y) \cdot p_k\cdot\nu^*_{\sm, Q}(y) \,\dd y-\int f(y)\cdot \mb E_{X^{(n)}}\big[\wh p_k\cdot \wt \nu_{\sm ,Q}(y)\big]\,\dd y\Big|\\
    &=\Big|\int f(y) \cdot \upsilon(y) \,\dd y-\int \int f(y)\cdot \frac{1}{h^d}\cdot \wt k(\frac{y-z}{h})\cdot \upsilon(z)\,\dd z\,\dd y\Big|\\
    &=\Big|\int \int f(y)\cdot \frac{1}{h^d}\cdot \wt k(\frac{y-z}{h})\cdot (\upsilon(z)-\upsilon(y))\,\dd z\,\dd y\Big|.\\
    \end{aligned}
\end{equation*}
When $\lfloor \tilde\alpha\rfloor$ is even, denote $s=\lfloor \tilde\alpha\rfloor$; when $\lfloor \tilde\alpha\rfloor$ is odd, denote $s=\lfloor \tilde\alpha\rfloor-1$. Then using Taylor's theorem, we can decompose 
\begin{equation*}
    \begin{aligned}
    \upsilon(z)-\upsilon(y)=\sum_{j\in \mb N_0^d\atop 1\leq |j|<s} \frac{\upsilon^{(j)}(y)}{j!}\cdot(z-y)^j+\sum_{j\in \mb N_0^d\atop |j|=s} \frac{s}{j!}\int_0^1 (1-t)^{s-1}\upsilon^{(j)}(y+t(z-y))\,\dd t\cdot (z-y)^j.
    \end{aligned}
\end{equation*}
Using the fact that for any $j\in \mb N_0^d$ with $1\leq |j|\leq \lfloor\tilde\alpha\rfloor$, $\int_{\mb R^d} \wt k(z) \cdot z^j\,\dd z=0$ and $\wt k(\cdot)=\wt k(-\cdot)$, we can obtain 
\begin{equation*}
\begin{aligned}
  &\int \int f(y)\cdot \frac{1}{h^d}\cdot \wt k(\frac{y-z}{h})\cdot \sum_{j\in \mb N_0^d\atop |j|<s} \frac{\upsilon^{(j)}(y)}{j!}\cdot(z-y)^j\,\dd z\,\dd y\\
  &\,=\int \int f(y)\cdot    \wt k(z)\cdot \sum_{j\in \mb N_0^d\atop |j|<s} \frac{\upsilon^{(j)}(y)}{j!}\cdot z^j\cdot h^{|j|}\,\dd z\,\dd y=0,
\end{aligned}
\end{equation*}
and 
\begin{equation*}
    \begin{aligned}
   & \int \int f(y)\cdot \frac{1}{h^d}\cdot \wt k(\frac{y-z}{h})\cdot  \sum_{j\in \mb N_0^d\atop |j|=s} \frac{s}{j!}\int_0^1 (1-t)^{s-1}\upsilon^{(j)}(y)\cdot (z-y)^j\,\dd t\,\dd z\,\dd y\\
    &=   \int \int f(y)\cdot   \wt k(z)\cdot  \sum_{j\in \mb N_0^d\atop |j|=s} \frac{s}{j!}\int_0^1 (1-t)^{s-1}\upsilon^{(j)}(y)\cdot z^j\cdot h^{s}\,\dd t\,\dd z\,\dd y=0.
    \end{aligned}
\end{equation*}
Therefore, we have 
\begin{equation*}
    \begin{aligned}
    &\int \int f(y)\cdot \frac{1}{h^d}\cdot \wt k(\frac{y-z}{h})\cdot (\upsilon(z)-\upsilon(y))\,\dd z\,\dd y\\
    &=\int \int f(y)\cdot \frac{1}{h^d}\cdot \wt k(\frac{y-z}{h})\cdot  \sum_{j\in \mb N_0^d\atop |j|=s} \frac{s}{j!}\int_0^1 (1-t)^{s-1}\big(\upsilon^{(j)}(y+t(z-y))-\upsilon^{(j)}(y)\big)\cdot (z-y)^j\,\dd t\,\dd z\,\dd y\\
    &\overset{(i)}{=}\int \int f(y)\cdot \frac{1}{t^dh^d}\cdot \wt k(\frac{x-y}{th})\cdot  \sum_{j\in \mb N_0^d\atop |j|=s} \frac{s}{j!}\int_0^1 (1-t)^{s-1}\big(\upsilon^{(j)}(x)-\upsilon^{(j)}(y)\big)\cdot \big(\frac{x-y}{t}\big)^j\,\dd t\,\dd x\,\dd y,\\
    \end{aligned}
\end{equation*}
where $(i)$ uses the change of variable $x=y+t(z-y)$ and $\wt k(\cdot)=\wt k(-\cdot)$. Then by switching the variable $x$ and $y$, using the facts that  $s$ is a even number and $k$ is an even function,  we have 
\begin{equation*}
    \begin{aligned}
    &\int \int f(y)\cdot \frac{1}{t^dh^d}\cdot \wt k(\frac{x-y}{th})\cdot  \sum_{j\in \mb N_0^d\atop |j|=s} \frac{s}{j!}\int_0^1 (1-t)^{s-1}\big(\upsilon^{(j)}(x)-\upsilon^{(j)}(y)\big)\cdot \big(\frac{x-y}{t}\big)^j\,\dd t\,\dd x\,\dd y\\
    &=\int \int f(x)\cdot \frac{1}{t^dh^d}\cdot \wt k(\frac{y-x}{th})\cdot  \sum_{j\in \mb N_0^d\atop |j|=s} \frac{s}{j!}\int_0^1 (1-t)^{s-1}\big(\upsilon^{(j)}(y)-\upsilon^{(j)}(x)\big)\cdot \big(\frac{y-x}{t}\big)^j\,\dd t\,\dd x\,\dd y\\
    &=-\int \int f(x)\cdot \frac{1}{t^dh^d}\cdot \wt k(\frac{x-y}{th})\cdot  \sum_{j\in \mb N_0^d\atop |j|=s} \frac{s}{j!}\int_0^1 (1-t)^{s-1}\big(\upsilon^{(j)}(x)-\upsilon^{(j)}(y)\big)\cdot \big(\frac{x-y}{t}\big)^j\,\dd t\,\dd x\,\dd y.
    \end{aligned}
\end{equation*}
Therefore, when $\lfloor\tilde \alpha\rfloor$ is even, we can obtain 
\begin{equation*}
    \begin{aligned}
    &\Big|\int \int f(y)\cdot \frac{1}{h^d}\cdot \wt k(\frac{y-z}{h})\cdot (\upsilon(z)-\upsilon(y))\,\dd z\,\dd y\Big|\\
    &=\Big|\frac{1}{2}\int \int (f(y)-f(x))\cdot \frac{1}{t^dh^d}\cdot \wt k(\frac{x-y}{th})\cdot  \sum_{j\in \mb N_0^d\atop |j|=s} \frac{s}{j!}\int_0^1 (1-t)^{s-1}\big(\upsilon^{(j)}(x)-\upsilon^{(j)}(y)\big)\cdot \big(\frac{x-y}{t}\big)^j\,\dd t\,\dd x\,\dd y\Big|\\
    &=\Big|\frac{1}{2}\int \int (f(y)-f(y+wth))\cdot \wt k(w)\cdot  \sum_{j\in \mb N_0^d\atop |j|=s} \frac{s}{j!}\int_0^1 (1-t)^{s-1}\big(\upsilon^{(j)}(y+wth)-\upsilon^{(j)}(y)\big)\cdot \big(wh\big)^j\,\dd t\,\dd w\,\dd y\Big|\\
    &\overset{(ii)}{\lesssim} h^{1+\tilde\alpha}\cdot\sum_{j\in \mb N_0^d\atop |j|=s} \frac{s}{j!}  \int_{[-L-h,L+h]^d} \int_{[-1,1]^d}  \|w\|^{1+\tilde\alpha} \cdot \wt k(w)\cdot  \int_0^1 (1-t)^{s-1}\,\dd t\,\dd w\,\dd y\\
    &\lesssim n^{-\frac{1+\tilde \alpha}{2\tilde \alpha+d}},
    \end{aligned}
\end{equation*}
where $(ii)$ uses $\upsilon(\cdot)=p_k\cdot \nu^*_{\sm ,Q}(\cdot)$ has support contained in $[-L,L]^d$ and $\wt k(\cdot)$ has support contained in $[-1,1]^d$.  On the other hand, when $\lfloor\tilde \alpha\rfloor$ is odd, recall $s=\lfloor\tilde \alpha\rfloor-1$ is even, we have
 \begin{equation*}
    \begin{aligned}
    &\Big|\int \int f(y)\cdot \frac{1}{h^d}\cdot \wt k(\frac{y-z}{h})\cdot (\upsilon(z)-\upsilon(y))\,\dd z\,\dd y\Big|\\
    &=\Big|\frac{1}{2}\int \int (f(x)-f(y))\cdot \frac{1}{t^dh^d}\cdot \wt k(\frac{x-y}{th})\cdot  \sum_{j\in \mb N_0^d\atop |j|=s} \frac{s}{j!}\int_0^1 (1-t)^{s-1}\big(\upsilon^{(j)}(x)-\upsilon^{(j)}(y)\big)\cdot \big(\frac{x-y}{t}\big)^j\,\dd t\,\dd x\,\dd y\Big|\\
    &\leq \bigg|\frac{1}{2}\int \int (f(x)-f(y))\cdot \frac{1}{t^dh^d}\cdot \wt k(\frac{x-y}{th})\\
    &\qquad\cdot  \sum_{j\in \mb N_0^d\atop |j|=s} \frac{s}{j!}\int_0^1 (1-t)^{s-1}\Big(\upsilon^{(j)}(x)-\upsilon^{(j)}(y)-\sum_{q\in \mb N_0^d\atop |q|=1} \upsilon^{(j+q)}(y)\cdot (x-y)^q\Big)\cdot \big(\frac{x-y}{t}\big)^j\,\dd t\,\dd x\,\dd y\bigg|\\
    &+\bigg|\frac{1}{2}\int \int (f(x)-f(y))\cdot \frac{1}{t^dh^d}\cdot \wt k(\frac{x-y}{th})\cdot  \sum_{j\in \mb N_0^d\atop |j|=s} \frac{s}{j!}\int_0^1 (1-t)^{s-1}\sum_{q\in \mb N_0^d\atop |q|=1} \upsilon^{(j+q)}(y)\cdot (x-y)^q\cdot \big(\frac{x-y}{t}\big)^j\,\dd t\,\dd x\,\dd y\bigg|\\
    &\leq C\,h^{1+\tilde \alpha}\\
    &\quad+\underbrace{\bigg|\frac{1}{2}\int \int (f(x)-f(y))\cdot \frac{1}{t^dh^d}\cdot \wt k(\frac{x-y}{th})\cdot  \sum_{j\in \mb N_0^d\atop |j|=s} \frac{s}{j!}\int_0^1 (1-t)^{s-1}\sum_{q\in \mb N_0^d\atop |q|=1} \upsilon^{(j+q)}(y)\cdot (x-y)^q\cdot \big(\frac{x-y}{t}\big)^j\,\dd t\,\dd x\,\dd y\bigg|}_{(D)}.\\
    \end{aligned}
\end{equation*} 
Then for the term $(D)$,  using Taylor's theorem and $f\in \bold{C}_1^1(\mb R^d)$, we can write 
\begin{equation}\label{taylorf}
   f(x)-f(y)=\sum_{j\in \mb N_0^d\atop |j|=1} \int_0^1 f^{(j)}(y+t(x-y))\,\dd t\cdot (x-y)^j. 
\end{equation}
So using the fact that for any $j\in \mb N_0^d$ with $1\leq |j|\leq \lfloor\tilde\alpha\rfloor+1$, $\int_{\mb R^d} \wt k(z) \cdot z^j\,\dd z=0$, we can obtain
\begin{equation*}
    \begin{aligned}
    & \int \int \sum_{l\in \mb N_0^d\atop |l|=1} f^{(l)}(y) \cdot (x-y)^l\cdot \frac{1}{t^dh^d}\cdot \wt k(\frac{x-y}{th})\cdot  \sum_{j\in \mb N_0^d\atop |j|=s} \frac{s}{j!}\int_0^1 (1-t)^{s-1}\sum_{q\in \mb N_0^d\atop |q|=1} \upsilon^{(j+q)}(y)\cdot (x-y)^q\cdot \big(\frac{x-y}{t}\big)^j\,\dd t\,\dd x\,\dd y \\
    &=\sum_{l\in \mb N_0^d\atop |l|=1}\sum_{j\in \mb N_0^d\atop |j|=s} \sum_{q\in \mb N_0^d\atop |q|=1} \int_0^1\frac{1}{t^dh^d}\cdot   \frac{s}{j!}\cdot\int \int f^{(l)}(y) \cdot (x-y)^l\cdot \wt k(\frac{x-y}{th}) (1-t)^{s-1} \upsilon^{(j+q)}(y)\cdot (x-y)^q\cdot \big(\frac{x-y}{t}\big)^j\,\dd x\,\dd y \,\dd t\\
    &=0.
    \end{aligned}
\end{equation*}
Therefore,
\begin{equation*}
    \begin{aligned}
    (D)&=\bigg|\frac{1}{2}\int \int \sum_{l\in \mb N_0^d\atop |l|=1} \int_0^1 \big(f^{(l)}(y+\omega(x-y))-f^{(l)}(y)\big)\,\dd \omega\cdot (x-y)^l \\
   &\quad\cdot \frac{1}{t^dh^d}\cdot \wt k(\frac{x-y}{th})\cdot  \sum_{j\in \mb N_0^d\atop |j|=s} \frac{s}{j!}\int_0^1 (1-t)^{s-1}\sum_{q\in \mb N_0^d\atop |q|=1} \upsilon^{(j+q)}(y)\cdot (x-y)^q\cdot \big(\frac{x-y}{t}\big)^j\,\dd t\,\dd x\,\dd y\bigg|\\
   &\overset{(iii)}{=} \bigg|\frac{1}{2}\int \int\int_0^1\int_0^1 \sum_{l\in \mb N_0^d\atop |l|=1}   \big(f^{(l)}(\theta)-f^{(l)}(y)\big)\cdot \big(\frac{\theta-y}{\omega}\big)^l \\
   &\quad\cdot \frac{1}{\omega^dt^dh^d}\cdot \wt k(\frac{\theta-y}{\omega th})\cdot  \sum_{j\in \mb N_0^d\atop |j|=s} \frac{s}{j!} (1-t)^{s-1}\sum_{q\in \mb N_0^d\atop |q|=1} \upsilon^{(j+q)}(y)\cdot \big(\frac{\theta-y}{\omega}\big)^q\cdot \big(\frac{\theta-y}{\omega t}\big)^j\,\dd \omega\,\dd t\,\dd \theta\,\dd y\bigg|,\\
    \end{aligned}
\end{equation*}
where $(iii)$ uses the change of variable $y+\omega(x-y)=\theta$. By switching the variable $\theta$ and $y$, we have 
\begin{equation*}
    \begin{aligned}
    &\int \int\int_0^1\int_0^1 \sum_{l\in \mb N_0^d\atop |l|=1}   \big(f^{(l)}(\theta)-f^{(l)}(y)\big)\cdot \big(\frac{\theta-y}{\omega}\big)^l \\
   &\quad\cdot \frac{1}{\omega^dt^dh^d}\cdot \wt k(\frac{\theta-y}{\omega th})\cdot  \sum_{j\in \mb N_0^d\atop |j|=s} \frac{s}{j!} (1-t)^{s-1}\sum_{q\in \mb N_0^d\atop |q|=1} \upsilon^{(j+q)}(y)\cdot \big(\frac{\theta-y}{\omega}\big)^q\cdot \big(\frac{\theta-y}{\omega t}\big)^j\,\dd \omega\,\dd t\,\dd \theta\,\dd y\\
   &= \int \int\int_0^1\int_0^1 \sum_{l\in \mb N_0^d\atop |l|=1}   \big(f^{(l)}(y)-f^{(l)}(\theta)\big)\cdot \big(\frac{y-\theta}{\omega}\big)^l \\
   &\quad\cdot \frac{1}{\omega^dt^dh^d}\cdot \wt k(\frac{y-\theta}{\omega th})\cdot  \sum_{j\in \mb N_0^d\atop |j|=s} \frac{s}{j!} (1-t)^{s-1}\sum_{q\in \mb N_0^d\atop |q|=1} \upsilon^{(j+q)}(\theta)\cdot \big(\frac{y-\theta}{\omega}\big)^q\cdot \big(\frac{y-\theta}{\omega t}\big)^j\,\dd \omega\,\dd t\,\dd \theta\,\dd y\\
   &=-\int \int\int_0^1\int_0^1 \sum_{l\in \mb N_0^d\atop |l|=1}   \big(f^{(l)}(\theta)-f^{(l)}(y)\big)\cdot \big(\frac{\theta-y}{\omega}\big)^l \\
   &\quad\cdot \frac{1}{\omega^dt^dh^d}\cdot \wt k(\frac{\theta-y}{\omega th})\cdot  \sum_{j\in \mb N_0^d\atop |j|=s} \frac{s}{j!} (1-t)^{s-1}\sum_{q\in \mb N_0^d\atop |q|=1} \upsilon^{(j+q)}(\theta)\cdot \big(\frac{\theta-y}{\omega}\big)^q\cdot \big(\frac{\theta-y}{\omega t}\big)^j\,\dd \omega\,\dd t\,\dd \theta\,\dd y,\\
    \end{aligned}
\end{equation*}
which leads to 
\begin{equation*}
    \begin{aligned}
 &\bigg|\frac{1}{2}\int \int\int_0^1\int_0^1 \sum_{l\in \mb N_0^d\atop |l|=1}   \big(f^{(l)}(\theta)-f^{(l)}(y)\big)\cdot \big(\frac{\theta-y}{\omega}\big)^l \\
   &\quad\cdot \frac{1}{\omega^dt^dh^d}\cdot \wt k(\frac{\theta-y}{\omega th})\cdot  \sum_{j\in \mb N_0^d\atop |j|=s} \frac{s}{j!} (1-t)^{s-1}\sum_{q\in \mb N_0^d\atop |q|=1} \upsilon^{(j+q)}(y)\cdot \big(\frac{\theta-y}{\omega}\big)^q\cdot \big(\frac{\theta-y}{\omega t}\big)^j\,\dd \omega\,\dd t\,\dd \theta\,\dd y\bigg|,\\
   &=\bigg|\frac{1}{4}\int \int\int_0^1\int_0^1 \sum_{l\in \mb N_0^d\atop |l|=1}   \big(f^{(l)}(\theta)-f^{(l)}(y)\big)\cdot \big(\frac{\theta-y}{\omega}\big)^l \\
   &\quad\cdot \frac{1}{\omega^dt^dh^d}\cdot \wt k(\frac{\theta-y}{\omega th})\cdot  \sum_{j\in \mb N_0^d\atop |j|=s} \frac{s}{j!} (1-t)^{s-1}\sum_{q\in \mb N_0^d\atop |q|=1} \big(\upsilon^{(j+q)}(y)-\upsilon^{(j+q)}(\theta)\big)\cdot \big(\frac{\theta-y}{\omega}\big)^q\cdot \big(\frac{\theta-y}{\omega t}\big)^j\,\dd \omega\,\dd t\,\dd \theta\,\dd y\bigg|\\
   \lesssim h^{1+\tilde\alpha},
    \end{aligned}
\end{equation*}
where the last inequality uses the fact that $s=\lfloor\tilde \alpha\rfloor-1$ and $\upsilon$ is $\tilde \alpha$ is smooth. We can then obtain that 
\begin{equation*}
    (A)\lesssim  n^{-\frac{\tilde \alpha+1}{2\tilde \alpha+d}}.
\end{equation*}
It remains to bound term $(B)$,  which is 
\begin{equation*}
    \begin{aligned}
   (B)&= \Big|\int f(y)\cdot \mb E_{X^{(n)}}\big[\wh p_k\cdot\wt \nu_{\sm ,Q}(y)\big]\,\dd y-\int f(y)\cdot \wh p_k\cdot \wt \nu_{\sm ,Q}(y)\,\dd y\Big|\\
   &=\bigg|\frac{1}{n} \sum_{i=1}^n \int f(y)\cdot \frac{1}{h^d}\cdot \wt k\big(\frac{y-Q(X_i)}{h}\big)\cdot \rho_k(X_i)\,\dd y-\mb{E}_{\mu^*}\Big[\int f(y)\cdot \frac{1}{h^d}\cdot \wt k\big(\frac{y-Q(X)}{h}\big)\cdot \rho_k(X)\,\dd y\Big]\bigg|.
    \end{aligned}
\end{equation*}
By standard symmetrization, we can get 
\begin{equation*}
\begin{aligned}
& \mathbb{E}\Bigg[\underset{f\in \bold{C}_1^1(\mathbb{R}^d)\atop Q\in \wt{\mathcal{Q}}_k}{\sup} \bigg|\frac{1}{n} \sum_{i=1}^n \int f(y)\cdot \frac{1}{h^d}\cdot \wt k\big(\frac{y-Q(X_i)}{h}\big)\cdot \rho_k(X_i)\,\dd y-\mathbb{E} \bigg[\int f(y)\cdot \frac{1}{h^d}\cdot k\big(\frac{y-Q(X)}{h}\big)\cdot \rho_k(X)\,\dd y\bigg]\bigg|\Bigg]\\
&\leq 2 \mathbb{E}\bigg[ \underset{f\in \bold{C}_1^1(\mathbb{R}^d)\atop Q\in \wt{\mathcal{Q}}_k}{\sup} \bigg|\frac{1}{n}  \sum_{i=1}^n \varepsilon_i \int f(y)\cdot \frac{1}{h^d}\cdot \wt k\big(\frac{y-Q(X_i)}{h}\big)\cdot \rho_k(X_i)\,\dd y\bigg|\bigg],
\end{aligned}
\end{equation*}
 where $\{\varepsilon_i\}_{i=1}^n$ are $n$ i.i.d. copies from Rademacher distribution, i.e. $P(\varepsilon_i = 1) = P(\varepsilon_i= -1) = 0.5$. Define function set 
 \begin{equation*}
 \mathcal{F}=\left\{f: \m M\to \mathbb{R}\,|\, f(x)=\int g(y)\cdot \frac{1}{h^d}\cdot \wt k\big(\frac{y-Q(x)}{h}\big)\cdot \rho_k(x)\,\dd y; \,g\in \bold{C}_1^1(\mb R^d);\, Q\in\wt{\mathcal{Q}}_k\right\}.
 \end{equation*}
Since there exists a constant $L$ so that ${\rm supp}(Q)=[-L,L]^d$, we can first consider the function set 
 \begin{equation*}
 \mathcal{F}_1=\left\{f:  [-L,L]^d\to \mathbb{R}\,|\,  f(z)=\int g(y)\cdot \frac{1}{h^d}\cdot \wt k\big(\frac{y-z}{h}\big)\,\dd y;\,g\in \bold{C}_1^1(\mb R^d)\right\}.
  \end{equation*}
 For any $f\in \m F_1$ and $j\in \mb N_0^d$ with $1\leq |j|\leq \lfloor\frac{d}{2}\rfloor$, we have 
  \begin{equation*}
  \begin{aligned}
     f^{(j)}(z)&=\int g(y)\frac{1}{h^{d+|j|}} \wt k^{(j)}(\frac{z-y}{h})\,\dd y\\
     &=\frac{1}{h^{|j|}}\int g(z-ht) \wt k^{(j)}(t)\,\dd t.\\
  \end{aligned}
  \end{equation*}
 Since $|j|\geq 1$, there exists $j_1\in \mb N_0^d$ with $|j_1|=1$ so that every element in $j-j_1$ is non-negative. Then we have 
  \begin{equation*}
  \begin{aligned}
     |f^{(j)}(z)|&=\Big|\frac{1}{h^{|j|-1}}\int g^{(j_1)}(z-ht) \wt k^{(j-j_1)}(t)\,\dd t\Big|\lesssim \frac{1}{h^{|j|-1}},\\
  \end{aligned}
  \end{equation*}
  where the last inequality uses $g\in \bold{C}_1^1(\mb R^d)$.  Moreover, when $d\geq 2$, for any  $j\in \mb N_0^d$ with $ |j|= \lfloor\frac{d}{2}\rfloor$  and $z_1,z_2\in \mb R^d$ it holds that 
  \begin{equation*}
  \begin{aligned}
     |f^{(j)}(z_1)-f^{(j)}(z_2)|&
     =\big|\frac{1}{h^{\lfloor\frac{d}{2}\rfloor-1}}\int g^{(j_1)}(z_1-ht) \wt k^{(j-j_1)}(t)\,\dd t-\frac{1}{h^{\lfloor\frac{d}{2}\rfloor-1}}\int g^{(j_1)}(z_2-ht) \wt k^{(j-j_1)}(t)\,\dd t\big|\\
    &=\Big|\frac{1}{h^{\lfloor\frac{d}{2}\rfloor-1}}\int \frac{1}{h^d} g^{(j_1)}(y) \big(\wt k^{(j-j_1)}(\frac{z_1-y}{h})-\wt k^{(j-j_1)}(\frac{z_2-y}{h})\big)\,\dd y\Big|\\
  \end{aligned}
  \end{equation*}
  If $\|z_1-z_2\|\leq h$, then 
   \begin{equation*}
  \begin{aligned}
     |f^{(j)}(z_1)-f^{(j)}(z_2)|&=\Big|\frac{1}{h^{\lfloor\frac{d}{2}\rfloor-1}}\int \frac{1}{h^d} g^{(j_1)}(y) \big(\wt k^{(j-j_1)}(\frac{z_1-y}{h})-\wt k^{(j-j_1)}(\frac{z_2-y}{h})\big)\,\dd y\Big|\\
     &\leq C\, \frac{1}{h^{\lfloor\frac{d}{2}\rfloor}}\|z_1-z_2\|\leq  C\, \frac{1}{h^{\frac{d}{2}-1}}\|z_1-z_2\|^{\frac{d}{2}-\lfloor\frac{d}{2}\rfloor};
  \end{aligned}
  \end{equation*}
  If $\|z_1-z_2\|\geq h$, then 
  \begin{equation*}
  \begin{aligned}
     |f^{(j)}(z_1)-f^{(j)}(z_2)|\leq  |f^{(j)}(z_1)|+|f^{(j)}(z_2)|\leq C\, \frac{1}{h^{\lfloor\frac{d}{2}\rfloor-1}}  \leq  C\, \frac{1}{h^{\frac{d}{2}-1}}\|z_1-z_2\|^{\frac{d}{2}-\lfloor\frac{d}{2}\rfloor}.
  \end{aligned}
  \end{equation*}
  So we can get when $d\geq 2$,   there exists a constant $C$ such that for any $f\in \mathcal{F}_1$,  it holds that 
\begin{equation*}
 C\, h^{\frac{d}{2}-1} f\in C^{\frac{d}{2}}_1 ( [-L,L]^d).
\end{equation*}
  Similarly, when $d=1$, we have 
for any $z_1,z_2\in \mb R^d$ it holds that 
  \begin{equation*}
  \begin{aligned}
     |f(z_1)-f(z_2)|&
     =\big|\int (g(z_1-ht)-g(z_2-ht))\cdot  \wt k\big(t\big)\,\dd y\big|\lesssim \|z_1-z_2\|.\\
  \end{aligned}
  \end{equation*}
  Therefore, we can conclude that 
  \begin{equation*}
    \m F_1\in C^{\frac{d}{2}}_{L_1\, h^{-(\frac{d}{2}-1)_{+}} }( [-L,L]^d),
  \end{equation*}
  for a constant $L_1$, where $(a)_{+}=\max(a,0)$. So  for any $\epsilon>0$, we can find a set $N^f_{\epsilon}\subseteq \mathcal{F}_1$ such that $\sqrt{\log |N^f_{\epsilon}|}\lesssim \frac{  h^{-(\frac{d}{2}-1)_{+}}}{\epsilon}$ and for any $f\in \mathcal{F}_1$, there exists $\wt{f}\in N^f_{\epsilon}$ such that 
\begin{equation*}
\underset{y\in[-L,L]^d}{\sup}|f(y)-\wt{f}(y)|\leq \epsilon. \end{equation*}
Then, to derive an $\epsilon$-covering number for $\m F$, we introduce the following lemma. 
\begin{lemma}\label{coveringmanifold}
(Lemma 12 of~\cite{tang2022minimax}) Let $\mathcal{X}_G=\big\{x\in \mathbb{R}^D:\, x=G(z), z\in \mb B_1^d\big\}$ be a $d$-dimensional submanifold induced by a Lipschitz continuous map $G:\,\mb R^d\to \mb R^D$, then it holds for any $\widetilde \gamma>0$ that
\begin{equation*}
\log \bold{N} \big(C_1^{\widetilde{\gamma}}(\mathbb{R}^D), \,\|\cdot\|_{L^{\infty}(\mathcal{X}_{G})},\,\epsilon \big)\leq C\,\epsilon^{-\frac{d}{\widetilde{\gamma}}}, \quad \forall \epsilon>0,
\end{equation*}
where $\bold{N}(\mathcal{F},\,\widetilde{d}, \,\epsilon)$ denotes the $\epsilon$-covering number of function space $\mathcal{F}$ with respect to pseudo-metric $\widetilde{d}$, and $\|f\|_{L^{\infty}(\mathcal{X}_{G})}=\underset{x\in \mathcal{X}_{G}} {\sup }\big|f(x)\big|$ denotes the functional supreme norm constrained on set $\m X_G$.
\end{lemma}
Then since $\Omega_k=Q_\sm^\ast(\m M\cap S_k)$ is compactly supported and $\m M \cap S_k= G_\sm^\ast(\Omega_k)$,  for any $\epsilon>0$, we can find a function set $N^Q_{\epsilon}$ such that 
 $\sqrt{\log |N^Q_{\epsilon}|} \lesssim (\frac{1}{\epsilon})^{\frac{d}{2\beta}}$  and for any $Q\in \wt{\mathcal{Q}}_k$, there exists $\wt{Q}\in N^Q_{\epsilon}$ such that 
 \begin{equation*}
 \underset{x\in \m M\cap S_k}{\sup}\|\wt{Q}(x)-Q(x)\|\leq \epsilon.
 \end{equation*}
  Then for any  $Q\in\wt{\mathcal{Q}}_k$ and $f\in\mathcal{F}_1$, there exists $\wt{Q}\in N^Q_{\epsilon}$, $\wt{f}\in N^f_{\epsilon}$ and a constant $c$  such that 
 \begin{equation*}
 \begin{aligned}
& \left|f(Q(x))\rho_k(x)-\wt{f}(\wt{Q}(x))\rho_k(x)\right|\\
 &\leq  \left|f(Q(x))\rho_k(x)-\wt{f}(Q(x))\rho_k(x)\right|+\left|\wt{f}(Q(x))\rho_k(x)-\wt{f}(\wt{Q}(x))\rho_k(x)\right|\\
 &\leq  c\epsilon.
 \end{aligned}
\end{equation*}
So we can get 
\begin{equation*}
\log \bold{N}(\mathcal{F},\epsilon, \|\cdot\|_{\infty})\lesssim \left(\frac{1}{\epsilon}\right)^{\frac{d}{\beta}}+\left(\frac{h^{-(\frac{d}{2}-1)_{+}}}{\epsilon}\right)^2. 
\end{equation*}
Choose $\delta=\big(\frac{1}{n}\big)^{\frac{\tilde\alpha+1}{2\tilde\alpha+d}} \vee \big(\frac{1}{n}\big)^{ \frac{\beta}{d}}$, we can get 
\begin{equation*}
 \begin{aligned}
&\frac{1}{\sqrt{n}}\int_{\delta}^1 \left[\left(\frac{1}{\epsilon}\right)^{\frac{d}{2\beta}}+\frac{h^{1-\frac{d}{2}}\vee 1}{\epsilon}\right] \,\dd\epsilon \\
&\lesssim \frac{\log n}{\sqrt{n}}+  \big(\frac{1}{n}\big)^{\frac{\tilde\alpha+1}{2\tilde\alpha+d}}+ \big(\frac{1}{n}\big)^{ \frac{\beta}{d}}. 
 \end{aligned}
\end{equation*}
By Dudley's entropy integral bound (see for example, Theorem 5.22 of~\cite{wainwright_2019}),  it holds that 
\begin{equation*}
 \begin{aligned}
&\mathbb{E}\bigg[ \underset{f\in \bold{C}_1^1(\mathbb{R}^d)\atop Q\in \wt{\mathcal{Q}}_k}{\sup} \bigg|\frac{1}{n}  \sum_{i=1}^n \varepsilon_i \int f(y)\cdot \frac{1}{h^d}\cdot \wt k\big(\frac{y-Q(X_i)}{h}\big)\cdot \rho_k(X_i)\,\dd y\bigg|\bigg]\\
&\lesssim  \frac{\log n}{\sqrt{n}}+  \big(\frac{1}{n}\big)^{\frac{\tilde\alpha+1}{2\tilde\alpha+d}}+ \big(\frac{1}{n}\big)^{ \frac{\beta}{d}}.
 \end{aligned}
\end{equation*}
 The statement is then followed by Talagrand concentration inequality (see for example, Theorem 3.27 of ~\cite{wainwright_2019}) and the fact that $\tilde\alpha+1\leq \beta$.

\subsection{Proof of Lemma~\ref{leup1}: Wavelet estimator}
Fix an arbitrary $k\in [K]$. Since $\wt{\mathcal{Q}}_k\subseteq C^{\beta}_L(\mb R^D;\mb R^d)$, it holds that for any $Q\in \wt{\mathcal{Q}}_k$, $\text{supp}(\nu^*_{\sm,Q})\subseteq [-L, L]^d$, where $\nu^*_{\sm,Q}$ is the density of the  push-forward measure of $\frac{\mu^*\cdot\rho_k}{p_k}$ by map $Q$. Moreover, by 
$\nu^*_{\sm,Q}\in C^{\alpha}_{L} (\mathbb{R}^d)$ with support contained in $[-L, L]^d$ and $0<p_k=\mb{E}_{\mu^\ast}[\rho_k]\leq 1$, we can write $p_k\cdot\nu^*_{\sm,Q}(y)$ as
\begin{equation*}
p_k\cdot\nu^*_{\sm,Q}(y)=\sum_{m \in  \mathbb{S}}a^Q_m \phi_m(y) +\sum_{l=1}^{2^{d}-1}\sum_{j=0}^{+\infty}\sum_{m\in \mathbb{S}_{lj}}\theta_{ljm}^Q \psi_{ljm}(y),\\
\end{equation*}
where $\{\phi_m, \psi_{ljm}:\, l=1,\cdots, 2^{d}-1,j \in \mathbb{N}, m\in \mathbb{Z}^d\}$ is the orthonormal wavelet basis  for Besov space on $\mathbb{R}^d$ defined as $\phi_m(y)=\phi(y-m)$ and $\psi_{ljm}(y)=2^{\frac{jd}{2}}\psi_l (2^jy-m)$, and it holds that $\phi(\cdot)$ and $\psi_l(\cdot)$ for any $1\leq l\leq 2^d-1$ are compactly supported and have bounded $\beta$ order derivatives~\citep{doi:10.1080/03610926.2015.1019144}. Then there exists a constant $C$ such that $|\theta_{ljm}^Q|\leq C (2^{-dj})^{\frac{\alpha}{d}+\frac{1}{2}}$ and $a^Q_m\leq C$. Recall that 
\begin{equation*}
 \wh{p}_k\cdot\wt{\nu}_{\sm,Q}(y)=\sum_{m \in  \mathbb{S}}\wt{a}^Q_m \phi_m(y) +\sum_{l=1}^{2^{d}-1}\sum_{j=0}^J\sum_{m\in \mathbb{S}_{lj}}\wt{\theta}_{ljm}^Q \psi_{ljm}(y),\\
 \end{equation*}
 with
\begin{equation*}
\begin{aligned}
\wt{a}^Q_{m}=\frac{1}{n}\sum_{i=1}^n \phi_m(Q(X_i))\rho_k(X_i);\\
\wt{\theta}_{ljm}^Q=\frac{1}{n}\sum_{i=1}^n \psi_{ljm}(Q(X_i))\rho_k(X_i).
 \end{aligned}
\end{equation*}
We have 
\begin{equation*}
\begin{aligned}
&\mathbb{E} [\wt{a}^Q_m]= p_k\cdot\int \phi_m(y) \nu^*_{\sm,Q}(y)dy=a^Q_m,\quad \mathbb{E}[\wt{\theta}_{ljm}^Q]=p_k\cdot\int   \psi_{ljm}(y) \nu^*_{\sm,Q}(y) dy=\theta_{ljm}^Q.
\end{aligned}
\end{equation*}
Moreover, by  the fact that $\phi(\cdot)$ and $\psi_l(\cdot)$ are compactly supported, we can get that there exists a constant $C$ such that for any $1\leq l\leq 2^d-1$ and $j \in \mathbb{N}$, it holds that $|\mathbb S_{lj}|\leq C2^{dj}$ and $| \mathbb{S}|\leq C$.  Since $\nu^*_{\sm ,Q}$ and $\wt\nu_{\sm ,Q}$ are both compactly supported.  There exists a constant $C$ so that for any $Q\in \wt{\m Q}_k$,
 \begin{align*}
\underset{f\in {\rm Lip}_1(\mb R^d)}{\sup}\Big(\int f(y) \nu^*_{\sm,Q}(y) \,\dd y-\int f(y) \wt{\nu}_{\sm ,Q}(y)\,\dd y\Big)\leq C\, \underset{f\in C_1^1(\mb R^d)}{\sup}\Big(\int f(y) \nu^*_{\sm,Q}(y) \,\dd y-\int f(y) \wt{\nu}_{\sm ,Q}(y)\,\dd y\Big).\\
\end{align*}
Then we consider $f\in C_1^{1}(\mathbb{R}^d)$, similarly, we can rewrite 
\begin{equation*}
f(y)=\sum_{m \in \mathbb{Z}^d} b_m \phi_m(y) +\sum_{l=1}^{2^{d}-1}\sum_{j=0}^{+\infty}\sum_{m\in \mathbb{Z}^d}f_{ljm} \psi_{ljm}(y)
\end{equation*}
where $|f_{ljm}|\leq C_1(2^{-dj})^{\frac{1}{d}+\frac{1}{2}}$ and $|b_m|\leq C_1$. So we can get 
\begin{equation}
\begin{aligned}
&\int f(y) \nu^*_{\sm,Q}(y) dy-\int f(y) \wt{\nu}_{\sm ,Q}(y)dy\\
&=\frac{1}{\wh p_k}\Big(\int f(y) p_k\cdot\nu^*_{\sm, Q}(y) dy-\int \wh p_k \cdot f(y) \wt{\nu}_{\sm,Q}(y)dy\Big)+\int f(y) \nu^*_{\sm,Q}(y) dy\cdot\big(1-\frac{p_k}{\wh p_k}\big)\\
&=\frac{1}{\wh p_k}\int  f(y) \left( \sum_{m \in  \mathbb{S}}(\wt{a}^Q_m-\mathbb{E}\wt{a}_m^Q) \phi_m(y) +\sum_{l=1}^{2^{d}-1}\sum_{j=0}^J\sum_{m\in \mathbb{S}_{lj}}(\wt{\theta}_{ljm}^Q-\mathbb{E}\wt{\theta}_{ljm}^Q)\psi_{ljm}(y)\right) dy\\
&+\frac{1}{\wh p_k}\int   f(y) \left(\sum_{l=1}^{2^{d}-1}\sum_{j=J}^{\infty}\sum_{m\in \mathbb{S}_{lj}}\theta_{ljm}^Q\psi_{ljm}(y)\right)dy+\int f(y) \nu^*_{\sm,Q}(y) dy\cdot\big(1-\frac{p_k}{\wh p_k}\big)\\
&\leq  \frac{1}{\wh p_k}\underbrace{\left|\frac{1}{n} \sum_{i=1}^n \sum_{l=1}^{2^d-1}\sum_{j=0}^J \sum_{m\in \mathbb{S}_{lj}} f_{ljm} \psi_{ljm}\left(Q(X_i)\right)\rho_k(X_i)-\mathbb{E} \bigg[\sum_{l=1}^{2^d-1}\sum_{j=0}^J \sum_{m \in \mathbb{S}_{lj}} f_{ljm} \psi_{ljm}\left(Q(X) \right)\rho_k(X)\bigg]\right|}_{(A)}\\
&+\frac{1}{\wh p_k}\underbrace{\left|\frac{1}{n}\sum_{i=1}^n \sum_{m\in  \mathbb{S}}b_m\phi_m(Q(X_i))\rho_k(X_i)-\mathbb{E} \Big[\sum_{m\in  \mathbb{S}}b_m\phi_m(Q(X))\rho_k(X)\Big]\right|}_{(B)} + \frac{1}{\wh p_k} \underbrace{\sum_{l=1}^{2^d-1}\sum_{j=J}^{+\infty}\sum_{m\in \mathbb{S}_{lj}} f_{ljm}\theta^Q_{ljm}}_{(C)}+\underbrace{\big|1-\frac{p_k}{\wh p_k}\big|}_{(D)}.
\end{aligned}
\end{equation}
First for term $(D)$,by Bernstein's inequality, it holds with probability at least $1-n^{-3}$ that $|p_k-\wh p_k|\leq C\, \sqrt{\frac{\log n}{n}}$, then by $p_k>0$, for large enough $n$, we have  $\big|1-\frac{p_k}{\wh p_k}\big|\leq  C\, \sqrt{\frac{ \log n}{n}}$. Moreover, for term $(C)$, since $|f_{ljm}|\lesssim  (2^{-dj})^{\frac{1}{d}+\frac{1}{2}}$,  $|\theta^Q_{ljm}|\lesssim  (2^{-dj})^{\frac{\alpha}{d}+\frac{1}{2}}$ and $2^{dJ} \asymp n^{\frac{d}{2\alpha+d}}$, we can get 
\begin{equation*}
 \sum_{j=J+1}^{+\infty}  \sum_{l=1}^{2^d-1} \sum_{m\in \mathbb{S}_{lj}} f_{ljm}\theta^Q_{ljm}\lesssim n^{-\frac{\alpha+1}{2\alpha+d}}.
\end{equation*}
Then for term $(A)$, by standard symmetrization, we can get 
\begin{equation*}
\begin{aligned}
& \mathbb{E}\Bigg[\underset{f\in C_1^1(\mathbb{R}^d)\atop Q\in \wt{\mathcal{Q}}_k}{\sup} \bigg|\frac{1}{n} \sum_{i=1}^n \sum_{l=1}^{2^d-1}\sum_{j=0}^J \sum_{m\in \mathbb{S}_{lj}} f_{ljm} \psi_{ljm}\left(Q(X_i)\right)\rho_k(X_i)-\mathbb{E} \bigg[\sum_{l=1}^{2^d-1}\sum_{j=0}^J \sum_{m \in \mathbb{S}_{lj}} f_{ljm} \psi_{ljm}\left(Q(X) \right)\rho_k(X)\bigg]\bigg|\Bigg]\\
&\leq 2 \mathbb{E}\bigg[ \underset{f\in C_1^1(\mathbb{R}^d)\atop Q\in \wt{\mathcal{Q}}_k}{\sup} \bigg|\frac{1}{n} \sum_{l=1}^{2^d-1} \sum_{i=1}^n \varepsilon_i \sum_{j=0}^J  \sum_{m\in \mathbb{S}_{lj}} f_{ljm} \psi_{ljm}\left(Q(X_i) \right)\rho_k(X_i)\bigg|\bigg],
\end{aligned}
\end{equation*}
 where $\{\varepsilon_i\}_{i=1}^n$ are $n$ i.i.d. copies from Rademacher distribution, i.e. $P(\varepsilon_i = 1) = P(\varepsilon_i= -1) = 0.5$. Define function set 
 \begin{equation*}
 \mathcal{F}=\left\{f: \m M\to \mathbb{R}: f(z)=\sum_{l=1}^{2^d-1}\sum_{j=0}^J \sum_{m \in \mathbb{S}_{lj}} f_{ljm} \psi_{ljm}(Q(x)); |f_{ljm}|\leq (2^{-dj})^{\frac{1}{d}+\frac{1}{2}}; Q\in\wt{\mathcal{Q}}_k\right\}.
 \end{equation*}
First we consider the function set 
 \begin{equation*}
 \mathcal{F}_1=\left\{f:  [-L_3,L_3]^d\to \mathbb{R},  f(y)=\sum_{l=1}^{2^d-1}\sum_{j=0}^J \sum_{m \in \mathbb{S}_{lj}} f_{ljm} \psi_{ljm}(y), |f_{ljm}|\leq (2^{-dj})^{\frac{1}{d}+\frac{1}{2}}\right\}.
  \end{equation*}
If $\frac{1}{d}\leq\frac{1}{2}$, then there exists a constant $c$ such that for any $f\in \mathcal{F}_1$,  it holds that 
\begin{equation*}
c  (2^{dJ})^{\frac{1}{d}-\frac{1}{2}} f\in C^{\frac{d}{2}}_1 ( [-L,L]^d).
\end{equation*}
So  for any $\epsilon>0$, we can find a set $N^f_{\epsilon}\subseteq \mathcal{F}_1$ such that $\sqrt{\log |N^f_{\epsilon}|}\lesssim \frac{  (2^{dJ})^{(\frac{1}{2}-\frac{1}{d})_{+}}}{\epsilon}$ and for any $f\in \mathcal{F}_1$, there exists $\wt{f}\in N^f_{\epsilon}$ such that 
\begin{equation*}
\underset{y\in[-L,L]^d}{\sup}|f(y)-\wt{f}(y)|\leq \epsilon. \end{equation*}
 Then since $\Omega_k=Q_\sm^\ast(\m M\cap S_k)$ is compactly supported and $\m M \cap S_k= G_\sm^\ast(\Omega_k)$, by lemma~\ref{coveringmanifold}, for any $\epsilon>0$, we can find a function set $N^Q_{\epsilon}$ such that 
 $\sqrt{\log |N^Q_{\epsilon}|} \lesssim (\frac{1}{\epsilon})^{\frac{d}{2\beta}}$  and for any $Q\in \wt{\mathcal{Q}}_k$, there exists $\wt{Q}\in N^Q_{\epsilon}$ such that 
 \begin{equation*}
 \underset{x\in \m M\cap S_k}{\sup}\|\wt{Q}(x)-Q(x)\|\leq \epsilon.
 \end{equation*}
  Then for any  $Q\in\wt{\mathcal{Q}}_k$ and $f\in\mathcal{F}_1$, there exists $\wt{Q}\in N^Q_{\epsilon}$, $\wt{f}\in N^f_{\epsilon}$ and a constant $c$  such that 
 \begin{equation*}
 \begin{aligned}
& \left|f(Q(x))\rho_k(x)-\wt{f}(\wt{Q}(x))\rho_k(x)\right|\\
 &\leq  \left|f(Q(x))\rho_k(x)-\wt{f}(Q(x))\rho_k(x)\right|+\left|\wt{f}(Q(x))\rho_k(x)-\wt{f}(\wt{Q}(x))\rho_k(x)\right|\\
 &\leq  c\epsilon.
 \end{aligned}
\end{equation*}
So we can get 
\begin{equation*}
\log \bold{N}(\mathcal{F},\epsilon, \|\cdot\|_{\infty})\lesssim \left(\frac{1}{\epsilon}\right)^{\frac{d}{\beta}}+\left(\frac{ 2^{dJ(\frac{1}{2}-\frac{1}{d})_{+}}}{\epsilon}\right)^2. 
\end{equation*}

Choose $\delta=\big(\frac{1}{n}\big)^{\frac{\alpha+1}{2\alpha+d}} \vee \big(\frac{1}{n}\big)^{ \frac{\beta}{d}}$, we can get 
\begin{equation*}
 \begin{aligned}
&\frac{1}{\sqrt{n}}\int_{\delta}^1 \left[\left(\frac{\log n}{\epsilon}\right)^{\frac{d}{2\beta}}+\frac{( 2^{dJ(\frac{1}{2}-\frac{1}{d})}\log n)\vee 1}{\epsilon}\right] \,\dd\epsilon \\
&\lesssim \frac{\log n}{\sqrt{n}}+  \big(\frac{1}{n}\big)^{\frac{\alpha+1}{2\alpha+d}}+ \big(\frac{1}{n}\big)^{ \frac{\beta}{d}}. 
 \end{aligned}
\end{equation*}

By Dudley's entropy integral bound,  it holds that 
\begin{equation*}
 \begin{aligned}
&\mathbb{E} \underset{f\in C_1^1(\mathbb{R}^d)\atop Q\in \wt{\mathcal{Q}}_k}{\sup} \left|\frac{1}{n} \sum_{l=1}^{2^d-1} \sum_{i=1}^n \varepsilon_i \sum_{j=0}^J  \sum_{m\in \mathbb{S}_{lj}} f_{ljm} \psi_{ljm}\left(Q(X_i)) \right)\rho_k(X_i)\right|\\
&\lesssim  \frac{\log n}{\sqrt{n}}+  \big(\frac{1}{n}\big)^{\frac{\alpha+1}{2\alpha+d}}+ \big(\frac{1}{n}\big)^{ \frac{\beta}{d}}.
 \end{aligned}
\end{equation*}
Similarly we can get 
\begin{equation*}
 \begin{aligned}
&\mathbb{E} \underset{f\in C_1^1(\mathbb{R}^d)\atop Q\in \wt{\mathcal{Q}}_k}{\sup} \left|\frac{1}{n}  \sum_{i=1}^n \varepsilon_i   \sum_{m\in  \mathbb{S}} b_{m} \phi_{m}\left(Q(X_i) \right)\rho_k(X_i)\right|\\
&\lesssim  n^{-\frac{\beta}{d}}+ \frac{\log n}{\sqrt{n}}. 
 \end{aligned}
\end{equation*}
The statement is then followed by Talagrand concentration inequality (see for example, Theorem 3.27 of ~\cite{wainwright_2019}) and the fact that $\alpha+1\leq \beta$.

\subsection{Proof of Lemma~\ref{leup2}}
We fix an arbitrary $k\in [K]$ in the following analysis.  Since
    \begin{equation*}  
    \frac{1}{n}\sum_{i=1}^n \|X_i-\wh{G}_\sm(\wh{Q}_\sm(X_i))\|_2^2\cdot\rho_k(X_i) \leq n^{-\frac{2\beta}{d}-1},
    \end{equation*}
    we can get 
     \begin{equation*}  
     \begin{aligned}
    \frac{1}{n}\sum_{i=1}^n \|X_i-\wh{G}_\sm(\wh{Q}_\sm(X_i))\|_2\rho_k(X_i) &\leq \frac{1}{n}\sum_{i=1}^n \|X_i-\wh{G}_\sm(\wh{Q}_\sm(X_i))\|_2\sqrt{\rho_k(X_i)}\\
    &\leq \sqrt{\frac{1}{n}\sum_{i=1}^n \|X_i-\wh{G}_\sm(\wh{Q}_\sm(X_i))\|_2^2\cdot\rho_k(X_i)}\\
    &\leq n^{-\frac{\beta}{d}-\frac{1}{2}}.
      \end{aligned}
    \end{equation*}
     Define
    \begin{equation*}
 \begin{aligned}
 &{\mathcal{F}}_2=\{f=G\circ Q\,:\, G\in C^{\beta}_L(\mb R^d;\mb R^D), Q\in C^{\beta}_L(\mb R^D;\mb R^d)\}.
     \end{aligned}
\end{equation*} 
Then we have $\wh{G}\circ\wh{Q}\in {\mathcal{F}}_2$. Moreover, when $\sup_{x\in \m M\cap S_k}\|f_1(x)-f_2(x)\|_2\leq \epsilon$, it holds that 
\begin{equation*}
 \begin{aligned}
 &\underset{x\in \m M}{\sup}\big|\|x-f_1(x)\|_2\rho_k(x)-\|x-f_2(x)\|_2\rho_k(x)\big|\\
 &\leq \underset{x\in \m M\cap S_k}{\sup}\|f_2(x)-f_1(x)\|_2\\
 &\leq \epsilon.
     \end{aligned}
\end{equation*} 
So consider the function class $\wt{\mathcal{F}}_2=\{|\|x-f(x)\|_2\rho_k(x), f\in \mathcal{F}_2\}$, by Lemma~\ref{coveringmanifold}, it holds that 
$\log N(\wt{\mathcal{F}}_2, \|\cdot\|_{\m M},\epsilon)\leq \log N({\mathcal{F}}_2, \|\cdot\|_{\m M},\epsilon)\lesssim (\frac{1}{\epsilon})^\frac{d}{\beta}$. By Dudley’s entropy integral bound (see for example,~\cite{wainwright_2019}), we can get that 
\begin{equation*}
 \begin{aligned}
 \mathbb{E}\bigg[\underset{f\in \mathcal{F}_2} {\sup}\Big|\frac{1}{n}\sum_{i=1}^n \|X_i-f(X_i)\|_2\rho_k(X_i)-\mathbb{E}_{\mu^*} \big[\|X-f(X)\|_2\rho_k(X)\big]\Big|\bigg]\leq C\, n^{-\frac{\beta}{d}}\vee \frac{\log n}{\sqrt{n}}.
      \end{aligned}
\end{equation*} 
 Then by Talagrand concentration inequality (see for example,~\cite{wainwright_2019}), we can get that there exists a constant $c_2$, such that it holds with probability $1-n^{-3}$ that 
\begin{equation*}
 \begin{aligned}
 \mathbb{E}_{\mu^*} \big[\|X-\wh{G}_\sm\circ{\wh{Q}_\sm}(X)\|_2\cdot\rho_k(X)\big] \leq c_2 \,\big(n^{-\frac{\beta}{d}}\vee \frac{\log n}{\sqrt{n}}\big).
      \end{aligned}
\end{equation*} 
For the second statement, we first fix a small enough positive constant $r>0$
  that will be chosen later. Then for any $z\in \Omega_k=Q^\ast(\m M\cap S_k)$, there exists $\sigma(z)\in  \Omega_k$  so that $z\in \mb B_r(\sigma(z))$ and $\nu^*_{\sm}(\sigma(z))\geq g(r)>0$. Let $\m A_z=\{\sigma(z)\,:\, z\in \Omega_k\}$ and $\wh{f}_\sm=\wh{G}_\sm\circ \wh{Q}_\sm\circ G^\ast_\sm$,  we resort to the following lemma that provides an upper bound on $\|G^\ast_\sm(z)-\wh{f}_\sm(z)\|_2$ for all $z\in  \m A_z$.
  \begin{lemma}\label{lemma:pointwise_error}
  It holds with probability at least $1-c\,n^{-3}$ that for all $z\in \m A_z$,
  \begin{align}
      \sum_{j\in \mathbb{N}_0^d\atop |j|\leq \lfloor\beta\rfloor}\frac{1}{j!}\, \|\wh{f}_\sm^{(j)}(\widetilde{z})- G_\sm ^{*,(j)}(\widetilde{z})\|_{2}\, (\delta_n)^{|j|} \leq C\cdot\big(g(r)\big)^{-\frac{2\beta}{d}-1}\cdot \big(\frac{\log n}{  n }\big)^{\frac{\beta}{d}},
  \end{align}
  where $\delta_n=b_1\cdot (g(r))^{-\frac{2}{d}}\cdot\big(\frac{\log n}{n}\big)^{\frac{1}{d}}$ for a constant $b_1$ independent of $n$ and $r$.
\end{lemma}
So by Lemma~\ref{lemma:pointwise_error}, we can get
\begin{equation*}
 \begin{aligned}
&\underset{z\in {\mathcal{A}}_z}{\sup}\|G^*_\sm(z)-\wh{f}_\sm(z)\|_2\leq  C\cdot\big(g(r)\big)^{-\frac{2\beta}{d}-1}\cdot \big(\frac{\log n}{  n }\big)^{\frac{\beta}{d}}\\
&\underset{z\in {\mathcal{A}}_z}{\sup}\|  \bold{J}_{G^*_\sm}(z)-\bold{J}_{\wh{f}_\sm}(z)\|_F \leq  C_1\cdot\big(g(r)\big)^{-\frac{2\beta-2}{d}-1}\cdot \big(\frac{\log n}{  n }\big)^{\frac{\beta-1}{d}}.
      \end{aligned}
\end{equation*} 
Also by the fact that $G^*_\sm$ and $\wh{f}=\wh G_\sm\circ \wh Q_\sm\circ G^\ast_\sm$ are $\beta$-H\"{o}lder smooth with $\beta\geq 1+\tilde\alpha>1$, we have 
\begin{equation}\label{eqndensitysmooth}
 \begin{aligned}
&\underset{z\in \Omega_k}{\sup}\|G^*_\sm(z)-\wh{f}_\sm(z)\|_2\leq  C\cdot\big(g(r)\big)^{-\frac{2\beta}{d}-1}\cdot \big(\frac{\log n}{  n }\big)^{\frac{\beta}{d}}+C_2\,r\\
&\underset{z\in\Omega_k}{\sup}\|\bold{J}_{G^*_\sm}(z)-\bold{J}_{\wh{f}_\sm}(z)\|_F \leq C_1\cdot\big(g(r)\big)^{-\frac{2\beta-2}{d}-1}\cdot \big(\frac{\log n}{  n }\big)^{\frac{\beta-1}{d}}+C_3\, r.
      \end{aligned}
\end{equation} 
By the fact that for any $z\in \Omega_k$, it holds that $z=Q^{\ast}_\sm(G^{\ast}_\sm(z))$, we obtain $I_{d}=\bold{J}_{Q^{\ast}_\sm}(G^{\ast}_\sm(z)) \, \bold{J}_{G^{\ast}_\sm}(z)$. Since $Q^{\ast}_\sm$ is $L$-Lipschitz, $\bold{J}_{Q^{\ast}_\sm}(G^{\ast}_\sm(z))$ has bounded operator norm, which implies
 ${\rm det}({\bold{J}^T_{G^{\ast}_\sm}(z)}\,\bold{J}_{G^{\ast}_\sm}(z))\geq c$ for some positive constant $c>0$ and $z\in \Omega_k$. Moroever, by the fact that $G^\ast_\sm$ is $\beta$-H\"{o}lder smooth with $\beta>1$, there exists a positive constant $\epsilon$ so that for the $\epsilon$- enlargement of $\Omega_k$: $\Omega_{k,\epsilon}=\{y\in \mb B_\epsilon(z)\,:\, z\in \Omega_k \}$, it holds that for any $z\in \Omega_{k,\epsilon}$, ${\rm det}({\bold{J}^T_{G^{\ast}_\sm}(z)}\,\bold{J}_{G^{\ast}_\sm}(z))\geq \frac{c}{2}$. Therefore, the second display in~\eqref{eqndensitysmooth} and $\beta$-H\"older smooth of $\wh f_\sm$ implies that $\inf_{z\in \Omega_{k,\epsilon}}{\rm det}(\bold{J}^T_{\wh{f}_\sm}(z)\,\bold{J}_{\wh{f}_\sm}(z))\geq \frac{c}{4}$ for all sufficiently small $\epsilon$, $r$ and sufficiently large $n$.
 Now let $\wh l_\sm=\wh Q_\sm \circ G^\ast_\sm$ and by using the identity $\wh{f}_\sm=\wh G_\sm\circ \wh l_\sm$,
\begin{equation*}
 \begin{aligned}
 \bold{J}^T_{\wh{f}_\sm}(z) \,  \bold{J}_{\wh{f}_\sm}(z)
 &=\left(\bold{J}_{\wh{G}_\sm}(\wh{l}_\sm(z))  \, \bold{J}_{\wh{l}_\sm}(z)\right)^T\left(\bold{J}_{\wh{G}_\sm}(\wh{l}_\sm(z)) \, \bold{J}_{\wh{l}_\sm}(z)\right)\\
 &=\bold{J}_{\wh{l}_\sm}^T(z) \, \bold{J}_{\wh{G}_\sm}^T(\wh{l}_\sm(z))\, \bold{J}_{\wh{G}_\sm}(\wh{l}_\sm(z)) \, \bold{J}_{\wh{l}_\sm}(z),
      \end{aligned}
\end{equation*} 
 by taking determinant we further obtain (note that $\bold{J}_{\wh{l}_\sm}(z)$ is a square matrix)
\begin{equation*}
 \begin{aligned}
{\rm det}^2\left(\bold{J}_{\wh{l}_\sm}(z)\right) \cdot {\rm det}\left(\bold{J}_{\wh{G}_\sm}^T(\wh{l}_\sm(z)) \,  \bold{J}_{\wh{G}_\sm}(\wh{l}_\sm(z))\right)\geq \frac{c}{4}.
        \end{aligned}
\end{equation*} 
 Since both $\wh G_\sm$ and $\wh Q_\sm$ are $L$-Lipschitz, we can further deduce that $0<c_1\leq{\rm det}(\bold{J}_{\wh{l}_\sm}(z))\leq c_2$ for all $z\in \Omega_{k,\epsilon}$.  
 
 We claim that $\wh l_\sm$ is globally invertible over $\Omega_{k,\epsilon}$ when $\epsilon$, $r$ are small enough and $n$ is large enough. Otherwise, suppose there exist distinct $z_0$ and $z_1$ in $\Omega_{k,\epsilon}$ such that $\wh l_\sm(z_0)=\wh l_\sm(z_1)$.
 Since $0<c_1\leq{\rm det}(\bold{J}_{\wh{l}_\sm}(z))\leq c_2$ implies $\wh l_\sm$ to be locally invertible, meaning that there exists some constant $b_0>0$ independent of $\epsilon$ such that $\|z_0-z_1\|\geq b_0$. By the definition of $\Omega_{k,\epsilon}$ and the Lipschitzness of $\wh G_\sm$ and $\wh l_\sm$, there exist $\bar z_0$ and $\bar z_1$ in $\Omega_k$ such that (for sufficiently small $\epsilon$)
 \begin{align}
     \|\bar z_0 - \bar z_1\|\geq \frac{1}{2} b_0,\quad \|\wh{l}_\sm(\bar{z}_0)-\wh{l}_\sm(\bar{z}_1)\|_2\leq C \epsilon\quad \mbox{and}\quad 
     \|\wh f_\sm(\bar z_0)- \wh f_\sm(\bar z_1)\| \leq C\epsilon.
 \end{align}
 The third display above combined with the first display in~\eqref{eqndensitysmooth} implies $\|G^{\ast}_\sm(\bar{z}_0)-G^{\ast}_\sm(\bar{z}_1)\|_2\leq C_1(\epsilon+r)$. On the other hand, from the first display above and the Lipschitzness of $Q^\ast_\sm$, we have
 \begin{align*}
     \frac{1}{2} b_0 \leq \|\bar z_0 - \bar z_1\| = \|Q^{\ast}_\sm(G^{\ast}_\sm(\bar{z}_0)-Q_\sm^{\ast}(G_\sm^{\ast}(\bar{z}_1))\|_2\leq C\|G^{\ast}_\sm(\bar{z}_0)-G^{\ast}_\sm(\bar{z}_1)\|_2\leq CC_1(\epsilon+r),
 \end{align*}
 which is a contradiction when $\epsilon$, $r$ are chosen small enough.
 
 Let $\wh{l}_\sm^{-1}:  \wh{l}_\sm(\Omega_{k,\epsilon/2})\to\Omega_{k,\epsilon/2}$ be the inverse of $\wh{l}_\sm$ over ${\Omega_{k,\epsilon/2}}$. By using the inverse function theorem for H\"{o}lder space (see for example, Appendix A of ~\citep{Eldering2013}), we can conclude $\wh{l}_\sm^{-1} \in C^{\beta}_{C_0} (\wh{l}_\sm(\Omega_{k,\epsilon/2});\mathbb{R}^d)$ for some sufficiently large constant $C_0$.  Therefore, we can write the  expression of the density function of $\nu_{\sm,\wh Q_{\sm}}^\ast=[\wh Q_\sm]_\# (\frac{\rho_k\mu^\ast}{p_k})$ as 
 \begin{equation*}
     \nu_{\sm,\wh Q_{\sm}}^\ast(y)=\nu^\ast_\sm(\wh l_\sm^{-1}(y))\cdot \Big({\rm det}\big(\bold{J}^T_{\wh l_\sm^{-1}}(x)\bold{J}_{\wh l_\sm^{-1}}(x)\big)\Big)^{\frac{1}{2}}\cdot\bold{1}\big(y\in \wh{l}_\sm (\Omega_k)\big)
 \end{equation*}
by applying the change of variable of $y=\wh l_\sm(z)$ with $z\sim \nu^\ast_\sm$. Moreover, since  $\nu^\ast_\sm \in C_L^{\alpha}(\mb R^d)$, this together with $\wh{l}_\sm^{-1} \in C^{ \beta}_{C_0} (\wh{l}_\sm(\Omega_{k,\epsilon/2});\mathbb{R}^d)$ implies 
  ${\nu}^{\ast}_{\wh{Q}}\in C^{ \tilde\alpha}_{C_1} (\mathbb{R}^d)$ for some constant $C_1$ (recall $\tilde\alpha=\alpha\wedge (\beta-1)$).  
 
 %%%%%%%%%%%%%%%%%%%%%%%%%%%%%%%%%%%%%%%%%%%%%%%%%%%%%%%%%%

 \subsection{Proof of Lemma~\ref{lemma:pointwise_error}}
  The proof follows the analysis in~\cite{tang2022minimax}. Let $h_{n}=\big(\frac{\log n}{n}\big)^{ \frac{1}{d}}$ and $\m N_{h_{n}}\subset \mathcal{A}_z$ be a minimal $h_{n}$-covering set of $\mathcal{A}_z$ under the $\ell_2$ distance, where its cardinality satisfies $|\m N_{h_{n}}|\leq C \frac{n}{\log n}$. For any $\widetilde z\in \m N_{h_{n}}$, define $\delta_n=b \big(\frac{\log n}{n}\big)^{ \frac{1}{d}}$.

We claim that it suffices to show that for sufficiently large $b$, it holds with probability at least $1-n^{-3}$ that for any $\widetilde{z}\in \m N_{h_{n}}$, 
\begin{equation}\label{leup2eq2}
 \sum_{j\in \mathbb{N}_0^d\atop |j|\leq \lfloor\beta\rfloor}\frac{1}{j!}\, \|\wh{f}_\sm^{(j)}(\widetilde{z})- G^{*,(j)}_\sm(\widetilde{z})\|_{2}\, (\delta_n)^{|j|} \leq C \left(\frac{\log n}{n}\right)^{\frac{\beta}{d}}.
 \end{equation}
In fact, if this inequality holds, then we can apply a standard argument of approximation by the $h_{n}$-covering set. Concretely, for any $z\in \mathcal{A}_z$, there exists $\widetilde{z}\in \m N_{h_{n}}$ such that  $\|z-\widetilde{z}\|_2\leq h_{n}= (\frac{\log n}{ n})^{\frac{1}{d}}$, we can obtain by applying Taylor expansion to $G^{\ast}_\sm(z)-\wh{f}_\sm(z)$ that 
\begin{equation*}
 \begin{aligned}
 \|G^{\ast}_\sm(z)-\wh{f}_\sm(z)\|_2&\leq  C \sum_{j\in \mathbb{N}_0^d\atop |j| \leq \lfloor\beta\rfloor}
 \frac{1}{j!}\,\|\wh{f}_\sm^{(j)}(\widetilde{z})- G_\sm^{*,(j)}(\widetilde{z})\|_{2}\,\Big(\frac{\log  n}{ n}\Big)^{\frac{|j|}{d}} \,+ \, C\, \Big(\frac{\log  n}{ n}\Big)^{\frac{\beta}{d}}\\
 &\leq C\left(\frac{\log n}{n}\right)^{\frac{\beta}{d}}.
     \end{aligned}
\end{equation*}
 Now let us prove inequality~\eqref{leup2eq2}. Recall that 
\begin{equation*}
\frac{1}{n}\sum_{i=1}^n\|X_i-\wh G_\sm(\wh Q_\sm(X_i))\|^2\rho_k(X_i)\leq C \,n^{-\frac{2\beta}{d}-1}.
\end{equation*}
In particular, by restricting the sum to those $Q^\ast(X_i)$ in $\mb B_{\delta_n}(\widetilde{z})$ for a fixed $\widetilde{z}\in \m N_{h_{n}}$, we further obtain (recall that $\wh f_\sm = \wh{G}_\sm\circ\wh{Q}_\sm\circ G^{\ast}_\sm$)
\begin{align*}
      \frac{1}{n}\sum_{i=1}^n\|G_k^\ast(Q_k^\ast(X_i))-\wh f_\sm(Q_\sm^\ast(X_i))\|^2\cdot\rho_k(X_i) \cdot \bold{1}_{\mb B_{\delta_n}(\widetilde{z})}(Q^{\ast}_\sm(X_i))\leq C\, n^{-\frac{2\beta}{d}-1}.
\end{align*}
 By applying the Taylor expansion to $G^{\ast}_\sm(z)-\wh{f}_\sm(z)$ around $\widetilde z$ in the preceding display and using the fact that $G^{\ast}_\sm-\wh{f}_\sm \in C_{C_0}^\beta(\mb B_1^d;\,\mb R^D)$ with some sufficiently large constant $C_0$, we can get the following \emph{localized basic inequality} after some algebra calculation
\begin{equation}\label{leup2eq1}
   \begin{aligned}
      & U_n(\widetilde z,\,\wh{f}_\sm):\,= \\
      &\frac{1}{n}\sum_{i=1}^n  \bigg\|\sum_{j\in \mathbb{N}_0^d\atop |j|\leq \lfloor\beta\rfloor} \frac{1}{j!} \, \big(G_\sm^{*,(j)}(\widetilde{z}) - \wh{f}_\sm^{(j)}(\widetilde{z})\big) \, ( Q_\sm^{\ast}(X_i)-\widetilde{z})^{j}\bigg\|_2^2 \cdot \bold{1}_{\mb B_{\delta_n}(\widetilde{z})}(Q_\sm^{\ast}(X_i))  \cdot\rho_k(X_i)\\
 \leq &\, c\bigg((\delta_n)^{2\beta}+(\delta_n)^{\beta} \sum_{j\in \mathbb{N}_0^d\atop |j|\leq \lfloor\beta\rfloor}\frac{1}{j!}\, \big\|G^{*,(j)}_\sm(\widetilde{z})-\wh{f}_\sm^{(j)}(\widetilde{z})\big\|_2\,(\delta_n)^{|j|}\bigg) \cdot 
 \frac{1}{n}\sum_{i=1}^n   \bold{1}_{\mb B_{\delta_n}(\widetilde{z})}(Q_\sm^{\ast}(X_i))
 \cdot \rho_k(X_i).
\end{aligned} 
\end{equation}
The second factor on the right hand side of~\eqref{leup2eq1} can be bounded by applying a simple union bound argument and Bernstein's inequality for bounded function as follows. First, we can bound the expectation
\begin{equation}\label{eqnboundprob}
\begin{aligned}
 &\mathbb{E}_{\mu^{\ast}}\big[\bold{1}_{\mb B_{\delta_n}(\widetilde{z})}(Q_\sm^{\ast}(X)) \cdot \rho_k(X)\big]
   \\
 &\overset{(i)}{=} p_k\,\int_{\mb B_{\delta_n}(\widetilde{z})} \nu^\ast_\sm(z) \,\dd z
 \leq C\,p_k\,\delta_n^d \\
 &\leq C\, b^d\, \frac{\log  n}{ n} ,
 \end{aligned}
\end{equation}
where step (i) follows by the fact that  $\nu^\ast_\sm=(Q^\ast_k)_{\#}(\frac{\mu^*\cdot\rho_k}{p_k})$.
 Since the random variable $\bold{1}_{\mb B_{\delta_n}(\widetilde{z})}(Q_\sm^{\ast}(X)) \cdot \rho_k(X)$ is uniformly bounded by $1$, and inequality~\eqref{eqnboundprob} and $\rho_k\leq 1$ implies its variance to be bounded by $C_1\, b^d\, \frac{\log n}{n}$, we may apply the Bernstein inequality and a simple union bound argument over all $\widetilde z \in \m N_{h_{n}}$ (with $|\m N_{h_{n}}|\leq C\frac{n}{\log n}$) to obtain that with probability at least $1-n^{-c}$,
 \begin{equation}\label{eqn:union+Bernstein}
\underset{\widetilde{z}\in \m N_{h_{n}}}{\sup} \bigg| \frac{1}{n}\sum_{i=1}^n   \bold{1}_{\mb B_{\delta_n}(\widetilde{z})}(Q_\sm^{\ast}(X_i))
 \cdot \rho_k(X_i)-  \mathbb{E}_{\mu^{\ast}}\big[\bold{1}_{\mb B_{\delta_n}(\widetilde{z})}(Q_\sm^{\ast}(X)) \cdot \rho_k(X)\big]\bigg|\leq C_2 b^{\frac{d}{2}}\cdot\frac{\log n}{n},
\end{equation}
which together with~\eqref{eqnboundprob} leads to 
\begin{equation}\label{eqn:binomial_union}
\underset{\widetilde{z}\in \m N_{h_{n}}}{\sup}  \bigg[\frac{1}{n}\sum_{i=1}^n   \bold{1}_{\mb B_{\delta_n}(\widetilde{z})}(Q_\sm^{\ast}(X_i))\bigg]\leq C\, b^d\cdot \frac{\log n}{n}.
\end{equation}
To analyze the quantity $U_n(\widetilde z,\,\wh{f}_\sm)$ on the left hand side of the localized basic inequality~\eqref{leup2eq1}, we will resort to the following lemma.
\begin{lemma}\label{lemma:empirical_proc}
  With probability at least $1-n^{-3}$, the following inequality holds for any $\beta$-smooth function $f\in C_{L}^\beta(\mb B_1^d;R^D)$ and $\widetilde z \in \m N_{h_{\widetilde n}}$,
  \begin{align*}
      \big|U_n(\widetilde z,\,f) - \mb E_{\mu^\ast}[U_n(\widetilde z,\,f)]\big| \leq  C\, b^{\frac{d}{2}}\cdot\frac{\log n}{n}\cdot \bigg\{ \big(\frac{\log n}{  n}\big)^{\frac{2\beta}{d}}+\Big[ \sum_{j\in \mathbb{N}_0^d\atop |j|\leq \lfloor\beta\rfloor}\frac{1}{j!} \, \|{f}^{(j)}(\widetilde{z})- G^{*,(j)}_\sm(\widetilde{z})\|_{2}\,(\delta_n)^{|j|}\Big]^2\bigg\},
  \end{align*}
  where the expectation is taken with respect to the randomness in $\{X_i\}_{i=1}^n$.
\end{lemma}
\noindent Before applying this lemma, notice that for any $\widetilde{z}\in \m N_{h_{n}}$, we can bound the expectation $\mb E_{\mu^\ast}[U_n(\widetilde z,\,\wh{f}_\sm\,)]$, where $f$ has been plugged-in with $\wh{f}_\sm$\,, by
\begin{equation}\label{eqn:emp_exp}
 \begin{aligned}
& \quad \mb E_{\mu^\ast}[U_n(\widetilde z,\,\wh{f}_\sm\,)]\\
&=\mathbb{E}_{\mu^{\ast}} \bigg[\Big\|\sum_{j\in \mathbb{N}_0^d\atop |j|\leq \lfloor\beta\rfloor} \frac{1}{j!} \,\big(G^{*,(j)}_\sm(\widetilde{z}) - \wh{f}_\sm^{(j)}(\widetilde{z}) \big)\, (Q^{\ast}_\sm(X)-\widetilde{z})^{j}\Big\|_2^2 \cdot  \bold{1}_{\mb B_{\delta_n}(\widetilde{z})}(Q^{\ast}_\sm(X)) \cdot \rho_k(X)\bigg]\\
&\overset{(i)}{\geq} \underset{z\in \mb B_{\delta_n}(\widetilde{z})}{\inf} \nu^{\ast}_\sm(z) \,\int_{z\in \mb B_{\delta_n}(\widetilde{z})} \Big\|\sum_{j\in \mathbb{N}_0^d\atop |j|\leq \lfloor\beta\rfloor}\frac{1}{j!} \,\big( G^{*,(j)}_\sm(\widetilde{z})-\wh{f}_\sm^{(j)}(\widetilde{z})\big)
\, (z-\widetilde{z})^{j}\Big\|_2^2\,\dd z\\
 &\overset{(ii)}{=} \delta_n^d \, \underset{z\in \mb B_{\delta_n}(\widetilde{z})}{\inf} \nu^{\ast}_\sm(z)  \,\int_{\mb B_1^d}  \Big\|\sum_{j\in \mathbb{N}_0^d\atop |j|\leq \lfloor\beta\rfloor}\frac{1}{j!}\,\delta_n^j \big( G^{*,(j)}_\sm(\widetilde{z})-\wh{f}_\sm^{(j)}(\widetilde{z})\big)\, z^{j}\Big\|_2^2 \, \dd z \\
 &\geq C\, b^d\cdot \frac{\log n}{n} \cdot  g(r)\cdot\int_{\mb B_1^d}  \Big\|\sum_{j\in \mathbb{N}_0^d\atop |j|\leq \lfloor\beta\rfloor}\frac{1}{j!}\delta_n^j \,\big( G^{*,(j)}_\sm(\widetilde{z})-\wh{f}_\sm^{(j)}(\widetilde{z})\big)\,z^{j}\Big\|_2^2 \,\dd z,
   \end{aligned}
\end{equation}
where step (i) uses the fact that $Q_\sm^\ast(X)$ given $X\sim \frac{\mu^*\cdot\rho_k}{p_k}$  is distributed as $\nu^\ast_\sm$, step (ii) follows by applying the change of variable of $\frac{z-\widetilde z}{\delta_n} \to z$, and the last step follows by the fact that $\nu^*_\sm(\wt z)\geq g(r)$ for $\wt z\in \m A_z$ and the smoothness of $\nu^*_\sm$.
Now using the fact that for any $d$-variate polynomial $\mathcal{S}(y)=\sum_{j\in \mb N_0^d,\,|j|\leq k} a_{j} y^{j}$, $y\in\mb R^d$, there exists some positive constant $C(d,k)$ only depending on $(d,k)$ such that 
\begin{equation*}
\int_{\mb B_1^d} \mathcal{S}^2(y) \, \dd y \geq C(d,k) \sum_{j\in \mb N_0^d,\,|j|\leq k} a_{j}^2,
  \end{equation*}
we can obtain that
\begin{equation}\label{eqn:exp_low_b}
 \begin{aligned}
  \int_{\mb B_1^d}   \Big\|\sum_{j\in \mathbb{N}_0^d\atop |j|\leq \lfloor\beta\rfloor}\frac{1}{j!}\delta_n^j \,\big( G^{*,(j)}_\sm(\widetilde{z})-\wh{f}_\sm^{(j)}(\widetilde{z})\big)\,z^{j}\Big\|_2^2 \,\dd z
   \geq c\, \bigg(\sum_{j\in \mathbb{N}_0^d\atop |j|\leq \lfloor\beta\rfloor}\frac{1}{j!}\,\|\wh{f}_\sm^{(j)}(\widetilde{z})- G^{*,(j)}_\sm(\widetilde{z})\|_{2}\, (\delta_n)^{|j|}\bigg)^2.
     \end{aligned}
\end{equation}
Finally, by combining equations~\eqref{leup2eq1},~\eqref{eqnboundprob},~\eqref{eqn:emp_exp},~\eqref{eqn:exp_low_b} and Lemma~\ref{lemma:empirical_proc}, we obtain that with probability at least $1-cn^{-3}$, for any $\widetilde{z}\in \m N_{h_{n}}$, 
\begin{equation*}
 \begin{aligned}
 &\quad b^d \cdot \frac{\log n}{n} \cdot g(r)\cdot \bigg(\sum_{j\in \mathbb{N}_0^d\atop |j|\leq \lfloor\beta\rfloor}\frac{1}{j!}\, \|\wh{f}_\sm^{(j)}(\widetilde{z})- G_\sm ^{*,(j)}(\widetilde{z})\|_{2}\, (\delta_n)^{|j|}\bigg)^2\\
 &\leq C b^{\frac{d}{2}} \cdot \frac{\log n}{n} \cdot \bigg(\sum_{j\in \mathbb{N}_0^d\atop |j|\leq \lfloor\beta\rfloor}\frac{1}{j!}\, \|\wh{f}_\sm^{(j)}(\widetilde{z})- G_\sm ^{*,(j)}(\widetilde{z})\|_{2}\, (\delta_n)^{|j|}\bigg)^2 
 \, +\,  C b^{\frac{d}{2}}\cdot \Big(\frac{\log n}{n}\Big)^{\frac{2\beta}{d}}\cdot \frac{\log n}{n}\\
& \qquad +\, C b^d\cdot \frac{\log n}{n}\cdot  \bigg((\delta_n)^{2\beta}+(\delta_n)^{\beta} \sum_{j\in \mathbb{N}_0^d\atop |j|\leq \lfloor\beta\rfloor}\frac{1}{j!}\, \| G^{*,(j)}_\sm(\widetilde{z})-\wh{f}_\sm^{(j)}(\widetilde{z})\|_{2}(\delta_n)^{|j|}\bigg).
\end{aligned}
\end{equation*}
Consequently, the claimed inequality~\eqref{leup2eq2} follows from the above by choosing $b=b_1 (g(r))^{-\frac{2}{d}}$ with sufficiently large $b_1$ and the definition that $\delta_n=b \big(\frac{\log n}{n}\big)^{ \frac{1}{d}}$.

\subsection{Proof of Lemma~\ref{lemma:empirical_proc}}\label{sec:proof_lemma:empirical_proc}
The proof follows from the proof of Lemma 18 in~\cite{tang2022minimax}, we include it here for completeness. Since $f\in C_{L}^\beta(\mb B_1^d;R^D)$, for any $z\in \mb B_1^d$ and $j\in \mathbb{N}_0^d$ with $|j|\leq \lfloor\beta\rfloor$, it holds that $\|{f}^{(j)}(z)\|_2\leq \sqrt{D} L=C_0$. For any fixed $\widetilde{z}\in \m N_{h_{\wt{n}}}$ and $\widetilde\delta>0$, let 
\begin{equation*}
\begin{aligned}
\bar{\mathcal{T}}(\widetilde{\delta})=\Big\{T=\{T_j\}_{j\in \mathbb{N}_0^d, \, |j|\leq \lfloor\beta\rfloor}\in [-C_0,\,C_0]^{D\times \binom{d+\lfloor \beta\rfloor-1}{d}} :\, \sum_{j\in \mathbb{N}_0^d\atop |j|\leq \lfloor\beta\rfloor}\frac{1}{j!}\, \big\|T_j-G^{*,(j)}_\sm(\widetilde{z})\big\|_2\, (\delta_{\widetilde{z}})^{|j|}\leq \widetilde{\delta}\Big\}.
\end{aligned}
\end{equation*}
We also define the following supreme of an empirical process indexed by $T\in \bar{\mathcal{T}}(\widetilde{\delta})$,
\begin{equation*}
 \begin{aligned}
&\quad Z_n(\widetilde{\delta})=\\
 &\underset{T\in \bar{\mathcal{T}}(\widetilde{\delta}) }{\sup}\Bigg| \, \mathbb{E}_{\mu^{\ast}} \bigg[\Big\|\sum_{j\in \mathbb{N}_0^d\atop |j|\leq \lfloor\beta\rfloor} \frac{1}{j!}\,\big(G^{*,(j)}_\sm(\widetilde{z}) -T_j\big)\, (Q_\sm^{\ast}(X)-\widetilde{z})^{j} \Big\|_2^2\cdot  \bold{1}_{\mb B_{\delta_{\widetilde{z}}}(\widetilde{z})}(Q^{\ast}_\sm(X)) \cdot \rho_k(X)\bigg]\\
& \ \ - \frac{1}{n} \sum_{i=1}^n \bigg[\Big\|\sum_{j\in \mathbb{N}_0^d\atop |j|\leq \lfloor\beta\rfloor} \frac{1}{j!}\,\big(G^{*,(j)}_\sm(\widetilde{z}) -T_j\big)\, (Q_\sm^{\ast}(X_i)-\widetilde{z})^{j}\Big\|_2^2\cdot  \bold{1}_{\mb B_{\delta_{\widetilde{z}}}(\widetilde{z})}(Q^{\ast}_\sm(X)) \cdot  \rho_k(X) \bigg]\Bigg|,
  \end{aligned}
\end{equation*}
and $R_n(\widetilde{\delta})=\mathbb{E}_{\mu^\ast}\big[ Z_n(\widetilde{\delta}) \big]$. We will first prove a concentration inequality for a fixed radius $\widetilde{\delta}>0$, and then using the peeling technique to allow the radius to be random, which leads to the desired result.

To apply the Talagrand concentration inequality (see, for example, Theorem 3.27 of~\cite{wainwright_2019}) for bounding the difference $|Z_n(\widetilde{\delta}) - R_n(\widetilde \delta)|$ for a fixed $\widetilde{\delta}>0$, we notice that each additive component in the second empirical sum above has second moment uniformly bounded by
\begin{equation*}
 \begin{aligned}
& \mathbb{E}_{\mu^{\ast}} \bigg[\underset{T\in \bar{\mathcal{T}}(\widetilde{\delta}) }{\sup}\Big(\Big\|\sum_{j\in \mathbb{N}_0^d\atop |j|\leq \lfloor\beta\rfloor} \frac{1}{j!}\,\big(G^{*,(j)}_\sm(\widetilde{z}) -T_j\big)\, (Q_\sm^{\ast}(X)-\widetilde{z})^{j} \Big\|_2^4\cdot  \bold{1}_{\mb B_{\delta_{\widetilde{z}}}(\widetilde{z})}(Q^{\ast}_\sm(X)) \cdot  \rho_k(X)\Big)\bigg]\\
&\leq \underset{z\in \mb B_{\delta_{\widetilde{z}}}(\widetilde{z})\atop T\in \bar{\mathcal{T}}(\widetilde{\delta})} {\sup} \Big\|\sum_{j\in \mathbb{N}_0^d\atop |j|\leq \lfloor\beta\rfloor} \frac{1}{j!}\,\big(G^{*,(j)}_\sm(\widetilde{z}) -T_j\big)\, (z-\widetilde{z})^{j} \Big\|_2^4 \cdot \mathbb{E}_{\mu^{\ast}}\big[ \bold{1}_{\mb B_{\delta_{\widetilde{z}}}(\widetilde{z})}(Q^{\ast}_\sm(X)) \cdot  \rho_k(X)\big]\\
&\leq C \underset{T\in \bar{\mathcal{T}}(\widetilde{\delta}) }{\sup}\bigg(\sum_{j\in \mathbb{N}_0^d\atop |j|\leq \lfloor\beta\rfloor}\frac{1}{j!}\,\|T_j- G^{*,(j)}_\sm(\widetilde{z})\|_{2}\, (\delta_{\widetilde{z}})^{|j|}\bigg)^4\cdot  b_2^d \cdot \frac{\log n}{n}\\
&\leq C\, b_2^d\, \widetilde \delta^4\cdot \frac{\log n}{n},
  \end{aligned}
\end{equation*}
where we have used inequality~\eqref{eqnboundprob} to bound $\mathbb{E}_{\mu^{\ast}}\big[ \bold{1}_{\mb B_{\delta_{\widetilde{z}}}(\widetilde{z})}(Q^{\ast}_\sm(X)) \cdot  \rho_k(X)\big]$.
Moreover, each additive component can be almost surely bounded by 
\begin{equation*}
 \begin{aligned}
& \quad \underset{z\in \mb B_{\delta_{\widetilde{z}}}(\widetilde{z})}{\sup} \Big\|\sum_{j\in \mathbb{N}_0^d\atop |j|\leq \lfloor\beta\rfloor} \frac{1}{j!}\,\big(G^{*,(j)}_\sm(\widetilde{z}) -T_j\big)\, (Q_\sm^{\ast}(X)-\widetilde{z})^{j} \Big\|_2^2\\
&\leq C\,  \bigg(\sum_{j\in \mathbb{N}_0^d\atop |j|\leq \lfloor\beta\rfloor}\frac{1}{j!} \, \|T_j- G^{*,(j)}(\widetilde{z})\|_{2}\, (\delta_{\widetilde{z}})^{|j|}\bigg)^2
\leq C\, \widetilde \delta^2.
   \end{aligned}
\end{equation*}
Based on these two bounds, we can apply the Talagrand concentration inequality to obtain that for any $s\geq 0$,
\begin{equation}~\label{Talagrand}
\mathbb{P} \big(Z_n(\widetilde{\delta}) \geq R_n(\widetilde{\delta}) +s^2\big )\leq 2 \exp\left(-\frac{c\,ns^4}{s^2 \,\widetilde{\delta}^2 + b_2^d\,\widetilde{\delta}^4 \cdot \frac{\log n}{n}}\right).
   \end{equation}
It remains to bound the expectation $R_n(\widetilde{\delta})$ via the symmetrization technique and chaining. 
By a standard symmetrization, we can get 
\begin{equation*}
\begin{aligned}
&R_n(\widetilde{\delta})\leq \frac{2}{\sqrt{n}}\, \mathbb{E}\Bigg[\underset{T\in \bar{\mathcal{T}}(\widetilde{\delta})}{\sup}\\
&\qquad \Bigg|\frac{1}{\sqrt{n}} \sum_{i=1}^n \varepsilon_i\bigg[\Big\|\sum_{j\in \mathbb{N}_0^d\atop |j|\leq \lfloor\beta\rfloor} \frac{1}{j!}\,\big(G^{*,(j)}_\sm(\widetilde{z}) -T_j\big)\, (Q_\sm^{\ast}(X)-\widetilde{z})^{j} \Big\|_2^2\cdot \bold{1}_{\mb B_{\delta_{\widetilde{z}}}(\widetilde{z})}(Q_\sm^{\ast}(X_i)) \cdot \rho_k(X_i)\bigg]\Bigg|\Bigg],
\end{aligned}
\end{equation*}
where $\{\varepsilon_i\}_{i=1}^n$ are $n$ i.i.d.~copies from the Rademacher distribution, i.e.~$\mb P(\varepsilon_i = 1) = \mb P(\varepsilon_i= -1) = 0.5$. 
Since given $\{X_i\}_{i=1}^n$, the stochastic process inside the supreme is a sub-Gaussian process with intrinsic metric 
 \begin{equation*}
 \begin{aligned}
& \quad d_n^2(T,\,\widetilde{T})\\
&=\frac{1}{n} \sum_{i=1}^n  \bigg(\Big\|\sum_{j\in \mathbb{N}_0^d\atop |j|\leq \lfloor\beta\rfloor} \frac{1}{j!}\,\big(G^{*,(j)}_\sm(\widetilde{z}) -T_j\big)\, (Q_\sm^{\ast}(X_i)-\widetilde{z})^{j} \Big\|_2^2 \\
&\qquad\qquad\qquad - \Big\|\sum_{j\in \mathbb{N}_0^d\atop |j|\leq \lfloor\beta\rfloor} \frac{1}{j!}\,\big(G^{*,(j)}_\sm(\widetilde{z}) -\widetilde T_j\big)\, (Q_\sm^{\ast}(X_i)-\widetilde{z})^{j} \Big\|_2^2\bigg)^2 \cdot \bold{1}_{\mb B_{\delta_{\widetilde{z}}}(\widetilde{z})}(Q^{\ast}_\sm (X_i))) \cdot \rho_k(X_i)\\
 &\leq C\,\widetilde{\delta}^4\, \frac{1}{n} \sum_{i=1}^n   \bold{1}_{\mb B_{\delta_{\widetilde{z}}}(\widetilde{z})}(Q_\sm^{\ast}(X_i))) \cdot \rho_k(X_i),
 \end{aligned}
\end{equation*}
for any $T,\widetilde{T}\in\bar{\mathcal{T}}(\widetilde \delta)$, where the last step uses the definition of $\bar{\mathcal{T}}(\widetilde \delta)$. The above combined with inequality~\eqref{eqnboundprob} implies 
\begin{equation*}
\mathbb{E}_{\mu^\ast}\Big[ \underset{T,\widetilde{T}\in \bar{\mathcal{T}}(\delta)}{\sup} d_n^2(T,\widetilde{T})\Big]\leq C\,b_2^d\, \widetilde{\delta}^4 \cdot\frac{\log n}{n}\quad \mbox{and}\quad
d_n(T,\widetilde{T}) \leq C \widetilde{\delta} \sum_{j\in \mathbb{N}_0^d\atop |j|\leq \lfloor\beta\rfloor}\frac{1}{j!}\, \|T_j-\widetilde{T}_{j}\|_{2}\, \delta_{\widetilde{z}}^{|j|}.
 \end{equation*}
 Lastly, let $\m K_n(\delta)= \underset{T,\widetilde{T}\in \bar{\mathcal{T}}(\delta)}{\sup} d_n^2(T,\widetilde{T})$, by applying the standard chaining via Dudley's inequality, we can get 
\begin{equation}\label{dudley}
 \begin{aligned}
 R_n(\widetilde{\delta})&\leq C\, \frac{1}{\sqrt{n}}\, \mathbb{E}_{\mu^\ast} \Big[ \int_{0}^{ \m K_n(\widetilde\delta)}\sqrt{\log \frac{\widetilde{\delta}}{u}}\,\dd u\Big]\\
 &= C\, \frac{1}{\sqrt{n}}\, \mathbb{E}_{\mu^\ast} \Big[ \m K_n(\widetilde\delta)\cdot\int_{0}^{1}\sqrt{\log \frac{\widetilde{\delta}}{u\cdot \m K_n(\widetilde\delta)}}\,\dd u\Big]\\
 &= C\, \frac{1}{\sqrt{n}}\, \mathbb{E}_{\mu^\ast} \Big[   \m K_n(\widetilde\delta)\cdot \bold{1}( \m K_n(\widetilde\delta)\leq b_2^{\frac{d}{2}}\widetilde \delta^2\sqrt{\frac{\log n}{n}})\int_{0}^{1}\sqrt{\log \frac{\widetilde{\delta}}{u \cdot\m K_n(\widetilde\delta)}}\,\dd u\Big]\\
 &+C\, \frac{1}{\sqrt{n}}\, \mathbb{E}_{\mu^\ast} \Big[   \m K_n(\widetilde\delta)\cdot \bold{1}( \m K_n(\widetilde\delta)> b_2^{\frac{d}{2}}\widetilde \delta^2\sqrt{\frac{\log n}{n}})\int_{0}^{1}\sqrt{\log \frac{\widetilde{\delta}}{u \cdot\m K_n(\widetilde\delta)}}\,\dd u\Big]\\
 &\leq C_1\, b_2^{\frac{d}{2}} \cdot \frac{\log ({n}/{\widetilde\delta})}{n}\cdot \widetilde{\delta}^2,
    \end{aligned}
\end{equation}
where we have used the fact that the $u$-covering entropy of $\bar{\mathcal{T}}(\widetilde \delta)$ relative to metric $d_n$ is at most $C_2\log\frac{\widetilde\delta}{u}$ for $u\in(0,1)$ where $C_2$ depends on $(d,D)$ (at most polynomial dependence on $D$). By combining this with inequality~\eqref{Talagrand}, we obtain that for all $t\geq 1$,
\begin{align}\label{fix_delta_tala}
    \mathbb{P} \Big(Z_n(\widetilde{\delta}) \geq  C\, t^2\,b_2^{\frac{d}{2}} \cdot \frac{\log (n/\widetilde \delta)}{n}\cdot \widetilde{\delta}^2\Big )\leq 2 \exp\Big(-c\,t^2\, \log (n/\widetilde \delta)\Big).
\end{align}

Finally, we apply the peeling technique to extend the above high probability bound on $Z_n(\widetilde\delta)$ to the random radius $\widetilde\delta=\sum_{j\in \mathbb{N}_0^d\atop |j|\leq \lfloor\beta\rfloor}\frac{1}{j!}\, \big\|\wh f^{(j)}_\sm-G^{*,(j)}_\sm(\widetilde{z})\big\|_2\, (\delta_{\widetilde{z}})^{|j|}$. Specifically, we first set the basic level $\bar{\delta} =\big(\frac{\log n}{n}\big)^{ \frac{\beta}{d}}$, and for $s=1,\cdots, S$ with $S\leq C\log \frac{1}{\bar{\delta}}$, define sets
\begin{equation*}
\begin{aligned}
\widetilde{\mathcal{T}}_0&=\Big\{T=\{T_j\}_{j\in \mathbb{N}_0^d, |j|\leq \lfloor\beta\rfloor}\in [-C_0,\,C_0]^{D\times \binom{d+\lfloor \beta\rfloor-1}{d}}:\,  \sum_{j\in \mathbb{N}_0^d\atop |j|\leq \lfloor\beta\rfloor}\frac{1}{j!}\, \|T_j- G^{*,(j)}_k(\widetilde{z})\|_{2}\, (\delta_{\widetilde{z}})^{|j|}\leq \bar{\delta}\Big\};\\
\widetilde{\mathcal{T}}_s&=\Big\{T=\{T_j\}_{j\in \mathbb{N}_0^d, |j|\leq \lfloor\beta\rfloor}\in [-C_0,\,C_0]^{D\times \binom{d+\lfloor \beta\rfloor-1}{d}}:\,  2^{s-1}\bar{\delta} \leq\sum_{j\in \mathbb{N}_0^d\atop |j|\leq \lfloor\beta\rfloor}\frac{1}{j!}\,\|T_j- G^{*,(j)}_k(\widetilde{z})\|_{2}\,(\delta_{\widetilde{z}})^{|j|}\leq 2^s\bar{\delta}\Big\}.
\end{aligned}
\end{equation*}
 By applying inequality~\eqref{fix_delta_tala} to $\widetilde\delta = 2^s\bar\delta$ for $s\in[S]$ with sufficiently large constant $t>0$, as $C_1\leq-\log(2^s\bar\delta)\leq C_2\log n$,  we obtain that 
 \begin{equation*}
\mathbb{P}\left(Z_n(\bar{\delta})\geq C\,b_2^{\frac{d}{2}}\,\frac{\log n}{n} \,\bar{\delta}^2\right) + \sum_{s=1}^S \mathbb{P}\left(Z_n(2^s \bar{\delta})\geq C\,b_2^{\frac{d}{2}} \,\frac{\log n}{n}\, 4^s \bar{\delta}^2\right)\leq n^{-(c+1)}.
\end{equation*}
Note that for any $T\in \widetilde{\mathcal{T}}_s$ and any $s\in\{0\}\cup [S]$, the event $Z_n(2^s \bar{\delta})\leq C\,b_2^{\frac{d}{2}} \,\frac{\log n}{n}\, 4^s \bar{\delta}^2$ implies
 \begin{align*}
     &\Bigg| \, \mathbb{E}_{\mu^{\ast}} \bigg[\Big\|\sum_{j\in \mathbb{N}_0^d\atop |j|\leq \lfloor\beta\rfloor} \frac{1}{j!}\,\big(G^{*,(j)}_\sm(\widetilde{z}) -T_j\big)\, (Q_\sm^{\ast}(X)-\widetilde{z})^{j} \Big\|_2^2\cdot  \bold{1}_{\mb B_{\delta_{\widetilde{z}}}(\widetilde{z})}(Q^{\ast}_\sm(X)) \cdot \rho_k(X)\bigg]\\
& \ \ - \frac{1}{n} \sum_{i=1}^n \bigg[\Big\|\sum_{j\in \mathbb{N}_0^d\atop |j|\leq \lfloor\beta\rfloor} \frac{1}{j!}\,\big(G^{*,(j)}_\sm(\widetilde{z}) -T_j\big)\, (Q_\sm^{\ast}(X_i)-\widetilde{z})^{j}\Big\|_2^2\cdot  \bold{1}_{\mb B_{\delta_{\widetilde{z}}}(\widetilde{z})}(Q^{\ast}_\sm(X)) \cdot \rho_k(X) \bigg]\Bigg|\\
&\leq  c_1\,b_2^{\frac{d}{2}}\,\frac{\log n}{n}\, \Bigg\{\bar\delta^2 + \bigg( \sum_{j\in \mathbb{N}_0^d\atop |j|\leq \lfloor\beta\rfloor}\frac{1}{j!}\,\|T_j- G^{*,(j)}_\sm(\widetilde{z})\|_{2}\, (\delta_{\widetilde{z}})^{|j|}\bigg)^2\Bigg\}.
 \end{align*}
 Finally, since for any $f\in C_{L}^\beta(\mb B_1^d;R^D)$, $T_f:\,=\{T_{f,j}=f^{(j)}\}_{j\in \mathbb{N}_0^d,|j|\leq \lfloor\beta\rfloor}$ must belong to some $\widetilde {\m T}_s$, the claimed result is a consequence of the two preceding displays and a simple union bound over $\widetilde z \in \m N_{h_{ n}}$ where $|\m N_{h_{{n}}}|\leq C\, \frac{{n}}{\log {n}} \leq C\, n$.
 
\subsection{Proof of Lemma~\ref{empiricalmean}}
Firstly by Bernstein's inequality, it holds with probability at least $1-n^{-3}$ that $|p_k-\wh p_k|\leq C\, \sqrt{\frac{\log n}{n}}$, then by $p_k>0$, for large enough $n$, we have  $\big|1-\frac{p_k}{\wh p_k}\big|\leq  C\, \sqrt{\frac{ \log n}{n}}$. Thus
\begin{equation*}
    \begin{aligned}
    &\underset{Q\in C^{\beta}_L(\mb R^D;\mb R^d)}{\sup}\underset{f\in {\rm Lip}_1(\mb R^d)}{\sup}\Big(\frac{1}{p_k}\int f(Q(x))\rho_k(x)\, \dd \mu^*-\int f(z)\, \wt{\nu}_{k,Q}(z)\,\dd z \Big) \\
    &=\underset{Q\in C^{\beta}_L(\mb R^D;\mb R^d)}{\sup}\underset{f\in {\rm Lip}_1(\mb R^d)}{\sup}\Big(\frac{1}{p_k}\int f(Q(x))\rho_k(x)\, \dd \mu^*-\frac{1}{\wh p_kn}\sum_{i=1}^n f(Q(X_i))\rho_k(X_i) \Big) \\
    &\leq  C\, \sqrt{\frac{ \log n}{n}}+\frac{1}{p_k}\underset{Q\in C^{\beta}_L(\mb R^D;\mb R^d)}{\sup}\underset{f\in {\rm Lip}_1(\mb R^d)}{\sup}\Big(\int f(Q(x))\rho_k(x)\, \dd \mu^*-\frac{1}{n}\sum_{i=1}^n   f(Q(X_i))\rho_k(X_i) \Big)\\
    &{\leq} C\, \sqrt{\frac{ \log n}{n}}+C\,\underset{f\in {\rm Lip}_1(\mb R^D)}{\sup}\Big(\int f(x)\rho_k(x)\, \dd \mu^*-\frac{1}{n}\sum_{i=1}^n   f(X_i)\rho_k(X_i) \Big),\\
        \end{aligned}
 \end{equation*}
 where the last inequality is due to the assumption that $\beta\geq 1$. Then  consider the pseudo-metric for $f,f'\in {\rm Lip}_1(\mb R^D)$ 
 \begin{equation*}
     d_n(f,f')=\sqrt{\frac{1}{n}\sum_{i=1}^n\big( f(X_i)\rho_k(X_i)- f'(X_i)\rho_k(X_i)\big)^2}\leq \underset{x\in S_k\cap \m M}{\sup} |f(x)-f'(x)|=\underset{x\in G^*(Q^*(S_k\cap \m M))}{\sup} |f(x)-f'(x)|.
 \end{equation*}
 Then by Lemma~\ref{coveringmanifold}, we have $\log \bold{N}({\rm Lip}_1(\mb R^D),\|\cdot\|_{L^{\infty}( S_k\cap \m M)},\epsilon)\leq C\, \epsilon^{-d}$. Choose $\delta=(\frac{1}{n})^{\frac{1}{d}}$, we can get 
 \begin{equation*}
     \frac{1}{\sqrt{n}}\int_{\delta}^1\sqrt{\log \bold{N}({\rm Lip}_1(\mb R^D),\|\cdot\|_{L^{\infty}( S_k\cap \m M)},\epsilon)}\,\dd \epsilon\\
     \leq C\, n^{-\frac{1}{d}}\vee \frac{\log n}{\sqrt{n}}.
 \end{equation*}
 Thus similar as the analysis in the proof of Lemma~\ref{leup1}, using Dudley's entropy integral bound and Talagrand concentration inequality, we can obtain the desired result.
\section{Proof of Technical Details}
 \subsection{Proof of Lemma~\ref{lemmaB.1}}
  Consider an aritrary $\mu^*\in \m P^*(d,D,\alpha,\beta,L^*)$. Denote $\m M={\rm supp}(\mu^*)$, to begin with, we consider the following lemma.
  \begin{lemma}[Lemma 17 of~\cite{tang2022minimax}]\label{lemmaB.2}
  There exist positive constants $(\tau_1,L_1)$ such that for any $x_0\in \m M$, define $Q_{x_0}: \mathbb{R}^D\to \mathbb{R}^d$ as $Q_{x_0}(x)=W_{x_0}^T(x-x_0)$ where $W_{x_0}\in \mathbb{R}^{D\times d}$ is an arbitrary orthonormal basis of the tangent space of $\m M$ at $x_0$, then there exists a set $\widetilde U_{x_0} $ satisfying $\mb B_{\tau_1}(x_0)\cap \m M\subset \widetilde U_{x_0}\subset  \m M$ and function $G_{x_0}\in C_{L_1}^{\beta}(\mb R^d;\mb R^D)$ so that  
  \vspace{0.5em}
  
     \quad(1). $G_{x_0}(\mb B_1^d)=\widetilde U_{x_0}$ and for any $z\in \mb B_1^d$,    $Q_{x_0}(G_{x_0}(z))=z$; 
     
    \quad(2). $\mu^*\circ G_{x_0}|_{\mb B_1^d}\in C^{\alpha}_{L_1}(\mb B_1^d)$ and for any $z\in \partial \mb B_1^d$,  $\|G_{x_0}(z)-x_0\|\geq \tau_1$.
 
  \end{lemma}
\noindent By Sobolev extension theorem, there exists  a constant $L_2$ so that for any $x_0\in \m M$, there exists $\ov Q_{x_0}\in C^{\beta}_{L_2}(\mb R^D)$ so that  $\ov Q_{x_0}|_{\wt U_{x_0}}= Q_{x_0}|_{\wt U_{x_0}}$.  Since $J_{G_{x_0}}(z)^TJ_{G_{x_0}}(z)$ has uniformly lower bounded  eigenvalues on $z\in \mb B_{1}^d$ and $G_{x_0}$ is $\beta$-smooth with $\beta>1$ and uniformly bounded H\"{o}lder norm, there exists a small enough positive constant $r_0$ so that for any $x_0\in \m M$,
\begin{equation}\label{eqneigenlower}
   \underset{v\in \mb S_1^{d-1}=\{v\in \mb R^d\,:\, \|v\|=1\}}{\sup}\frac{\underset{z,z'\in \mb B_{r_0}^d} {\sup}\|(J_{G_{x_0}}(z)-J_{G_{x_0}}(z'))v\|}{  \| J_{G_{x_0}}(0) v\|}\leq \frac{1}{3}.
\end{equation}
We then choose $r^*\leq \tau_1/2$ to be a small enough positive constant  so that for any $x_0\in \m M$, $Q_{x_0}(\mb B_{r^*}(x_0)\cap \m M)\subset \mb B_{r_0}^d$. For an arbitrary $k \in [K]$, consider $x_0=a_k\in \m M$. Define $G_k=G_{x_0}$ and $Q_k=\ov Q_{x_0}$. Then we have $S_k\cap \m M\subset \mb B_{\tau_1}(x_0)\cap \m M\subset \wt U_{x_0}$, and for any $x\in \m M\cap S_k$, $x=G_k(Q_k(x))$.   Moreover, let $p_k=\mb{E}_{\mu^*}[\rho_k(X)]$, since the density of $\mu^*$ is uniformly bounded from below and $r_k\geq L^*$, we have $p_k$ is also uniformly bounded from below. Furthermore, we can write $\nu_k=(Q_k)_{\#}(\frac{\mu^*\rho_k}{p_k})$ as 
  \begin{equation*}
      \nu_k(z)=\left\{
      \begin{array}{cc}
         \frac{\mu^*(G_k(z))\cdot \rho_k(G_k(z))\sqrt{{\rm det}(J_{G_k}(z)^TJ_{G_k}(z))}}{\int_{Q_k(\m M\cap S_k)}\mu^*(G_k(z))\cdot \rho_k(G_k(z))\sqrt{{\rm det}(J_{G_k}(z)^TJ_{G_k}(z))}\,\dd z}  &  z\in \mb B_1^d;\\
          0 &  o.w.
      \end{array}
      \right.
  \end{equation*}
  Then by the $\alpha$-smoothness of $\wt \rho_k(\cdot)$ and the uniformly lower boundness of $\sum_{k \in [K]}\wt \rho_k(\cdot)$, we have     $\nu_k(z)|_{\mb B_1^d}\in  C^{\alpha}_{L}(\mb B_1^d)$ for some constant $L$. On the other hand, by the second statement of Lemma~\ref{lemmaB.2} and the Lipschitzness of $G_{m}$, there exists a positive constant $\epsilon$ so that $Q_k(\m M\cap S_k)\subset \mb B_{1-\epsilon}^d$.  Combined with the fact that $\nu_k(z)=0$ when $z\notin Q_k(\m M\cap S_k)$, we can obtain $\nu_k(z)\in  C^{\alpha}_{L^*}(\mb R^d)$. In addition, for any $z\in \Omega_k=Q_k(\m M\cap S_k)$ and any $r>0$, we will show that there exists $z'\in \mb B_{r}(z)$ so that $\nu_k(z')\geq c\,(r^{\gamma}\wedge 1)$.
  Firstly if $\|G_k(z)-x_0\| \leq r^*/2$, then we have 
  \begin{equation*}
       \nu_k(z)\geq c_1\,\rho_k(G_k(z))\geq c_1\, \frac{(\frac{3r_k^2}{4})^{\gamma}}{M (r^*)^{2\gamma}}.
  \end{equation*}
 On the other hand,  if  $\|G_k(z)-G_k(0)\|=\|G_k(z)-x_0\|\geq r^*/2$,  denote $z=av$ with $v=z/\|z\|$ and $a=\|z\|$. Then we have $a\leq r_0$ and
\begin{equation*}
   r^*/2\leq  \|G_k(z)-G_k(0)\|\leq c_2\, \|z\|=c_2\,|a|. 
\end{equation*}
 If $r\geq a$, then $z=av\in \mb B_r(0)$ and $\nu_k(0)\geq c$ for some positive constant $c$. If $r<a$, choose $z'=(a-r)v$, then $z'\in \mb B_r(z)$ and 
\begin{equation*}
    \begin{aligned}
    \|G_k(z')-G_k(0)\|&=\underset{l\in \mb S_1^{d-1}}{\sup} \Big(l^TG_k(z')-l^TG_k(0)\Big)\\
    &=\underset{l\in \mb S_1^{d-1}}{\sup} \Big(l^TG_k(z)-l^TG_k(0)+l^TG_k(z')-l^TG_k(z)\Big)\\
    &\overset{(i)}{=}\underset{l\in \mb S_1^{d-1}}{\sup} \Big(l^TJ_{G_k}(z_l)av-l^TJ_{G_k}(z'_l)rv\Big)\\
    &=\underset{l\in \mb S_1^{d-1}}{\sup} \Big(l^TJ_{G_k}(z_l)(a-r)v+l^T(J_{G_k}(z_l)-J_{G_k}(z'_l))rv\Big)\\
    &\overset{(ii)}{\leq}\underset{l\in \mb S_1^{d-1}}{\sup} \Big(l^TJ_{G_k}(z_l)(a-r)v+\frac{r}{2a}\|G_k(z)-G_k(0)\|)\\
    &= \frac{a-r/2}{a}\|G_k(z)-G_k(0)\|\\
    &\leq r_k-\frac{rr_k}{2r_0},
    \end{aligned}
\end{equation*}
where $(i)$ uses mean-value theorem and $(ii)$ uses equation~\eqref{eqneigenlower} and Taylor's theorem to obtain
\begin{equation*}
\begin{aligned}
\frac{r}{2a} \|G_k(z)-G_k(0)\|&=\frac{r}{2}\Big\|\int_{0}^1 J_{G_k}(tz)\,\dd t \cdot v\Big\|\\
&\geq \frac{r}{2}\| J_{G_k}(0) \cdot v\|- \frac{r}{2}\Big\|\int_{0}^1 J_{G_k}(0)-J_{G_k}(tz)\,\dd t\cdot  v\Big\|\\
&\geq  \frac{3r}{2}\underset{z,z'\in \mb B_{r_0}^d} {\sup}\|(J_{G_{x_0}}(z)-J_{G_{x_0}}(z'))v\|- \frac{r}{2}\int_{0}^1 \|(J_{G_k}(0)-J_{G_k}(tz))v\|\,\dd t 
\\
&\geq r \underset{z,z'\in \mb B_{r_0}^d} {\sup}\|(J_{G_{x_0}}(z)-J_{G_{x_0}}(z'))v\|\\
&\geq  \underset{l\in \mb S_1^{d-1}}{\sup} l^T(J_{G_k}(z_l)-J_{G_k}(z'_l))rv.\\
\end{aligned}
\end{equation*}
So we have 
 \begin{equation}\label{eqn:rho_kG_k}
     \wt \rho_k(G_k(z'))= (r_k^2-\|G_k(z')-x_0\|^2)^{\gamma}\cdot  \geq  (\frac{rr_k^2}{2r_0})^{\gamma}\geq (\frac{r_k^2}{2r_0})^{\gamma}r^{\gamma}\geq (\frac{(L_1^*)^2}{2r_0})^{\gamma}r^{\gamma}.
 \end{equation}
Thus there exists constant $c$ so that 
\begin{equation*}
    \nu_k(z')\geq c\, r^{\gamma}.
\end{equation*}
Therefore, we have Assumption A holds for $\mu^*$ with $G^*_k=G_k$, $Q^*_k=Q_k$ and $\nu^*_k=\nu_k$ with $k\in [K]$. For the second statement, note that by the $\beta$-smoothness of $G_k^*=G_k$, $Q_k^*=Q_k$ and the $\alpha$-smoothness of $\nu_k^*=\nu_k$, Assumption B trivially holds for approximation family $\m G=\m G_1$. Moreover,  consider
\begin{equation*}
    \ov \nu_k(z)=  \frac{\mu^*(G_k(z))\cdot \sqrt{{\rm det}(J_{G_k}(z)^TJ_{G_k}(z))}}{\int_{\mb B_1^d}\mu^*(G_k(z))\cdot \sqrt{{\rm det}(J_{G_k}(z)^TJ_{G_k}(z))}\,\dd z},\quad z\in \mb B_1^d.
\end{equation*}
Then we have $ \nu_k(z)=\frac{ \ov \nu_k(z)\cdot \rho_k(G_k(z))}{\mb{E}_{\ov\nu_k}[\rho_k(G_k(z))]}$, $\ov\nu_k(z)\in C^{\alpha}_L(\mb B_1^d)$ and ${\inf}_{z\in \mb B_1^d}\ov\nu_k(z)\geq L_3>0$. So there exists an $(\alpha+1)$-smooth invertible function $V_k:\mb B_1^d\to \mb B_1^d$ (see for example,~\citep{10.2307/2118564,Villani2009}) so that  $\nu_0={V_{m}}_{\#}\ov\nu_k$ and $\ov\nu_k={V_{m}^{-1}}_{\#}\nu_0$. Therefore,   $\m G_2$  suffices to model $\mu^*$. 
\quad\\

\noindent For the family $\m G_3$,  let $\ov V_k$ be an $(\alpha+1)$-smooth extension of   $V_k|_{\mb B_{1-\epsilon/2}^d}$ to $\mb R^d$. Note that  $V_k$ has $(\alpha+1)$-smooth inverse and $V_k(\mb B_{1-\epsilon/2}^d)\subset \mb B_{1-\epsilon_1}^d $ for some positive constant $\epsilon_1$.  We can consider $\ov{ V^{-1}_k}$ as an $\alpha$-smooth extension of  $ V^{-1}_k|_{V_k(\mb B_{1-\epsilon/2}^d)}$ to $\mb R^d$. Then we can define $G_k'=G_k\circ \ov{V_k^{-1}}$ and $Q_k'=\ov{V}_k\circ Q_k$, by the fact that  $Q_k(\m M\cap S_k)\subset \mb B_{1-\epsilon}^d$, we have for any $x\in \m M\cap S_k$, $G_k'(Q_k'(x))=x$. Moreover, let 
\begin{equation*}
    \nu'_k(z)=(Q'_k)_{\#}\frac{\mu^*\rho_k}{p_k}=\left\{
    \begin{array}{cc}
    \frac{\nu_0(z)\cdot \rho_k(G_k\circ V_k^{-1}(z))}{\int_{\mb B_1^d}\nu_0(z)\cdot \rho_k(G_k\circ V_k^{-1}(z))\,\dd z},     & z\in  V_k(\mb B_{1-\epsilon/2}^d), \\
      0,   & o.w.
    \end{array}
    \right.
\end{equation*}
  Using the fact that $V_k^{-1}|_{V_k(\mb B_{1-\epsilon/2}^d)}$ is $(\alpha+1)$-smooth with bounded H\"{o}lder norm and $\nu_k'(z)=0$ when $z\notin  V_k(\mb B_{1-\epsilon}^d)$, we have $\nu'_k\in C^{\alpha}_L(\mb R^d)$ for some constant $L$. In addition, recall that for any $z_0\in Q_k(\m M\cap S_k)$ and $r>0$, there exists $z'_0\in Q_k(\m M\cap S_k)$ so that  $z_0\in \mb B_r(z'_0)$ and $\wt\rho_k(G_k(z_0'))\geq c_1 (r^{\gamma}\wedge 1)$. Note that by the Lipschitzness of $V_k$, there exists a constant $L_4\geq 1$ so that
  \begin{equation*}
      \|V_k(z_0)-V_k(z'_0)\|\leq L_4\|z_0-z_0'\|.
  \end{equation*}
  Therefore, for any $z\in V_k(Q_k(\m M\cap S_k))$ and $r>0$, there exists $z'\in V_k(Q_k(\m M\cap S_k)) \cap \mb B_r(z)$, so that $\wt\rho_k(G_k\circ V_k^{-1}(z'))\geq \frac{c_1}{L_4^{\gamma}} (r^{\gamma}\wedge 1)$ and $\nu_k'(z)\geq c\,  (r^{\gamma}\wedge 1)$. Therefore, when $\alpha=\beta-1$. Assumption A holds with $G_k^*=G_k'$,  $Q_k^*=Q_k'$  and $\nu_k^*= \nu'_k$, and  Assumption B holds with $\m G=\m G_3$.
 \subsection{Proof of Lemma~\ref{lemmapou}}

Let $\m N_{\epsilon}$ be the minimal $\epsilon$-covering set of $\m M$, where $\epsilon$ is a number that will be chosen later, then 
by Lemma~\ref{lemmaB.2} and the compactness of $\m M$, we have $|\m N_{\epsilon}|\leq C_1\, (\frac{1}{\epsilon})^d$ where $C_1$ is a positive constant that only depends on $(d,D,\beta,L^*)$. Then if $\epsilon\leq \tau_1$, by Lemma~\ref{lemmaB.2}, we have for any $x_0\in \m M$
\begin{equation*}
    \m P_{\mu^*}(\mb B_{\epsilon}(x_0))=\int_{  Q_{x_0}(\mb B_{\epsilon}(x_0))}\mu^*(G_{x_0}(z))\sqrt{{\rm det}(J_{G_{x_0}}(z)^TJ_{G_{x_0}}(z))}\,\dd z\geq C_2\, \epsilon^d.
\end{equation*}
Then, by Bernstein's inequality and a simple union bound argument, it holds with probability at least $1-n_1^{-c}$ that for any $x_0\in \m N_{\epsilon}$,
\begin{equation*}
    \Big|\frac{1}{n_1}\sum_{i\in I_1}\bold{1}(\|X_i-x_0\|\leq \epsilon)- \m P_{\mu^*}(\mb B_{\epsilon}(x_0))\Big|\leq \frac{1}{3n_1}\log(\delta)+\sqrt{\frac{2C_2\epsilon^d\log (\delta)}{n_1}},\quad \delta=2C_1n_1^c(\frac{1}{\epsilon})^d.
\end{equation*}
Therefore, there exists a constant $C_3,C$ so that when $n_1\geq C$, by choosing $\epsilon=C_3\,(\frac{\log n_1}{n_1})^{\frac{1}{d}}$, we have it holds with probability at least $1-n_1^{-c}$ that for any $x_0\in \m N_{\epsilon}$,
\begin{equation*}
    \Big|\frac{1}{n_1}\sum_{i\in I_1}\bold{1}(\|X_i-x_0\|\leq \epsilon)- \m P_{\mu^*}(\mb B_{\epsilon}(x_0))\Big|\leq \frac{C_2}{2}\epsilon^d\leq \frac{1}{2}\m  P_{\mu^*}(\mb B_{\epsilon}(x_0)).
\end{equation*}
Therefore, for any $x_0\in \m N_{\epsilon}$, there exists $i\in I_1$ so that $\|X_i-x_0\|\leq \epsilon$.  We can then obtain that for any $x\in \m M$, there exists $i\in I_1$ so that $\|X_i-x_0\|\leq 2\epsilon$. Proof of the first statement is then completed. For the second statement, when $n_1$ is large enough, we have $\epsilon=C_3\,(\frac{\log n_1}{n_1})^{\frac{1}{d}}\leq \frac{r^*}{16}$.  Let $\wt{\m N}_{r^*/4}$ denote the minimal $r^*/4$-covering  set of $\bigcup_{i\in I_1}\mb B_{2\epsilon}(X_i)$.  Then $|\wt{\m N}_{r^*/4}|$ is controlled by the  minimal $r^*/8$-covering number of $\m M$.  For any $x_0\in |\wt{\m N}_{r^*/4}|$, there exists an index $i\in I_1$ so that $\mb B_{2\epsilon}(X_{i})\cap \mb B_{r^*/4}(x_0)\neq\emptyset$. Let $I_2$ be the set of such index  $i$ for $x_0\in |\wt{\m N}_{r^*/4}|$. Then for any $x\in\bigcup_{i\in I_1}\mb B_{2\epsilon}(X_i)$,  there exists $i\in I_2$ so that
\begin{equation*}
    \|x-X_i\|\leq r^*/4+r^*/4+2\epsilon\leq \frac{5r^*}{8}.
\end{equation*}
Therefore, set $M=|I_2|$ and $\{a_k\}_{k=1}^K=\{X_i\}_{i\in I_2}$, we have 
\begin{equation*}
    \bigcup_{i\in I_1}\mb B_{2\epsilon}(X_i)\subset  \bigcup_{k\in [K]}\mb B_{r^*}(a_k),
\end{equation*}
and 
\begin{equation*}
     \inf_{x\in \m M}\sum_{k\in [K]}\wt \rho_k(x)\geq  ((r^*)^2-(\frac{5r^*}{8})^2)^{\gamma}> (\frac{(r^*)^2}{2})^{\gamma}.
\end{equation*}
Proof is completed.
 \end{document}